\documentclass[dvipsnames, 12pt]{article}
\usepackage[margin=25mm]{geometry}
\usepackage{amsmath}\allowdisplaybreaks
\usepackage{amsfonts}
\usepackage{amssymb}
\usepackage{verbatim}
\usepackage{setspace}
\usepackage{rotating}
\usepackage{booktabs}
\usepackage{longtable}
\usepackage{tabularx}
\usepackage[x11names]{xcolor}
\usepackage{subcaption}
    \usepackage[flushleft]{threeparttable}
\usepackage{colortbl}
\usepackage{svg}
\usepackage[section]{placeins}
\usepackage{float}

\usepackage{etoolbox}\AtBeginEnvironment{thebibliography}{\linespread{1}\selectfont}
\newcommand{\zerodisplayskips}{%
  \setlength{\abovedisplayskip}{-5pt}%
  \setlength{\belowdisplayskip}{5pt}%
  \setlength{\abovedisplayshortskip}{-5pt}%
  \setlength{\belowdisplayshortskip}{5pt}}
\appto{\normalsize}{\zerodisplayskips}
\appto{\small}{\zerodisplayskips}
\appto{\footnotesize}{\zerodisplayskips}
\usepackage{tikz}
\usetikzlibrary{fit, arrows.meta, positioning}
\usetikzlibrary{shapes.geometric, arrows, decorations.pathreplacing,calc}

\tikzstyle{circleblock} = [circle, draw, minimum size=2.5cm, text centered]
\tikzstyle{rectangleblock} = [rectangle, draw, minimum width=4.5cm, minimum height=1.5cm, text centered]
\tikzstyle{smallrectangle} = [rectangle, draw, minimum width=3cm, minimum height=0.8cm, text centered]
\tikzstyle{arrow} = [thick,->,>=stealth]

\usepackage{amsthm}
\usepackage{bm}
\usepackage{dsfont}
\usepackage{lscape}
\usepackage{geometry}
\usepackage{float}
\usepackage{verbatim}
\usepackage[toc,page,header]{appendix}
\usepackage{minitoc}

\usepackage{dsfont}
\usepackage{tablefootnote}
\usepackage{multirow}
\usepackage{mathtools}
\usepackage{pdflscape}

\usepackage{algorithm2e}

\usepackage{hyperref}
\hypersetup{
    colorlinks=true,
    linkcolor=cyan,
    filecolor=cyan,      
    urlcolor=cyan,
    citecolor=cyan,
    breaklinks=true}

\usepackage[square,numbers]{natbib}
\usepackage{authblk}
\usepackage{caption}
\usepackage{framed}
 
\usepackage{listings}

\usepackage{tcolorbox}
\usepackage{varwidth}

\begin{document}
\doparttoc % Tell to minitoc to generate a toc for the parts
%\faketableofcontents % Run a fake tableofcontents command for the partocs

\title{PoSSUM: A \textbf{P}r\textbf{o}tocol for \textbf{S}urveying \textbf{S}ocial-media \textbf{U}sers with \textbf{M}ultimodal LLMs}

\author{Roberto Cerina \\\texttt{r.cerina@uva.nl}\\Institute for Logic, Language and Computation\\University of Amsterdam}

\date{}
\maketitle
\thispagestyle{empty}

\vspace{-30pt}

\begin{centering}
    \begin{tcolorbox}[hbox,colback=white, colframe=black, boxsep=0pt, boxrule=1pt, arc=0pt, outer arc=0pt, auto outer arc, left=0pt, right=0pt, top=0pt, bottom=0pt]
    \includegraphics[width = 200pt]{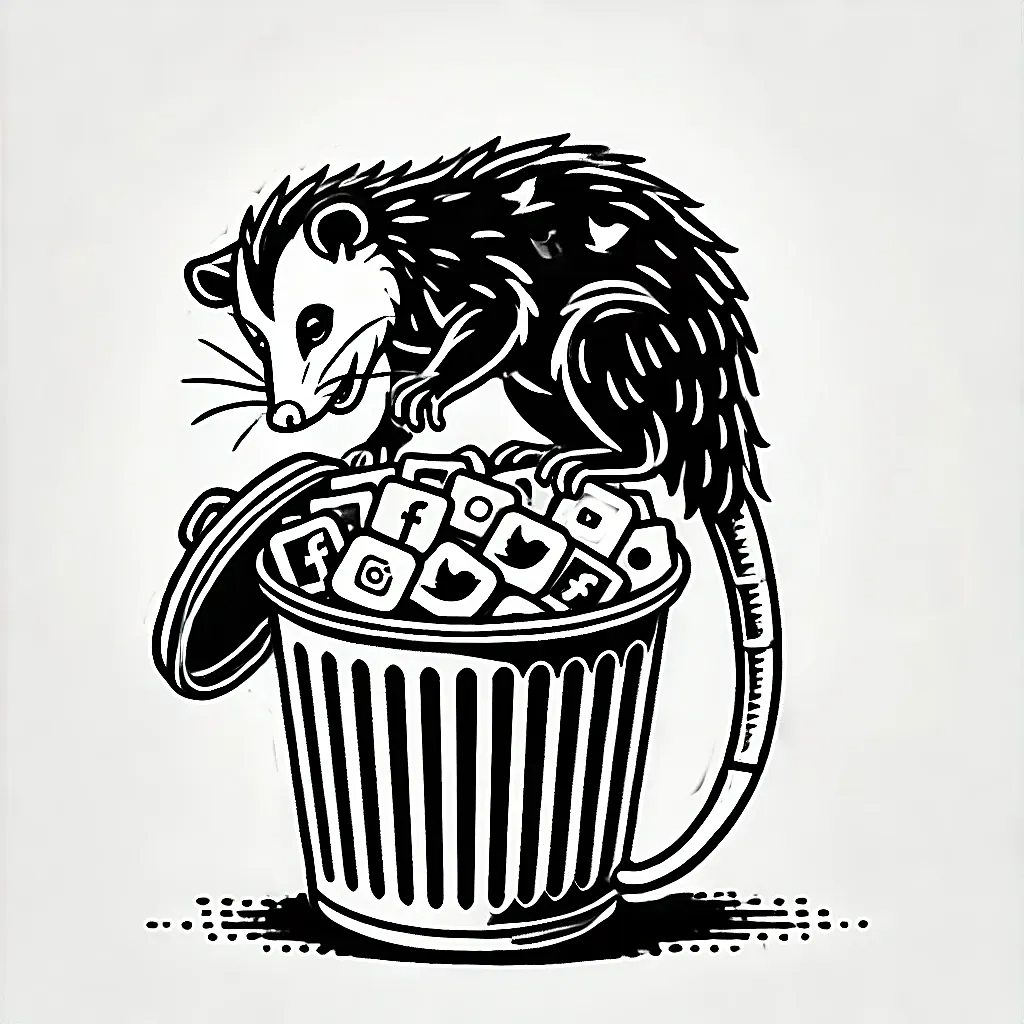}
    \end{tcolorbox}
\end{centering}

%\begin{abstract}
%\noindent 
%\end{abstract}

\begin{abstract}
\noindent
This paper introduces \texttt{PoSSUM}, an open-source protocol for unobtrusive polling of social-media users via multimodal Large Language Models (LLMs). \texttt{PoSSUM} leverages users' real-time posts, images, and other digital traces to create silicon samples that capture information not present in the LLM's training data. To obtain representative estimates, \texttt{PoSSUM} employs Multilevel Regression and Post-Stratification (MrP) with structured priors to counteract the observable selection biases of social-media platforms. The protocol is validated during the 2024 U.S.\ Presidential Election, for which five \texttt{PoSSUM} polls were conducted and published on GitHub and \(\mathbb{X}\). In the final poll, fielded October~17--26 with a synthetic sample of 1,054 \(\mathbb{X}\) users, \texttt{PoSSUM} accurately predicted the outcomes in 50 of 51 states and assigned the Republican candidate a win probability of 0.65. Notably, it also exhibited lower state-level bias than most established pollsters. These results demonstrate \texttt{PoSSUM}’s potential as a fully automated, unobtrusive alternative to traditional survey methods.
\end{abstract}

\section*{Acknowledgements}
The development of this protocol was partially funded by the \emph{Talking to Machines} initiative (\url{https://talkingtomachines.org}) at Nuffield College, University of Oxford. I am grateful to Prof. Raymond Duch for believing in this project and supporting my efforts to realise it.

\newpage
\pagenumbering{arabic} 
\section{Introduction}\label{intro}
\begin{figure}[htp!]
    \centering
\begin{tikzpicture}[
    node distance=3cm,
    box/.style={
        rectangle,
        draw,
        minimum height=1cm,
        minimum width=2cm,
        align=center
    },
    >={Stealth[round]},
    auto
]

% Nodes
\node (mould) [box] {Mould};
\node (prompt) [box, right of=mould] {Prompt};
\node (llm) [box, right of=prompt] {LLM};
\node (dtwin) [box, right of=llm] {Synthetic\\Response};

% Arrows
\draw[->] (mould) -> (prompt);
\draw[->] (prompt) -> (llm);
\draw[->] (llm) -> (dtwin);

\end{tikzpicture}
    \caption{A conceptual description of silicon sampling.}
    \label{fig::silicon}
\end{figure}
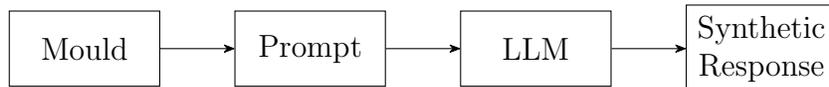

This article describes \texttt{PoSSUM}, an open source\footnote{ \url{https://github.com/robertocerinaprojects/PoSSUM}} protocol to poll social-media users unobtrusively using multimodal Large Language Models (LLMs). The protocol seeks to address the skepticism \cite{morris2024} surrounding Artificially Intelligent (AI) polling by establishing a methodology comparable to that used by traditional pollsters who use online panels \cite{twyman2008getting}. Concerns around AI polling are summarised aptly by this anonymous review to a related paper \cite{cerina2023artificially}: \textit{`... The goal of polling is quite simple: TO. LEARN. FROM. PEOPLE. NOW. I believe this model here does not learn, not from people, and not now'.} Three necessary conditions emerge from this animated critique -- to be a credible alternative to random digit dial (rdd) or self-selected online panels, Silicon samples \cite{argyle2023out} must enable \textit{novel learning} -- i.e. must contain more information than the \textit{mould}\footnote{Silicon samples as per Argyle et al. \cite{argyle2023out} can be elicited from LLMs by using high-quality, real-life survey responses of humans as a \textit{mould}. See Figure \ref{fig::silicon} for a graphical illustration of silicon sampling.} on which they are based; must be \textit{human-aligned} -- i.e. conditional on the same generating process, they must produce a distribution of responses which matches that of humans; must be \textit{time-sensitive} -- i.e. we must be able to learn about changes in preferences and attitudes over time by studying these samples, and these changes should be reflective of true societal dynamics, rather than artifacts of data engineering.\\

\noindent \texttt{PoSSUM} proposes to poll the public by inferring attitudes and preferences of real-life social-media users with multimodal LLMs. A key innovation of \texttt{PoSSUM} entails the use of real-time unstructured digital-trace data to inform a \emph{mould}. The unstructured nature of the data, in its free-flowing and unobtrusivly observable  nature, provides a much richer compendium of measurable and non-measurable information than the sterile tabular socio-demographic data that has come to define the literature \cite{argyle2023out,sanders2023demonstrations,bisbee2023synthetic}. LLMs can pick up on textual or visual cues that are not easily observable for humans \cite{wei2022emergent}. Importantly, the underlying unstructured data forms a unique digital footprint for a human existing in the world, hence allowing \texttt{PoSSUM} to allocate a unique mould to each individual whose preferences and attitudes we seek to simulate. Others have recently picked up on the potential for informing agents with large unstructured data generated by real human beings (e.g. interview transcripts \cite{park2024generative} ), but \texttt{PoSSUM} remains alone, at the time of writing, in having these unique moulds update dynamically over time as subjects offer up new text, images, video and other media to the web. This dynamic, time-sensitive mould contributes to solving issues around degradation of simulated agents as we move further away in time from the point at which the mould was created.\\

\noindent A second innovation pertains the generation of representative samples from unrepresentative pools of social media users \cite{alizadeh2024comparing}. Here too \texttt{PoSSUM} leverages the power of LLMs to digest unstructured user profiles and produce socio-demographic labels. This opens the door to traditional non-probability survey sampling techniques for social media users. I introduce a quota sampling module, where quotas are satisfied by matching the users' inferred characteristics to a stratification frame. Whilst quota sampling can itself produce unrepresentative samples under self-selection \cite{baker2013summary}, it is a valuable tool to partially curb platform selection effects and demographic imbalances \cite{pew2024socialmedia}. To further address selection and representation issues the protocol relies on Multilevel Regression and Post-Stratification (MrP) \cite{gelman1997poststratification,park2004bayesian} with structured priors \cite{gao2021improving} to analyse the resulting synthetic responses. \\

\noindent \texttt{PoSSUM}'s formulation therefore is set-up to satisfy the three necessary conditions outlined above. It enables \emph{novel learning}, in that the LLMs help us infer previously unknowable tabular survey responses from the unstructured flow of social media data for each user. The temporally updating mould addresses the \emph{time-sensitivity} of learning, clearly enabling the study of preferences changing over time -- \texttt{PoSSUM} can learn from synthetic panel data. Finally, efforts to tackle selection and representation via matching individuals to a stratification frame address \emph{human alignment}, under the condition that the LLMs can faithfully infer the distribution of social media users' preferences from their unstructured data\footnote{LLMs have been shown to be at least as capable as humans on this specific task in previous work \cite{cerina2023artificially}.}.\\

\noindent The central argument of this paper is that artificially intelligent polls generated using the \texttt{PoSSUM} protocol are valid instruments for measuring public opinion, on par with traditional polling methods. To test this proposition, \texttt{PoSSUM} is deployed during the $2024$ US Presidential election -- a setting that allows assessing LLMs' ability to generalise understandings of political preferences beyond their training data. In this study I employ \texttt{gpt-4o-2024-05-13}, whose training concluded in October $2023$. Hence the model faces a novel candidate choice-set and unforeseen demographic realignments, characteristic of the $2024$ election. Data leakage type criticisms do not apply to this paper as a result.\\

\noindent The paper is organized as follows. Section~\ref{sec:protocol} presents an overview of the \texttt{PoSSUM} protocol. Section~\ref{sec:get_pool} describes the \texttt{get\_pool} routine, which acquires an initial subject pool of social media users. Section~\ref{sec::prompting} introduces the modular prompting architecture employed throughout the protocol. Section~\ref{sec::filter} reviews the LLM-enabled filtering procedure that selects statistically informative users from the subject pool. Section~\ref{sec:inference} details the implementation of MrP with structured priors tailored to this framework. Section~\ref{sec:results} delineates the evaluation criteria for assessing \texttt{PoSSUM}’s performance, including a multi-dimensional appraisal of predictive accuracy, capacity for novel learning, degree of human alignment, and sensitivity to temporal shifts. Finally, Section~\ref{sec:discussion} provides a comprehensive discussion of key insights arising from the \texttt{PoSSUM} experiment, the limitations of the protocol, and directions for future research.

%\section{Establishing Performance Benchmarks}

%A challenge for any novel methodology to measure public opinion is to benchmark performance. At the individual level, it is generally impossible to measure the degree of agreement between the respondent's inferred characteristics (obtained by self-report or otherwise) and their true underlying attributes, which are typically unobservable. At the aggregate level, we lack a true measures of aggregate preferences, absent these having been collected as part of a census-type study. It is therefore difficult to find real-life applications to public opinion estimation which enable objective verification of performance. 

%Elections are a notable exception -- a rare case of mass collection of revealed preferences at the aggregate level. 
%The $2024$ US Presidential Election contest makes for a particularly comprehensive case study to test a methodology which uses LLMs to infer revealed preferences from digital trace data. Changes in the field of candidates since $2020$ (including seemingly consequential third-party candidates), as well as 

%make for a substantial challenge to LLMs' ability to generalise understandings of political preferences outside the training data; ii. temporal trends in the shape of public opinion over the last $2$ months of the election test the ability of the data collection and analysis protocol to study of opinion dynamics over time; iii. 

\pagebreak

\section{Protocol}\label{sec:protocol}

\noindent The goal of \texttt{PoSSUM} is to make granular inference about preferences and attitudes, for a given digital fieldwork period, which is representative of the true underlying population. What I describe is an approach tailored to the $\mathbb{X}$ API, which uses the digital trace of $\mathbb{X}$ users as the mould for LLM generation, but can be extended to any social-media which allows querying of a user panel via user- and content-level queries.\\

\noindent The protocol unfolds as follows: i) \emph{Generating a subject pool} -- For each period of interest \texttt{PoSSUM} generates a subject pool of $\mathbb{X}$ users. ii) \emph{Sampling desirable respondents} -- Iteratively, each user in the pool is screened for desirable characteristics. Users who do not pass data quality checks, or are unlikely to provide new information to the desired distribution of preferences, are discarded. iii) \emph{Augmenting the mould} -- For the surviving users, further information (e.g. their history of tweets) from their $\mathbb{X}$ timeline is elicited. The information is appended to the existing record of the subject, and compiled into a user-specific \emph{mould} -- an object containing unstructured multi-media data generated by the user on the social media platform. iv) \emph{Feature extraction} -- The mould is passed to a LLM in the form of a prompt. The LLM is given instructions to deduce a set of characteristics for each user according to their mould. The resulting LLM output is structured into tabular data. v) \emph{Hierarchical Bayes} --  The synthetic tabular data, representing the hypothetical responses to survey questions from the real-life set of individuals who are active on $\mathbb{X}$, is then analysed via Hierarchical Bayesian modeling to account for observable selection and address LLM biases. vi) \emph{Post-stratification} -- The predictions of the Bayesian multilevel model are post-stratified to generate estimates of the distribution of preferences at the national, state and crosstab levels of analysis. \\

\begin{figure}[ht!]
    \centering
\begin{tikzpicture}

% Nodes
    \node [circle, draw, inner sep=2pt] (X) {$\mathbb{X}$};

    \node [below of=X, node distance=0.75cm, align=center,inner sep=0pt] (query_tweets) {\scriptsize \texttt{tweets/search/recent}};

    \node [rectangle, draw, right of=query_tweets, node distance=2.75cm, align=center] (pool) {\scriptsize Subject \\[-1ex] \scriptsize Pool};

    \node [below of=pool, node distance=1.15cm, align=center, inner xsep=10pt, inner ysep=0pt] (temporal) {\scriptsize \hspace{-8pt}\texttt{temporal\_filter}};
    
    \node [below of=temporal, node distance=.5cm, align=center,inner sep=0pt] (geographic) {\scriptsize \hspace{-16.5pt}\texttt{geographic\_filter}};

    \node [below of=geographic, node distance=.5cm, align=center,inner sep=0pt] (entity) {\scriptsize \texttt{entity\_filter}};
        
    \node [below of=entity, node distance=.5cm, align=center,inner sep=0pt] (quota) {\scriptsize \hspace{2pt}\texttt{quota\_filter}};
    
    \node [rectangle, draw, below of=quota, node distance=1cm, align=center] (sample) {
    \scriptsize Sample\\[-1ex] \scriptsize (\emph{Basic Mould})};

    \node [below of=sample, node distance=0.8cm, align=center,inner sep=0pt] (query_user) {\scriptsize \texttt{users/\{user\_id\}/tweets}};

    \node [rectangle, draw, below of=query_user, node distance=1.05cm, align=center] (sample+) {\scriptsize Sample\\[-1ex] \scriptsize (\emph{Augmented Mould})};

    \node [right of=query_user, node distance=3.5cm, align=center] (empty) { \\ };

    \node [rectangle, draw, below of=empty, node distance=0.9cm, align=center] (prompt) {\scriptsize  \hspace{10pt} \emph{Prompt} \hspace{10pt} \\ \\ \\ };
    
    \node [rectangle, draw, right of=sample+, node distance=3.5cm, align=center] (mould) {\scriptsize \hspace{7.35pt} Mould \hspace{7.35pt}};

    \node [rectangle, draw, above of= mould, node distance=0.5cm, align=center] (context) { \scriptsize \hspace*{-0.95pt}Background \hspace*{-1.95pt}};

    \node [rectangle, draw, below of=mould, node distance=0.485cm, align=center] (instructions) { \scriptsize Instructions};

    \node [circle, draw, below of=prompt, node distance = 1.65cm, align=center, inner sep=1pt] (LLM) {\hspace{-2pt}\includegraphics[width = 0.5cm]{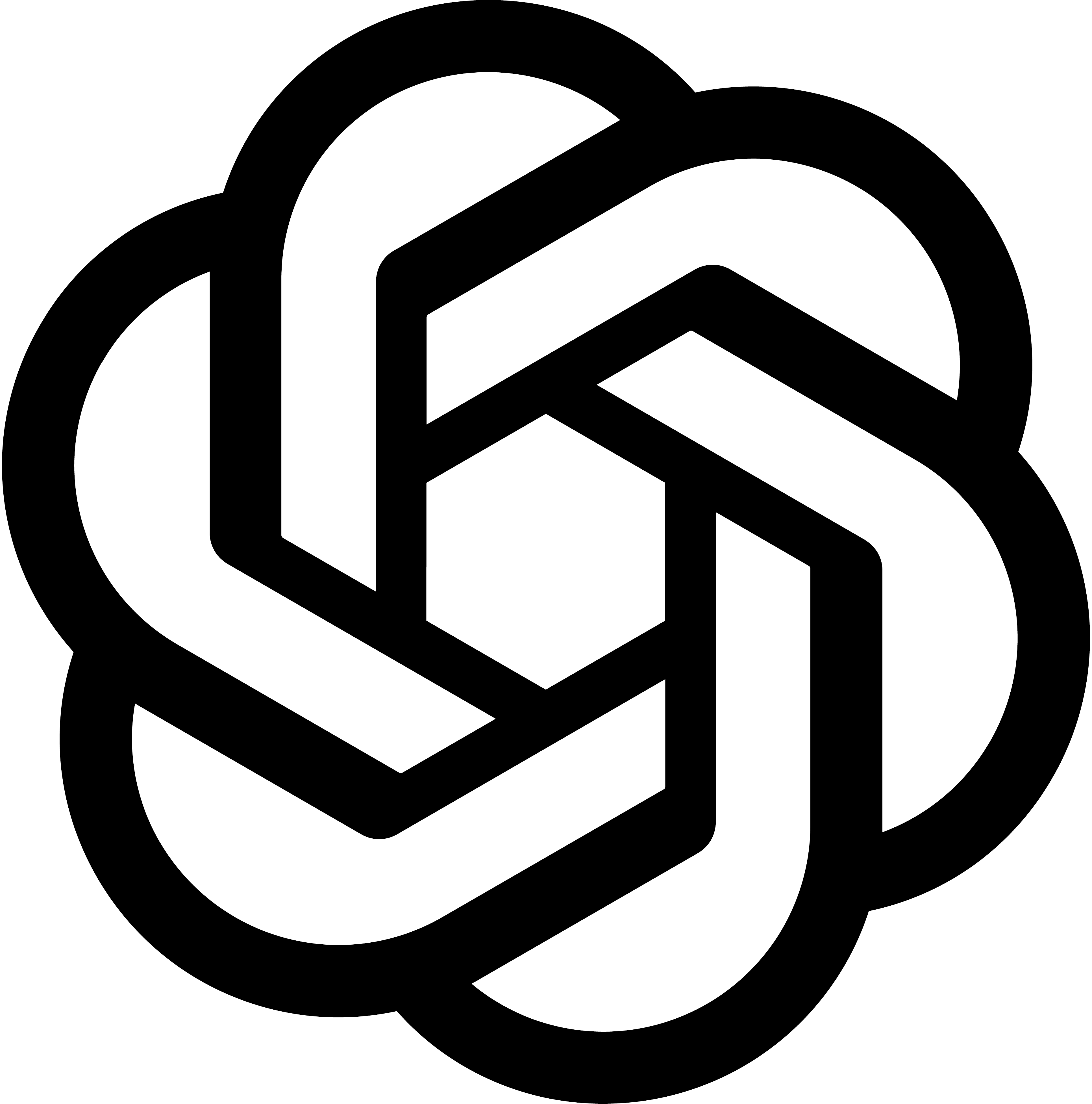} \hspace{-10pt} };
    
    \node [rectangle, draw, below of=LLM, node distance = 1.25cm, align=center] (response) { \scriptsize Synthetic \\[-1ex] \scriptsize Survey \\[-1ex] \scriptsize Data};

    % Decision node
    %\node [below of=response, node distance=1.5cm, align=center] (decision) {\scriptsize $n = n^{\star}$ ?};

    \node [circle, draw, right of=quota, node distance = 5.75cm,  align=center, inner sep=2pt] (MrP) { \scriptsize \texttt{MrP}};

    \node [rectangle, draw, right of=MrP, node distance = 1.75cm,  align=center] (inference) { \scriptsize Representative \\[-1ex] \scriptsize Estimates};

    \node [rectangle, draw, left of=MrP, node distance = 2.25cm,  align=center] (frame) { \scriptsize Stratification \\[-1ex] \scriptsize Frame};

    \node [rectangle, draw, above of=MrP, node distance = 1.1cm,  align=center] (prior) { \scriptsize Prior \\[-1ex] \scriptsize Structure};

    \node [draw, rectangle, dotted, fit=(X)(query_tweets)(pool), inner sep=0.5em, label={[xshift=-3.25em, yshift=0em]north east:\scriptsize \texttt{get\_pool}}] {};

\node [left of=sample, node distance=2.25cm, align=center] (empty2) { \\ };

    \node [draw, rectangle, dotted, fit=(temporal)(geographic)(entity)(quota)(sample)(query_user)(sample+)(prompt)(context)(mould)(instructions)(LLM)(response)(empty2)(frame), inner sep=0.5em, label={[xshift=-4em, yshift=0em]north east:\scriptsize \texttt{poll\_users}}] {};
       
    \node [draw, rectangle, dotted, fit=(MrP)(frame)(prior)(inference), inner sep=0.5em, label={[xshift=-5.5em, yshift=0em]north east:\scriptsize \texttt{make\_inference}}] {};
    
% Arrows
\draw[-] (X) -- (query_tweets);
\draw[->] (query_tweets) -- (pool);

\draw[-] (pool) -- (temporal); % Mixed segments: horizontal and vertical
\draw[-] (temporal) -- (geographic);
\draw[-] (geographic) -- (entity);
\draw[-] (entity) -- (quota);
\draw[->] (quota) -- (sample);
\draw[-] (sample) -- (query_user);
\draw[->] (query_user) -- (sample+);
\draw[->] (sample+) -- (mould); % Mixed segments: horizontal and vertical
\draw[-] (prompt) -- (LLM);
\draw[->] (LLM) -- (response);
%\draw[-] (response) -- (decision);
\draw[->] (MrP) -- (inference);
\draw[->] (frame) -- (MrP);
\draw[->] (prior) -- (MrP);

% Decision connections
%\draw[->] (decision.west) -- ++(-1cm,0) node[midway, above] {\scriptsize \texttt{No}} -| ([yshift=-6pt]pool.south);
\draw[->] (response.east) -- ++(0.8cm,0) -| (MrP.south);
\draw[->] (frame.west) |- (quota.east);

% Flipped curly bracket and OpenAI symbol
\draw [decorate, decoration={brace, mirror, amplitude=5pt}]
    ([shift={(-18pt,0)}] geographic.north west) -- ([shift={(-19pt,0)}] quota.south west)
    node [midway, left=5pt] {\includegraphics[width=0.5cm]{openai_2.pdf}};

% Labels
\end{tikzpicture}
    \caption{An overview of the \texttt{PoSSUM} protocol.}
    \label{fig:protocol}
\end{figure}

\noindent \texttt{PoSSUM} is therefore composed of three principal routines: \texttt{get\_pool} (Pseudo-code \ref{alg:get_pool}) is used to identify a subject pool according to keywords matching recent tweets; \texttt{poll\_users} (Pseudo-code \ref{alg:poll_users}) is designed to implement a series of inclusion checks on the users in the pool, and infer socio-demographics, attitudes and preferences based on their most recent activity on the platform;  \texttt{make\_inference} performs a MrP with structured priors on the synthetic survey data, and generates representative estimates of preferences and attitudes. Figure \ref{fig:protocol} provides and overview of the protocol\footnote{The protocol is implemented in \texttt{R}, leveraging the \texttt{openai} package \cite{rudnytskyi2023} to call the OpenAI api and prompt the \texttt{gpt-4o-2024-05-13} model \cite{openai2024gpt4o,openai2023gpt4}. Although new versions of the model were released during the course of the study, they were not fit for use to extract user characteristics from social media data. The new models use an instruction hierarchy \cite{wallace2024instruction} which flags certain model inputs as high-risk and pushes the model to decline following the instructions. This new model feature generated unacceptable rejection rates, and I therefore opted to keep working with the older model for every \texttt{PoSSUM} poll conducted during the $2024$ US election campaign.}.

\section{\texttt{get\_pool}}\label{sec:get_pool}

\noindent I wish to sample a maximally informative set of US adults amongst $\mathbb{X}$ users, to be part of a poll-specific subject pool, from which I can generate a representative sample. Given the time-sensitive nature of the inference I wish to make (what are people's attitudes \textbf{today} ?) I search for these users amongst those who are currently active on the platform. \\

\noindent The $\mathbb{X}$ API \emph{enterprise} and \emph{pro} tiers can be prohibitively expensive, hence I will assume users of \texttt{PoSSUM} have access to the \emph{basic} tier, and cannot collect a simple-random-sample of Tweets from a given day\footnote{I use the \emph{basic} tier $\mathbb{X}$ api, meaning I pay \$ $100$ for downloading $10k$ tweets per month. I can pay this price multiple times a month, and each payment allows another $10k$ tweets. The allowance is reset to the original $10k$ at the end of every month.}. Even if such sampling were possible, it may not be the default choice. As I demonstrate below, $\mathbb{X}$ API queries can be tuned to target specific sub-populations of active users, potentially mitigating platform-wide selection effects. I propose to use a combination of search queries for the \texttt{tweets/search/recent} endpoint\footnote{See \url{https://developer.x.com/en/docs/twitter-api/tweets/search/introduction} for more details.}, and obtain a series of users who have tweeted the search terms on the platform, up to seven days prior to query-time\footnote{I implement a set of functions, available in the \texttt{GitHub} repository, in the file \texttt{X.api.v2\_function.R}, reminiscent of the now-defunct \texttt{rtweet} \cite{kearney2019rtweet} and \texttt{academictwitteR} \cite{barrie2021academictwitter}, in order to specify an appropriate set of queries.}. \\

\noindent \textbf{High-Attention Subjects} \hspace{10pt} I need the content produced by the selected users to be informative of their political beliefs and attitudes. One way to ensure this is to use political search terms in the $\mathbb{X}$ query. To perform US $2024$ pre-election polling we could use a query such as that in Listing \ref{lst:search_query_politics}. Notice this is a joint query for all the candidates. This is preferable to independent queries per candidate, as these would yield estimates of support subject to selection on the dependent variable. The independent approach ignores the distribution of the search terms across tweets, and over-samples supporters of each candidate, distorting the distribution of support in favour of smaller parties\footnote{On the other hand, this sampling is very efficient per party -- if you have access to selection-correction terms in the style of King \& Zeng \cite{king2001logistic,cerina2023artificially}, this approach would allow for the most sampling-efficient analysis.}. I assign a \emph{weight} (maximum number of tweets extracted) to this query of size $\omega$. \\

\noindent \textbf{Low-Attention Subjects} \hspace{10pt} Individuals who talk about politics on $\mathbb{X}$ are still unlike their counterparts in the general population. In particular, these are high political attention individuals, who are significantly more likely to vote than their population counterparts. To alleviate selection on political-attention I rely on a second set of queries, which are more likely to sample \emph{normies}. I extract a random sample of $L$ \emph{trending topics} in the US (obtainable via \url{https://trends24.in/united-states/}), and produce a separate query for each topic. Each trending query is assigned a weight of $\frac{\omega}{L}$, such that the queries seeking to capture high-attention individuals and those capturing \emph{normies} are assigned the same weight. Note this is arbitrary -- I noticed this worked well in the US, but in general $\omega$ and $L$ are hyper-parameters that need tuning. I end up with a set of queries $\bm{q}$, which is an object of size $K = L+1$, and a corresponding set of weights $\bm{w} = \{\omega,\frac{\omega}{L},...,\frac{\omega}{L}\}$. \\

\noindent I execute each of these queries in a loop, and for each I obtain a tweet-user object $(\bm{\mathcal{T}},\bm{\upsilon})_{kt}$ containing at most $w_k$ tweets, and their associated $\mathbb{X}$ user profiles. %We are uninterested in the tweets at this stage, and temporarily store the unique user-ids responsible for the tweets. 
The result of \texttt{get\_pool} is a dated subject pool, containing the profile information about each user (e.g. self-reported description, location, profile picture, etc.), the date on which this user was included in the pool, the search-query used to capture them, and the set of query-related tweets which the user is responsible for. 

\newgeometry{left=0.75cm, right=2cm, top=2cm, bottom=1.5cm}  % Adjust the values as needed

\begin{landscape}
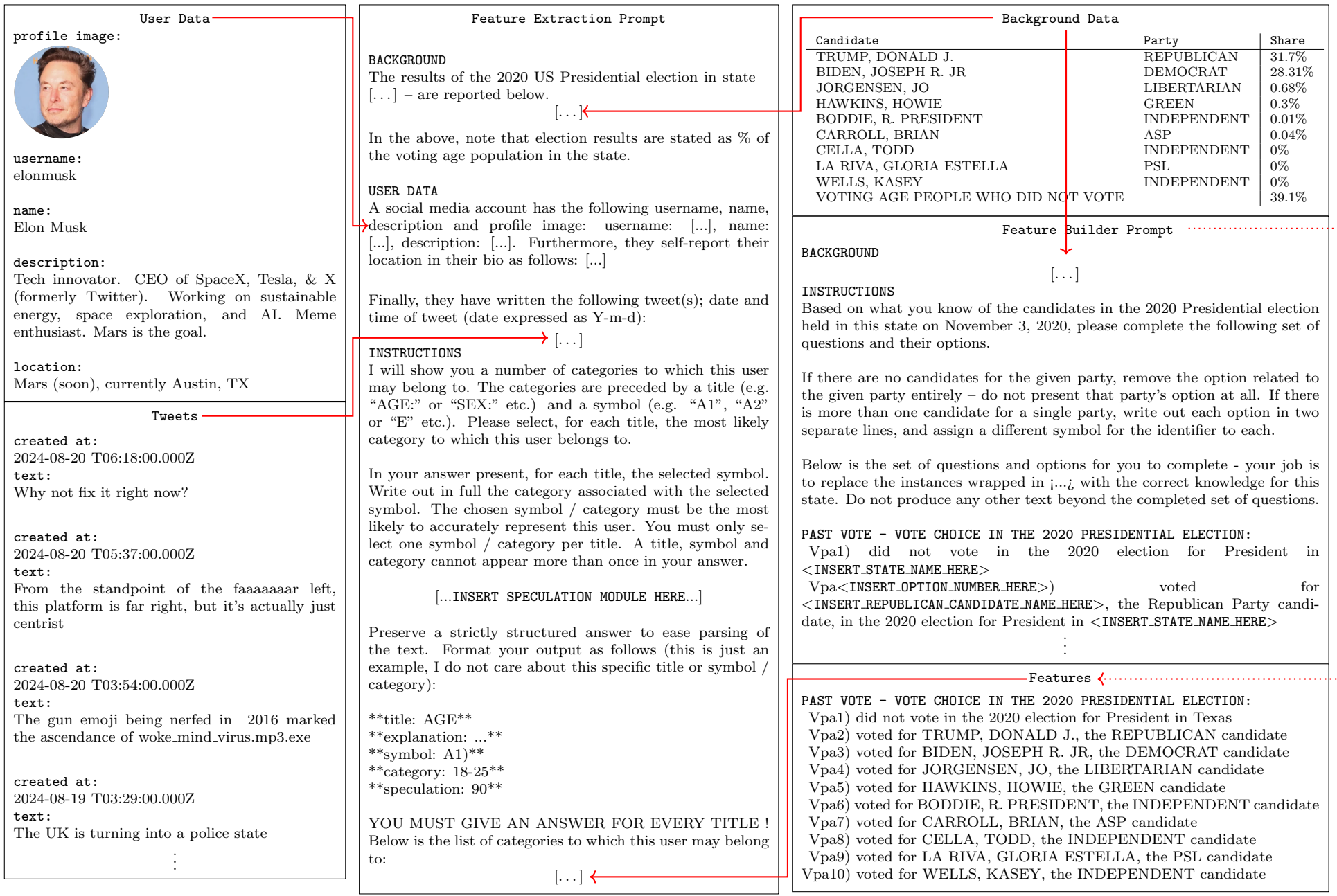
\begin{figure}
    \centering

\begin{tikzpicture}[font=\scriptsize]

% Overall width/height setup for three columns:
\def\totalwidth{27cm}
\def\totalheight{18cm}

% Number of columns and gaps:
\def\numcols{3}
\def\colgap{0.25cm}
\def\colwidthA{6.5cm} % Width of first column
\def\colwidthB{8cm}  % Width of second column
\def\colwidthC{10.25cm}    % Width of third column
% Vertical space allocations:
\def\headheight{6cm}      % height for user data area (Column 1 top)
\def\tweetsheight{6cm}    % height for tweets area (Column 1 bottom)
\def\promptheight{12cm}    % height for prompt (Column 2 top)
\def\backgroundheight{4cm}% height for background data (Column 2 bottom)
\def\featuresheight{4cm} % height for features data (Column 3 full column)

%%%%%%%%%%%%%%%%%%%%%%%%%
% COLUMN 1: User Data & Tweets
%%%%%%%%%%%%%%%%%%%%%%%%%

% User Data box (Column 1 top)
\node[draw,anchor=north west,minimum width=\colwidthA,minimum height=\headheight,inner sep=5pt] (userDataBox) at (0,0) {\begin{minipage}[t]{\colwidthA-1em}

\begin{centering}
    \textbf{\texttt{User Data}} \\
\end{centering} 

\texttt{profile image:}\\
% Generic user icon
\hspace*{-1.5cm}\vspace*{-0.9cm}\includegraphics[width = 5cm]{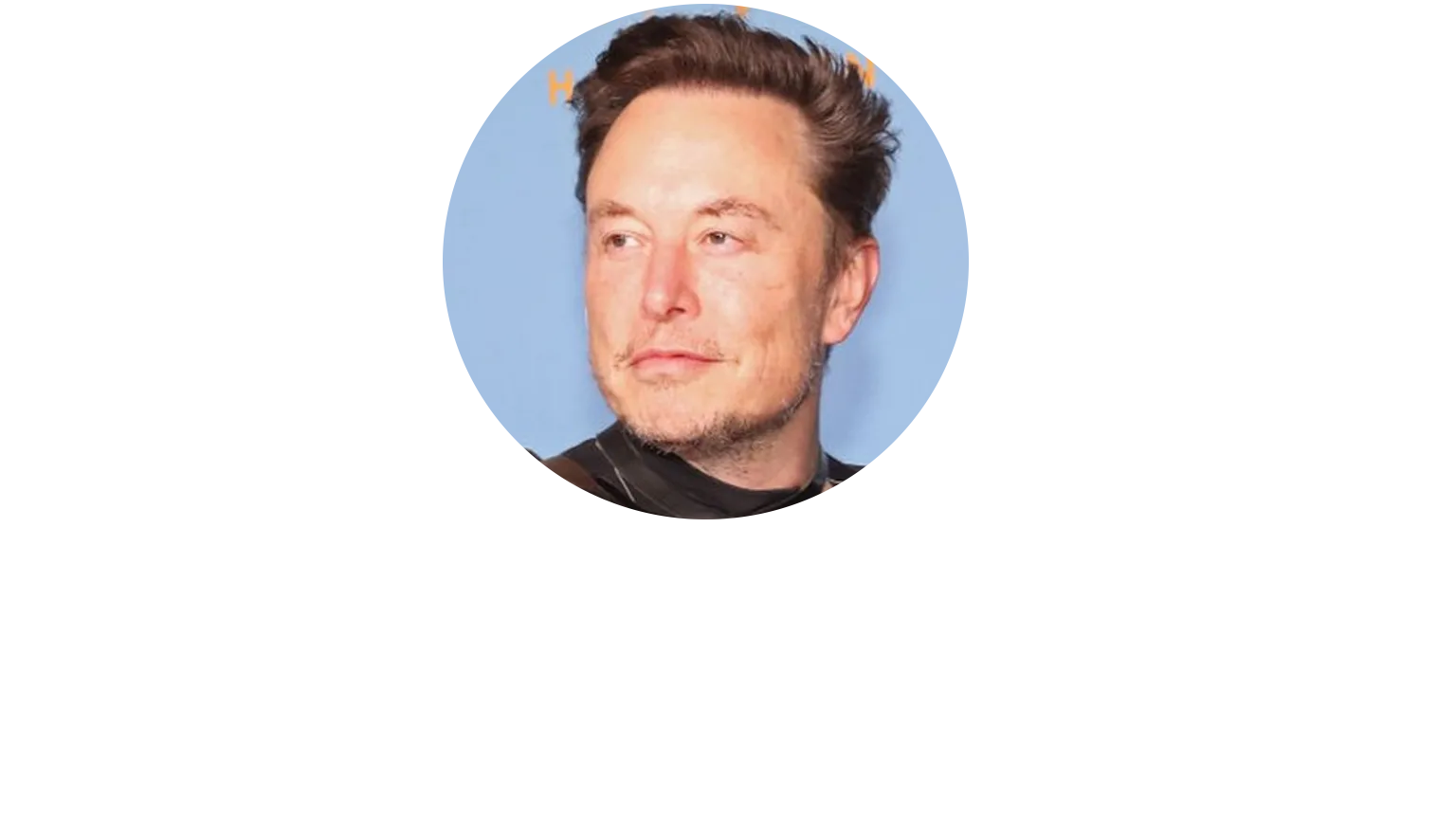}
\\
\texttt{username:}

 elonmusk \\

\texttt{name:}

Elon Musk \\

\texttt{description:}

Tech innovator. CEO of SpaceX, Tesla, \& X (formerly Twitter). Working on sustainable energy, space exploration, and AI. Meme enthusiast. Mars is the goal. \\

\texttt{location:}

Mars (soon), currently Austin, TX
\end{minipage}};

% Tweets box (Column 1 bottom), placed below userDataBox
\node[draw,anchor=north west,minimum width=\colwidthA,minimum height=\tweetsheight,inner sep=5pt] (tweetsBox) at (userDataBox.south west) {\begin{minipage}[t]{\colwidthA-1em}
\begin{centering}
    \textbf{\texttt{Tweets}} \\
\end{centering} 

\vspace{4.2pt}

\texttt{created at:}\\ 2024-08-20 T06:18:00.000Z\\
\texttt{text:}\\ Why not fix it right now? \\[5pt]

\texttt{created at:}\\ 2024-08-20 T05:37:00.000Z\\
\texttt{text:}\\ From the standpoint of the faaaaaaar left, this platform is far right, but it's actually just centrist \\[5pt]

\texttt{created at:}\\ 2024-08-20 T03:54:00.000Z\\
\texttt{text:}\\ The gun emoji being nerfed in ~2016 marked the ascendance of woke\_mind\_virus.mp3.exe \\[5pt]

\texttt{created at:}\\ 2024-08-19 T03:29:00.000Z\\
\texttt{text:}\\ The UK is turning into a police state 

\begin{centering}
    $\vdots$ \\
\end{centering} 

\end{minipage}};

%%%%%%%%%%%%%%%%%%%%%%%%%
% COLUMN 2: Prompt 
%%%%%%%%%%%%%%%%%%%%%%%%%

% Prompt box (Column 2 top)
\node[draw,anchor=north west,minimum width=\colwidthB,minimum height=\promptheight,inner sep=5pt]
(promptBox) at ($(userDataBox.north east) + (\colgap,0)$) {\begin{minipage}[t]{\colwidthB-1em}
\begin{center}
    \textbf{\texttt{Feature Extraction Prompt}}\\[3pt]
\end{center}
\texttt{BACKGROUND}\\
The results of the 2020 US Presidential election in state -- [\dots] -- are reported below.
\vspace{-10pt}
\begin{center}
    [\dots] 
\end{center}
\vspace{-5pt}
\noindent In the above, note that election results are stated as \% of the voting age population in the state.\\

\noindent\texttt{USER DATA}\\
A social media account has the following username, name, description and profile image: username: [...], name: [...], description: [...]. Furthermore, they self-report their location in their bio as follows:  [...]\\[3pt]

Finally, they have written the following tweet(s); date and time of tweet (date expressed as Y-m-d): 
\vspace{-7.5pt}
\begin{center}
    [\dots] 
\end{center}
\vspace{-12.5pt}

\texttt{INSTRUCTIONS}\\
I will show you a number of categories to which this user may belong to.
The categories are preceded by a title (e.g. ``AGE:'' or ``SEX:'' etc.) and a symbol (e.g. ``A1'', ``A2'' or ``E'' etc.).
Please select, for each title, the most likely category to which this user belongs to.\\

In your answer present, for each title, the selected symbol. Write out in full the category associated with the selected symbol. The chosen symbol / category must be the most likely to accurately represent this user. You must only select one symbol / category per title. A title, symbol and category cannot appear more than once in your answer.
\begin{center}
    [...\texttt{INSERT SPECULATION MODULE HERE}...] 
\end{center}
Preserve a strictly structured answer to ease parsing of the text. Format your output as follows (this is just an example, I do not care about this specific title or symbol / category):\\

**title: AGE**\\
**explanation: ...**\\
**symbol: A1)**\\
**category: 18-25**\\
**speculation: 90**\\

YOU MUST GIVE AN ANSWER FOR EVERY TITLE !
Below is the list of categories to which this user may belong to: 
\vspace{-9.7pt}
\begin{center}
    [\dots] 
\end{center}
\end{minipage}};

%%%%%%%%%%%%%%%%%%%%%%%%%
% COLUMN 3: Background and Features 
%%%%%%%%%%%%%%%%%%%%%%%%%

% Background Data box (Column 3top)
\node[draw,anchor=north west,minimum width=\colwidthC,minimum height=\backgroundheight,inner sep=5pt]
(bgDataBox) at ($(promptBox.north east) + (\colgap,0)$) {\begin{minipage}[t]{\colwidthC-1em}
\begin{center}
    \textbf{\texttt{Background Data}}
\end{center}
\vspace{-10pt}
\scalebox{0.9}{
\begin{tabular}{l l |l}
\texttt{Candidate} & \texttt{Party} & \texttt{Share} \\
\hline
TRUMP, DONALD J. & REPUBLICAN & 31.7\% \\
BIDEN, JOSEPH R. JR & DEMOCRAT & 28.31\% \\
JORGENSEN, JO & LIBERTARIAN & 0.68\% \\
HAWKINS, HOWIE & GREEN & 0.3\% \\
BODDIE, R. PRESIDENT & INDEPENDENT & 0.01\% \\
CARROLL, BRIAN & ASP & 0.04\% \\
CELLA, TODD & INDEPENDENT & 0\% \\
LA RIVA, GLORIA ESTELLA & PSL & 0\% \\
WELLS, KASEY & INDEPENDENT & 0\% \\
VOTING AGE PEOPLE WHO DID NOT VOTE & & 39.1\% \\
\end{tabular}
}
\end{minipage}};

% Features Data box (Column 3 full height)
\node[draw,anchor=north west,minimum width=\colwidthC,minimum height=\featuresheight,inner sep=5pt]
(featuresPrompt) at (bgDataBox.south west) 
{\begin{minipage}[t]{\colwidthC-1em}

\begin{center}
    \hspace*{30pt}\textbf{\texttt{Feature Builder Prompt}}\\
\end{center}
\vspace{-10pt}

\texttt{BACKGROUND}\\
\vspace{-17.6pt}
\begin{center}
    \hspace*{5pt}[\dots] 
\end{center}
\vspace{-10pt}
\texttt{INSTRUCTIONS}\\
Based on what you know of the candidates in the 2020 Presidential election held in this state on November 3, 2020, please complete the following set of questions and their options.\\

If there are no candidates for the given party, remove the option related to the given party entirely -- do not present that party's option at all.
If there is more than one candidate for a single party, write out each option in two separate lines, and assign a different symbol for the identifier to each.\\

Below is the set of questions and options for you to complete - your job is to replace the instances wrapped in <...> with the correct knowledge for this state.
Do not produce any other text beyond the completed set of questions.\\

\texttt{PAST VOTE - VOTE CHOICE IN THE 2020 PRESIDENTIAL ELECTION:}\\
\hspace*{4pt}Vpa1) did not vote in the 2020 election for President in \texttt{$<$INSERT\_STATE\_NAME\_HERE$>$}\\
\hspace*{4pt}Vpa\texttt{$<$INSERT\_OPTION\_NUMBER\_HERE$>$}) voted for \texttt{$<$INSERT\_REPUBLICAN\_CANDIDATE\_NAME\_HERE$>$}, the Republican Party candidate, in the 2020 election for President in \texttt{$<$INSERT\_STATE\_NAME\_HERE$>$}\\
\vspace{-20pt}
\begin{center}
    \hspace*{5pt}$\vdots$
\end{center}
\vspace{-10pt}

\end{minipage}};

% Features Data box (Column 3 full height)
\node[draw,anchor=north west,minimum width=\colwidthC,minimum height=\featuresheight,inner sep=5pt]
(featuresBox) at (featuresPrompt.south west) 
{\begin{minipage}[t]{\colwidthC-1em}

\begin{center}
    \textbf{\texttt{Features}}\\
\end{center}
\vspace{-10pt}
\texttt{PAST VOTE - VOTE CHOICE IN THE 2020 PRESIDENTIAL ELECTION:}\\
 \hspace*{4pt}Vpa1) did not vote in the 2020 election for President in Texas \\
 \hspace*{4pt}Vpa2) voted for TRUMP, DONALD J., the REPUBLICAN candidate \\
 \hspace*{4pt}Vpa3) voted for BIDEN, JOSEPH R. JR, the DEMOCRAT candidate \\
 \hspace*{4pt}Vpa4) voted for JORGENSEN, JO, the LIBERTARIAN candidate \\
 \hspace*{4pt}Vpa5) voted for HAWKINS, HOWIE, the GREEN candidate \\
 \hspace*{4pt}Vpa6) voted for BODDIE, R. PRESIDENT, the INDEPENDENT candidate \\
 \hspace*{4pt}Vpa7) voted for CARROLL, BRIAN, the ASP candidate \\
 \hspace*{4pt}Vpa8) voted for CELLA, TODD, the INDEPENDENT candidate \\
 \hspace*{4pt}Vpa9) voted for LA RIVA, GLORIA ESTELLA, the PSL candidate \\
Vpa10) voted for WELLS, KASEY, the INDEPENDENT candidate
\end{minipage}};

% Draw a thick red line connecting the User Data box to the "USER DATA" section in the prompt
\draw[thick,red,->] (4,-0.25) -- ++(2.7cm,0) |- ++(0,-4cm) -- ++(0.3cm,0);

\draw[thick,red,->] (3.8,-7.9) -- ++(2.9cm,0) |- ++(0,1.5cm) -- ++(3.75cm,0);

\draw[thick,red,->] (19,-0.25) -- ++(-3.7cm,0) -- ++(0,-1.8cm) -- ++(-4.2cm,0);

\draw[thick,red,->]  (20.4,-0.5) to (20.4,-4.8);

\draw[thick,dotted,red,->]  (22.75,-4.3) -- ++(3cm,0) |- (21,-12.95);

\draw[thick,red,->] (19.65,-12.95) -- ++(-4.6cm,0) |- ++(0,-3.8cm) -- ++(-3.8cm,0);

\end{tikzpicture}

\caption{Toy example showing the composition of a prompt under the \texttt{PoSSUM} framework. Red arrows pointing to `[...]' indicate instances where modular prompt components are slotted in. The dotted arrow indicates the LLM generation conditional on a given prompt. The above toy example showcases a single feature ($2020$ vote choice), though multiple features can generally be extracted simultaneously. Not every prompt contains all of the elements indicated in this Figure.}
    \label{fig:prompt_structure}
\end{figure}
\end{landscape}

\restoregeometry

\section{Prompting Architecture}\label{sec::prompting}
%Prompts are objects constructed of tokens which are used to pass information to, and engender a response from, LLMs. Prompts can contain multimedia data -- in this work I limit myself to images and text, though general purpose LLMs can as easily include tokenised video or audio as data in the prompt \cite{team2024gemini}.
\texttt{PoSSUM} prompts the LLM for a variety of reasons. An initial reason is the filtering of users -- non-persons, non-members of the population of interest, and over-sampled users are identified with the aid of LLMs and discarded. I will cover these set of prompts in Section \ref{sec::filter}.

The principle task of the protocol is \emph{feature extraction}, where the LLM is given a user's timeline to read and infer a set of socio-demographic characteristics, preferences and attitudes. This is performed in a \texttt{feature extraction prompt}. An example of a feature extraction prompt is provided in Figure \ref{fig:prompt_structure}.

The customisation of downstream feature extraction prompts conditional on some user-specific characteristics can also be achieved with a \texttt{feature builder prompt}. Figure \ref{fig:prompt_structure} presents an example of this in reference to the feature ``$2020$ vote choice''. The user in the example lives in Texas, hence the ``$2020$ vote choice'' feature-set available to this user should only include candidates which had ballot-access in the state. This is implemented via a two-step prompting strategy: first I generate a prompt that retrieves up-to-date $2020$ ballot-access data, and provides a clear example of how to structure a feature-set. The model responds to the prompt by generating a user-specific feature-set; second, the feature set is passed on to the feature extraction prompt (Pseudo-code \ref{alg:bg_informed_prompt}).\\

\noindent It follows that generally, within \texttt{PoSSUM}, prompts have a standard modular form -- they are composed of: i. \emph{background information}; ii. a \emph{mould} based on the available user data; iii. and a set of \emph{instructions}, usually dependent on a set of features of interest. Pseudo algorithm \ref{alg:build_prompt} describes the building of the prompt.\\

%The \emph{background} element includes any necessary auxiliary information which could help the LLM improve its decision making for a given task. For example, when prompting a LLM to deduce ``$2020$ past vote'' for a given user, we might wish to provide it with the election results by candidate for the state of residency of the user. The \emph{mould} describes a given user in terms of the information which we have derived via the $\mathbb{X}$ api -- i.e. the users' profile picture, name, username, self-reported description, and the full set of time-stamped tweets written by the user which are available to us at this stage. It is helpful to think of the mould as a function of the user data, $\mathcal{M}(\bm{\Upsilon})$, which returns a natural language description of the user, structured in some useful manner. The \emph{instructions} element describes a given task. It is again useful to think of an operation as a function which takes a list of features as input, $\mathcal{I}(\bm{\mathcal{F}})$, and returns a structured natural language task, whereby these features are embedded. For example, an entity recognition operation would be roughly described in terms of the following words: \emph{``is this user a person ?''}; the feature of interest in this example would then be a dichotomous ``personhood'' variable. Pesudo-code \ref{alg:build_prompt} presents the steps involved in the prompt-building routine.

\noindent \textbf{Instructing a Neutral Annotator} \hspace{10pt} \texttt{gpt-4o-2024-05-13} is tasked with annotating unstructured social media profiles of selected users. LLMs are capable general purpose task solvers \cite{brown2020language,achiam2023gpt}. Annotation of unstructured data is one such task in which LLMs have shown superhuman performance \cite{gilardi2023chatgpt,tornberg2024large}. Despite some efforts to provide basic best practices around prompting LLMs for this kind of task \cite{tornberg2024best}, the degrees of freedom around prompt building are simply too large, and the literature too young, to have had the full spectrum of practices systematically tested. There have been attempts at standardised, systematic testing of prompts \cite{barrie2024prompt} though these have largely focused on prompt stability rather than scoring of wholly different architectures. Systematic architecture testing requires the ability prompt the LLM with the same information, but under different architectures, many times over -- a prohibitive enterprise when the mould is large. A consequence of this is that the \emph{agentic} approach to silicon sampling \cite{argyle2023out, argyle2023leveraging, sanders2023demonstrations, bisbee2023synthetic}, where the LLM is asked to role-play, or impersonate a given individual according to a set of characteristics, has not been systematically tested against a simple approach where the AI is a neutral annotator. Moreover the literature is afflicted by baffling findings regarding the sensitivity of prompting strategies to any number of arbitrary tricks. Studies have reported that introducing emotional stimuli in prompts can improve their performance on benchmarks \cite{li2023large}, or that treating the AI as a method actor, and providing ``dramatic scene settings and role definition'' \cite{doyle2024llms} outperforms traditional prompting styles.\\

\noindent I propose a modular prompting architecture, which builds on the Chain of Thoughts (CoT) \cite{wei2022chain,wang2024strategic} approaches to address the social media feature extraction task. I treat the LLM as a \emph{neutral annotator}, purposely avoiding imbuing it with specific personalities. This is similar to a silicon forecaster \cite{schoenegger2024wisdom}, though the LLM is not explicitly instructed to play the role of an expert, forecaster, or any other.\\

%The instructions are tailored to address potential issues with symbolic binding \cite{robinson2022leveraging}.

%\subsubsection{General Feature Extraction Framework}\label{sec:general_feature_extraction}

%Implementing the quota inclusion criteria requires a defined set of features of interest $\bm{\mathcal{F}}$. %These features should be extracted from the user profiles and should have direct counterparts in the stratification frame, where possible, such that we are able to stop sampling from a given cell once it becomes fully populated.\\

\noindent Examples of relevant instructions are available in Figure \ref{fig:prompt_structure}. Key to the feature-extraction exercise are ``feature objects'', which define a specific choice-set. Feature objects are assigned a standard modular structure: each object contains a \emph{title}, which describes a survey question; a set of \emph{categories}, which represent the potential response-set; and each category is identified by a unique \emph{symbol}.

%The feature builder prompt merely necessitates an empty feature object, with clearly defined placeholders. The LLM can then fill these placeholders with information it has access to via the background data and / or its pre-existing knowledge, in order to create an actionable, complete feature object. 

%The feature extraction prompt is more complex. 
The feature object structure described above is made explicit to the LLM via a set of instructions, in an attempt to pre-emptively address issues around symbolic binding \cite{robinson2022leveraging}. Strong language is used to encourage the LLM to provide rigidly structured and consistent output, temperature notwithstanding. 

An important caveat specific to LLM feature extraction pertains the order in which text is presented in the LLM's prompts and outputs. The auto-regressive nature of LLMs \cite{lecun2023large} implies that when text is generated in response to a given prompt, earlier tokens will affect the next-token-probabilities downstream \cite{liu2023pre}. To encourage the LLM to provide answers which are conditional on some degree of reasoning, rather than the product of post-hoc justification, I enforce an output structure requiring the LLM to provide an explanation before selecting a symbol / category for a given feature. I am limited in the extent of reasoning I can stimulate by the OpenAI API's limit on output tokens ($4,096$).  \\

%\noindent Having defined the set of features $\bm{\mathcal{F}}$ to extract, we need to define a feature extraction operation $O$ to complete our prompt. Listing \ref{lst:feature_extraction_op} presents \texttt{PoSSUM}'s standardised feature extraction operation. The chosen $O(\bm{\mathcal{F}})$ endows the prompt with a set of standard instructions, and an example output which enforces a desirable structure to the LLM generation.\\

\noindent When multiple features are to be extracted simultaneously, the respective feature objects are appended to the prompt. The feature extraction operation then considers all features jointly, and prompts the LLM to produce a joint set of imputed features for the given user. I find for most tasks, simultaneous feature extraction is preferable to a set of independent prompts, one for each attribute of interest. Separating prompts is an intuitively attractive choice due to its preservation of full-independence between extracted features. But this is extremely inefficient in terms of tokens, given that each prompt has to re-describe the background, the mould and the operations of interest. Prompting the LLM to extract all features simultaneously, by including the full list of desired features in a single prompt, is generally a productive approach. To minimise the effects of auto-regression on the generated survey-object, we can randomise the order of all features in the feature extraction prompt, so that order effects on the overall sample cancel-out with a large enough number of observations.\\

\noindent \textbf{Feature Objects} \hspace{10pt} Listing \ref{lst::features_example} presents an example of a multi-features object, to be appended to the instructions module of the prompt.

Categories in the features object can seem needlessly verbose -- there are two reasons for this: i. detailed descriptions of the categories can help reduce the ``neutrality bias''  of the LLM -- namely the tendency for the LLM to systematically prefer a more ``neutral'', ``majority class'', or ``wide-net'' option under uncertainty; ii. associating each category with a unique text string helps ensure the unique parsing of the LLM output, especially when dealing with long prompts which include multiple questions sharing the same answer-text. 

It can be helpful at this stage to categorise features of interest in two distinct sets: those whose distribution in the population is known, and is directly available to us via auxiliary surveys, census data, election results, etc., we consider \emph{independent} variables; those whose distribution in the population is unknown, we consider \emph{dependent} variables. When ascertaining if a given feature belongs to either group, a simple rule of thumb is used: could I weight the poll by the marginal distribution of this variable ? if so, this is an independent variable; if not, I consider it dependent. Dependent variables should always come after the independent variables in the prompt, so that their distribution can be conditionally inferred.\\

\noindent \textbf{Handling Prompting Strategy Uncertainty} \label{sec:moe} \hspace{10pt} There is unresolvable uncertainty around LLM queries to infer voting preferences for $2024$ (inferred prospective vote) and $2020$ (inferred past vote). Small changes in prompt wording can lead to large variations in results \cite{barrie2024prompt}. In the context of inferring voting preferences, at least four discernible strategies emerge based on the prompting framework outlined above:

a. \emph{minimally informative} prompting offers a standard choice-set to the LLM, applicable to all users in the pool. The resulting prompt is unconditional of any user characteristic, and vote choice is inferred independently of other feature-extraction tasks. This is desirable if we want the inferred vote to be the result of an exclusive analysis of the mould, and be unaffected by other sources of information; 

b. \emph{moderately informative} prompting uses the feature-builder module to update the general choice-set, reflecting candidate options available to the user conditional on some characteristics. In the context of vote choice, the choice-set is conditional on the unser's state of residency; 

c. \emph{highly informative} prompting includes the moderately informative provisions, as well as leveraging relevant area-level election results. These past election results are included in the feature extraction prompt to directly influence LLM inference. Conditioning on relevant background election results induces a behaviour similar to raking, in that the inferred individual-level distribution of the vote will be somewhat constrained by the available area-level marginal distributions; 

d. \emph{joint socio-demographic} prompts do not use background data, and rely on the choice-set approach of the minimally informative prompting style. The key difference is that vote choice is estimated contemporaneously as the other features of interest, effectively conditioning LLM inference on on sequentially inferred socio-demographics. This approach can be useful if the LLM can infer auxiliary characteristics with little error, and if the LLM's understanding of the relationship between these inferred auxiliary characteristics and the vote is accurate enough to provide useful inferential constraints. Again I wish for the LLM to implicitly perform raking to its internal vote choice representation -- the inferred vote distribution is to conform to the underlying marginal distributions of the vote by the inferred socio-demographics, which exist in the LLM's silicon mind.  \\

\noindent In absence of a clear preference amongst prompting strategies, I rely on a wisdom-of-the-synthetic-crowd approach \cite{schoenegger2024wisdom,t2024fake,murr2011wisdom}. The heuristic here is that prompt heterogeneity can make LLM learners more uncorrelated, and aggregates of uncorrelated learners typically have desirable properties \cite{goodfellow2016deep,graefe2014combining,graefe2015limitations}. I apply a classic majority voting algorithm, breaking ties at random. %Speculation levels for this aggregated measure are simply the average speculation across candidate measurements.  

%\noindent Listing \ref{lst:ind.features} presents a comprehensive list of independent features. Notice this is a more expansive list when compared to those in Table \ref{tab:frame}. This is deliberate, as we wish to retain the option of further weighting the sample after the data collection process has concluded, and we may wish to target a wider spectrum of marginal/joint distributions than those of the variables used to determine the quotas. 

\section{Filters} \label{sec::filter}

$\mathbb{X}$ accounts can make for noisy subjects, and not all accounts are deserving of resource allocation for profile augmentation with timeline data. A large number of accounts are not extensions of real existing individual persons, but rather  represent organisations \cite{wang2019demographic}, bots \cite{ferrara2016rise}, parody accounts or other non-person entities. A small number of accounts is responsible for a disproportionate amount of activity on the platform \cite{kwak2010twitter}, and could dominate inference in absence of appropriate mitigation measures. Many accounts contain no discernible information about the location of the user \cite{malik2015population}, a key feature for being able to make representative inference at the area-level. Relative to the population of interest, select socio-demographic groups are likely to be over-sampled \cite{pew2024socialmedia,mellon2017twitter}. \\

\noindent The \texttt{poll\_users} routine implements a number of filters to decide which accounts are most deserving of attention -- which are most valuable at a given moment in the digital fieldwork period to construct a representative sample of US adults. Some of these filters are \emph{mechanical}, in that they simply apply rules to user meta-data to discard or retain profiles. Others are \emph{intelligent}, in that they leverage AI to deduce some key features of the profile to make a value determination. The filtering routines which follow are applied to the data generated by the \texttt{get\_pool} routine, so the user-profile data along with usually a single tweet related to the query of interest.\\

%Much like in a traditional survey, several checks to ensure data quality and representation must be put in place.\\ %An initial set of checks is purely mechanical, in that it does not make use of any ``intelligence'' beyond simple object manipulation. A second

\noindent \textbf{Temporal Filter} (Pseudo-code \ref{alg:temporal_filter}): A routine limiting the number of synthetic survey responses we wish to obtain from a single user within a given time-frame. For example, in the context of pre-election polling, we may wish to collect new data on a given user only once every $30$ days -- if the digital fieldwork is spread over a full month -- despite their more frequent content creation. The routine to implement the temporal exclusion criteria involves: i. tallying the users that have been processed up to now; ii. identifying which of those have been processed within the last $30$ days (or whatever the exclusion criteria); iii. removing those users from the \emph{fresh} pool generated by the \texttt{get\_pool} routine.\\

\noindent \textbf{Null Geography Filter} (Pseudo-code \ref{alg:nullgeo_filter}): This is a relatively simple data-quality check. Geography is a fundamental part of pre-election opinion polling -- we must be able to place individuals within the given geographic boundary we wish to make inference for. If the user has no self-reported location, we exclude the user a-priori. Intelligent geographic filtering is in principle possible when an explicit location field is absent, by prompting the LLM to infer a location from other content generated by the user -- this tends to be less accurate and more expensive due to the larger amounts of input-tokens necessary.\\

\noindent \textbf{Entity Filter} (Listing \ref{lst:entity_filter}):  Consists of defining the kinds of social-media profiles we want to include in our analysis. For pre-election opinion polling, we would wish to exclude $\mathbb{X}$ accounts related to organisations (e.g. news outlets, NGOs), bots, and focus solely on real-life persons \cite{wang2019demographic}. 
\begin{lstlisting}[caption={Entity Filter prompt.}  ,captionpos=t,label={lst:entity_filter},framexleftmargin=0cm,xrightmargin=0cm]
Is this the account of a real-life existing Person, or of another kind of entity ? 
Respond either with "P" for Person or "O" for Other.
\end{lstlisting}

\noindent \textbf{Intelligent Geographic Filter} (Listing \ref{lst:geo_inclusion}) : %In \texttt{PoSSUM} I use a three-levels of geography system, whereby Level 1 geography constitutes the broadest boundary within which the population falls -- e.g. residents of the US fall within any state or territory of the US; Level 2 would be the given state or territory; Level 3 would be a congressional district, county, etc. which falls within the Level 2 geography. This system can be adapted to smaller areas, so If I were interested in polling residents of Kentucky exclusively, the geographic system would respond as follows: Level 1 -- state; Level 2 -- congressional district; Level 3 -- precinct; etc. Notice however that the geographic information available on $\mathbb{X}$ is rarely detailed enough to assign individuals to very small areas, and beyond the $2^{nd}$ level of geography \texttt{PoSSUM} will tend to produce noisy estimates. 
This filter helps exclude users who are unlikely to reside in the Level 1 geography. Level 1 constitutes the broadest boundary within which individuals belonging to the population of interest fall. For US pre-election polling I set this to the ``United States of America''. Level 2 geography is then intended to be the ``State'', and Level 3 is the relevant ``Congressional District'', and so on. It is efficient to use a prompt which allows rejection of users who fail the Level 1 inclusion criteria, and simultaneously extracts the Level 2 information. Listing \ref{lst:geo_inclusion} presents an implementation of the geographic extraction prompt. \texttt{PoSSUM} rejects users who are \emph{``Not from a state in the USA''}. The great advantage of using $\mathbb{X}$ relative to other platforms is the relatively high rate of available self-reported locations, which makes geographically-bound polling possible.

\begin{lstlisting}[caption={Intelligent Geographic Filter prompt.}  ,captionpos=t,label={lst:geo_inclusion},framexleftmargin=0cm]
Which state of the USA do they live in?
If they do not specify a state, but are still from the United States, write "USA".
If they are not from a state in the USA, write "Not from a state in the USA".
Write out just the full name of the state.
If they are from the District of Columbia, also known as Washington D.C., write "District of Columbia".
\end{lstlisting}

\noindent \textbf{Quota Filter}: The population of $\mathbb{X}$ users is notoriously unrepresentative of the US population \cite{pew2024socialmedia}. It is nevertheless a very large pool of US residents, accounting for around $22\%$ of the US population. Whilst some categories -- namely higher educated and higher income individuals -- are extraordinarily over-represented, the pool is ``deep enough'' that we could expect to eventually find a number of representatives for most relevant socio-demographic groups in the population. It follows that implementing quotas is liable to make sampling more efficient. \\

\label{quota_details}

\noindent \texttt{PoSSUM} implements quota sampling as follows: i. define a stratification frame (e.g. Table \ref{tab:frame}) which describes the number of individuals $\omega^\star_c$ from each ``cell'' $c \in \{1,...,C\}$, which we would expect to capture in a random sample of target size $\Omega^\star$ users -- we could for instance set $\Omega^\star = 1,500$ to produce polls of a somewhat traditional sample size; ii. a feature extraction operation is deployed to infer the values of the relevant variables for the user at hand. At this stage the LLM does not make use of any background information, and it utilises the same user-level information as the other intelligent inclusion criteria; iii. if the user belongs to a cell in the stratification frame for which the number of sampled users $\omega^\prime_c$ is smaller than the number of desired users $\omega^\star_c$, I retain the user and update the quota counter -- otherwise I exclude the user from the analysis. Pseudo-code \ref{alg:poll_users} contains a symbolic description of the quota exclusion criteria implemented here.\\

\begin{table}[ht]
\centering
\scalebox{0.8}{
\begin{tabular}{r|ccccc|ll}
  \hline
  \hline
Cell & Sex & Age & Household Income & Race/Ethnicity & Vote $2020$ & Quota & Counter \\ 
  \hline
  1 & male & 65 or older & up to 25k & black & D &   2 & 0 \\ 
    2 & female & 25 to 34 & between 25k and 50k & white & D &   3 & 3 \\ 
    3 & male & 35 to 44 & between 75k and 100k & hispanic & D &   2 & 2 \\ 
    4 & female & 45 to 54 & between 75k and 100k & white & D &   6 & 6 \\ 
    5 & female & 35 to 44 & between 25k and 50k & black & D &   1 & 1 \\ 
  \vdots &\vdots &\vdots &\vdots &\vdots &\vdots &\vdots & \vdots\\
  430 & female & 25 to 34 & between 25k and 50k & asian & stayed home &   1 & 0 \\ 
  431 & female & 65 or older & between 50k and 75k & hispanic & stayed home &   1 & 0 \\ 
  432 & female & 18 to 24 & more than 100k & asian & stayed home &   1 & 0 \\ 
  433 & male & 18 to 24 & between 50k and 75k & native & stayed home &   1 & 0 \\ 
  434 & female & 55 to 64 & between 50k and 75k & asian & stayed home &   1 & 0 \\ 
  435 & male & 18 to 24 & between 50k and 75k & asian & stayed home &   1 & 0 \\ 
   \hline
   \hline
\end{tabular}
}
\caption{Example implementation of a stratification frame with quota counter, for a target sample size $\Omega^\star = 1,500$. This is a snapshot taken with $647$ respondents still to be collected.}
\label{tab:frame}
\end{table}

\noindent Surviving user profiles are sufficiently information-rich, representing a real-life person in the Level 1 geography. Their latest digital trace is recent, at most $1$ week old from the moment the \texttt{get\_pool} routine is initiated. It is then efficient to expend resources to ``survey'' these profiles. The LLM is prompted under the general feature extraction framework described in Section \ref{sec::prompting}, with two important differences: i. the digital trace available for each user is expanded further by querying their timeline for their last $m$ tweets, further augmenting their respective mould; ii. we impute a complete set of \emph{independent} and \emph{dependent} characteristics conditional on this new expanded mould.\\

\noindent When expanding profiles, I distinguish between users captured via \emph{trending topics}, as opposed to \emph{political talk}, queries. A very small number of tweets is necessary to estimate the preferences of those who talk politics on $\mathbb{X}$. Conversely, users discussing trending topics on $\mathbb{X}$ can be totally enigmatic with respect to their politics -- their last $m$ tweets could never mention anything remotely useful to indicate political preferences. As a result I set two distinct values of $m$ for these two sets of subjects: $m^{tredning} = \lambda \times m^{politics}, \mbox{ } \forall \lambda >1$. I use $\lambda = 2$ and $m = 20$, but this is open to further tuning. What is generally true is that, where resources permit, ``more is better'' in terms of information used to generate or expand a user's mould.

\section{\texttt{make\_inference}}\label{sec:inference}

The goal of \texttt{PoSSUM} is to make representative inference at the population-level, as well as for granular socio-demographic and geographic segments of the population of interest. Each poll will seldom be large enough to make such granular inference, in that crosstabs will be scarcely populated, and not immediately generalisable. Moreover selection effects are still likely to plague the sample due to social media selection, attention selection, and other socio-demographic sample imbalances.\\

\noindent The \texttt{make\_inference} routine implements a weighting strategy to generalise the findings from silicon samples to the population of interest. The weighting method of choice here is Multilevel Regression with Post-Stratification (MrP) \cite{gelman1997poststratification,park2004bayesian,lauderdale2020model}. The explicit knowledge of unfilled quotas prompts a treatment of these cells as having missing dependent variables. We can then use a hierarchical model, under the ignorability assumption \cite{van2018flexible}, to estimate the dependent values for the incomplete cells, and stratify these estimates to obtain national and state-level estimates. This also allows a comprehensive treatment of uncertainty at the cell-level, which is liable to provide more realistic intervals on the poll's topline than traditional adjustments. 

Structured priors \cite{gao2021improving}, as well as deliberate model selection, are crucial -- expecting a biased sample and noisy crosstabs, I leverage informative priors and theoretical knowledge of the functional form of the dependent variable to ``direct'' learning towards a useful configuration. The model is estimated using the Bayesian probabilistic programming language \texttt{Stan} \cite{carpenter2017stan}. Post-stratification is performed at various levels of analysis -- predictions from this model are made for every `cell' in the stratification frame, and these are then aggregated at the national, crosstab (e.g. by categories of age, gender, ethnicity, etc.) and state levels respectively, according to the weight associated with each cell (see Cerina \& Duch \cite{cerina2023artificially} for a comprehensive theoretical model connecting unrepresentative social media samples to post-stratified estimates). \\

\subsection{Extending the Stratification Frame} \label{sec::extend}
To improve the MrP estimates I use a modified MrsP \cite{leemann2017extending} procedure (Smooth MrsP)\cite{cerina2020thesis,cerina2022machine}. The goal of this procedure is to extend the stratification frame, which is derived from the $2021$ American Community Survey \cite{uscensus_acs2021}, to include the joint distribution of $2020$ Vote Choice as derived from an auxiliary survey. It differs from traditional MrsP in that it doesn't use the auxiliary survey crosstabs to augment the frame, but rather it fits a model to smooth the crosstabs first, and then projects these onto the existing frame. This approach can help generate more plausible estimates for `noisy' cells, when the number of cells in the frame is large and the sample-size-per-cell in the auxiliary survey is relatively small.  I use the $2022$ Cooperative Election Study (CES) \cite{DVN/PR4L8P_2023} as the auxiliary survey to get estimates of $2020$ recall vote\footnote{I use this dataset for the following reasons: a. it is a large sample of $60k$ subjects, affording greater scope for estimating interaction effects between demographic attributes; b. the alternative (ANES) was much too biased in favour of the Democratic candidate in $2020$; c. it allows me flexibility to extend the frame further by $2022$ vote, using the same dataset, if it is reasonable to do so at a later stage.}. I fit a deep-MrP \cite{ghitza2013deep,goplerud2024re} model using \texttt{Stan} \cite{carpenter2017stan} to generate estimates of past-vote which leverage interactions between demographics as much as possible, in order to avoid attenuation bias in the estimated cell-level distribution. The likelihood of the model is categorical, and SoftMax is used as the link-function. The \emph{depth} of the Bayesian Hierarchical model is given by the inclusion of marginal effects of sex, age, ethnicity, education, household income and state, as well as all two- and three- way interactions. All effects are estimated as random effects under non-centered parametrisation and recommended weakly-informative priors \cite{gelman2024default}. The \texttt{Stan} code for this model is available in the GitHub repository under name \texttt{model\_ai.survey\_SmoothMrsP.stan}. The resulting frame is then raked to the known state-level distribution of demographics and past vote, using the \texttt{anesrake} procedure \cite{pasek2018package}. The quota-frames used for the quota-filter are samples of size $\Omega^\star$ from this ``mother-frame'', where a new ``daughter-frame'' is sampled to generate targets for each new poll. 
\begin{table}[ht]
\centering
\scalebox{0.665}{
\begin{tabular}{cllcc|cc}
\hline
\hline

 \emph{predictor} & \emph{level} & \emph{description} & \emph{index} & \emph{domain} & \emph{parameter} & \emph{prior correlation structure} \\[0.15cm]
 \hline
\hline
$\mathbf{1}$ & global & / & / & / & $\alpha_j$ &
\begin{tabular}{@{}l@{}} iid \end{tabular} \\[0.15cm]
\hline 

/  & state & state\_id & $l$ & $\{1,\dots,54\}$ & $\lambda_{sj}$ & \begin{tabular}{@{}l@{}} spatial (BYM2) \end{tabular} \\[0.15cm]
\hline 

%/  & poll & poll\_id & $p$ & $\{1,\dots,P\}$ & $\eta^{P}_{tj}$ & random-walk \\[0.15cm]
%\hline

/ & \multirow{5}{*}{individual} & age\_id & $a$ & $\{1,\dots,6\}$ & $\eta^{A}_{aj}$ & random-walk \\[0.15cm]

/ & & income\_id & $h$ & $\{1,\dots,5\}$ & $\eta^{H}_{hj}$ & random-walk \\[0.15cm]

/ & & sex\_id & $g$ & $\{1,2\}$ & $\gamma^{G}_{gj}$ & unstructured + shared variance \\[0.15cm]

/ & & race\_id & $r$ & $\{1,\dots,6\}$ & $\gamma^{R}_{rj}$ &  unstructured + shared variance  \\[0.15cm]

/ & & vote20\_id &  $v$  & $\{1,\dots,5\}$ & $\gamma^{V}_{vj}$ &  unstructured + shared variance  \\[0.15cm]
\hline

$z_1$ & \multirow{9}{*}{state}  &  $2020$ $R$ share & $  \multirow{9}{*}{/} $ &  \multirow{9}{*}{$\mathbb{R}$} & $\beta_{1j=\mbox{R}}$ & \multirow{9}{*}{iid}  \\[0.15cm]

$z_{2}$ &  &  On ballot: R.F.K. Jr.  & & &  $\beta_{1j=\mbox{K}}$ &  \\[0.15cm]

$z_{3}$ & &  On ballot: Jill Stein  & & & $\beta_{1j=\mbox{G}}$ & \\[0.15cm]
$z_{4}$ & &  $2020$ $G$ share  & & & $\beta_{2j=\mbox{G}}$ & \\[0.15cm]

$z_{5}$ & &  On ballot: Chase Oliver  & & &  $\beta_{1j=\mbox{L}}$&\\[0.15cm]

$z_{6}$ & &  $2020$ $L$ share  & & & $\beta_{2j=\mbox{L}}$ & \\[0.15cm]

$z_{7}$ &  & On ballot: Cornel West & & & $\beta_{1j=\mbox{W}}$ &  \\[0.15cm]

$z_{8}$ & &  $2020$ ``stay home'' share  & & & $\beta_{1j=\mbox{stay\_home}}$ & \\[0.15cm]

\hline

/ & individual : state & vote20\_id $\times$ $2020$ j share & / & / & $\bm{\zeta}_{v,j}$ & unstructured + shared variance\\
 \hline
 \hline
\end{tabular}
}
\caption{Predictors and Parameters for the $2024$ vote-choice model. `iid' refers to fully independent parameters, or `fixed' effects \cite{gelman2013bayesian}. `unstructured + shared variance' priors refers to classic random-intercepts. Note: the Democrat choice ``D'' is taken as the reference category, hence it has no associated predictor.}
\label{tab:params}
\end{table}

\subsection{Hierarchical Bayes to Model Silicon Samples} 

The final hierarchical model used to generate smoothed estimates of the dependent variable of interest is a simple MrP with structured priors \cite{gao2021improving}. The ``structure'' of the model plays an important role here, as it can help smooth the learned effects of a model trained on AI generated data in a sensible way. LLMs can leverage stereotypes in making their imputations \cite{choenni2021stepmothers}, which can translate to exaggerated relationships between covariates and dependent variables. Adding structured smoothing to the model allows us to correct for this phenomena, to some degree.

I regress the dependent variable, which is assigned a categorical likelihood with SoftMax link, onto sex, age, ethnicity, household income and $2020$ vote.  Sex and ethnicity effects are estimated as unstructured random effects; state\footnote{Because I have an interest in being able to estimate the number of electoral votes won by each candidate, I treat the congressional districts of Nebraska and Maine as separate states.} effects are assigned a BYM2 prior \cite{donegan2022flexible,morris2018spatial,besag1991bayesian}; income and age effects are given random-walk priors. Separate area-level predictors are created for each dependent variable of interest. Table \ref{tab:params} presents the covariates and parameters used in the model for $2024$ vote choice.\\

\pagebreak
\noindent %The covariates used for other models involving different dependent variables are very similar, and available in the GitHub repository. 
I present the full %\footnote{Note: this model aggregates learning from multiple polls conducted during a given period -- hence the poll-level random effects $\eta^{P}$. These poll effects are ordered in time, according to the fieldwork dates of the poll, and as such are assigned random-walk priors to allow for temporal smoothing. The most recent estimate of preferences can then be obtained by setting the \texttt{poll\_id} to the most recent poll. The model fit to weight individual polls is identical, with the only exception being the absence of poll-level effects.} 
Hierarchical Bayesian model below -- see \cite{cerina2023artificially} for a more attentive explanation of each model component. I describe the generation of given choice $j \in \{1,...,J\}$, made by a sampled user $i \in \{1,...,n\}$, as follows:

\begin{align*}
y_{i} \sim & \mbox{Categorical}(\pi_{i1},...,\pi_{iJ}) && \mbox{\texttt{likelihood}}\\[0.5cm]
\pi_{ij} = & \frac{\exp (\mu_{ij})}{\sum_j \exp(\mu_{i,j})}; && \mbox{\texttt{softmax link}}\\[0.5cm]
\mu_{ij} = & \alpha_j + \\
 & \lambda_{\mbox{\tiny state\_id}[i],j} + \\
 & %\eta^{P}_{\mbox{\tiny poll\_id}[i],j} + 
 \eta^{A}_{\mbox{\tiny age\_id}[i],j} + \eta^{H}_{\mbox{\tiny income\_id}[i],j} + \\
 & \gamma^{G}_{\mbox{\tiny sex\_id}[i],j} + \gamma^{R}_{\mbox{\tiny race\_id}[i],j} + \gamma^{V}_{\mbox{\tiny vote20\_id}[i],j} + \\
 & \sum_{k_j} \beta_{\{k_j,j\}} \times  z_{ \{\mbox{\tiny state\_id[i]},k_j\}} +  && \mbox{\texttt{state-level predictor}}\\
& \zeta_{\{\mbox{\tiny vote20\_id[i]},j\}} \times \nu_{\{\mbox{\tiny state\_id[i]},j\}} ; && \mbox{\texttt{ind. by state interactions }}\\[0.5cm]
 \alpha_j \sim & \mbox{N}(0,1); && \mbox{\texttt{intercept}}\\[0.5cm]
\lambda_{sj} =& \mbox{ }\sigma^{\lambda}_j \left( \phi_{sj}\sqrt{(1-\xi_j)} + \psi_{sj}\sqrt{(\xi_j/\epsilon)}  \right); && \mbox{\texttt{BYM2 effects}}\\
\phi_{sj} \sim &  N(0,1); && \mbox{\texttt{unstr. state-level effects}}\\
\psi_{sj}  \mid \psi_{{s}^\prime j}  \sim & N\left(\frac{\sum_{{l} ^{\prime} \neq l} \psi_{{s} ^\prime j} }{\nu_{l}},\frac{1}{\sqrt{\nu_{l}}} \right); && \mbox{\texttt{conditional auto-reg. effects}}\\
\xi_j \sim & \mbox{Beta}\left(\frac{1}{2},\frac{1}{2}\right); && \mbox{\texttt{mixing weights}}\\
\sigma_j^{\lambda} \sim & N^{+}(0,1); && \mbox{\texttt{state-level scale}} \\[0.5cm]
\gamma^U_{uj}  \mid \gamma^U_{u-1\mbox{ }j} \dots  \gamma^U_{1j} \sim & N( \gamma^U_{u-1\mbox{ }j}, \sigma_j^{U}),\mbox{ }  \hspace{20.5pt} \forall \mbox{ }  U \in  \{A,H\};&& \mbox{\texttt{random walk effects}} \\
\gamma^U_{uj} \sim&  N(0,\sigma_j^U), \mbox{ } \hspace{45pt}  \forall \mbox{ } U \in \{G,R,V\}; && \mbox{\texttt{unstructured effects}}\\
\sigma_j^{U} \sim & N^{+}(0,1); && \mbox{\texttt{random effect scales}}\\[0.5cm]
\bm{\beta}_j \sim & N(0,1). && \mbox{\texttt{fixed state cov. effects}}\\[0.5cm]
\bm{\zeta}_j \sim & N(0,\sigma^{\zeta}_j). && \mbox{\texttt{unstr. ind. by state effects}}\\
\sigma_j^{\zeta} \sim & N^{+}(0,1); && \mbox{\texttt{ind. by state scale}} \\
\end{align*}

\pagebreak 

\subsubsection{Learning from Stateless Users} \label{sec:nostate}
\texttt{PoSSUM}'s geographic filtering ensures users who are selected for analysis are most likely based in the US. The protocol does however allow for the inclusion in the sample of users whose state of residency within the US -- their $2^{nd}$ level geography -- is unknown. Learning from these users can bring to bear evidence pertaining the relationship between individual-level attributes, such as age, gender, education, past-vote, etc., and $2024$ voting preferences.\\ %How to best allow for this in the hierarchical model described above ? \\

\noindent I consider an approach that uses two separate linear predictors, one for the observations missing a state, and one for those observations which are complete. For the latter, the linear predictor is exactly as described above; for the former, the following linear predictor is used:

\begin{align*}
 \mu^{s'}_{ij} = & \alpha_j + \\
 & \Xi_{j} + \\ 
 & \lambda_{\mbox{\tiny state\_id}[i],j} + \\
 & %\eta^{P}_{\mbox{\tiny poll\_id}[i],j} + 
 \eta^{A}_{\mbox{\tiny age\_id}[i],j} + \eta^{H}_{\mbox{\tiny income\_id}[i],j} + \\
 & \gamma^{G}_{\mbox{\tiny sex\_id}[i],j} + \gamma^{R}_{\mbox{\tiny race\_id}[i],j} + \gamma^{V}_{\mbox{\tiny vote20\_id}[i],j}\\
\Xi \sim & N(0,1).
\end{align*}

\noindent where $\Xi_j$ is a `no-state' fixed effect, which captures the average difference between the baseline level of support relative to the average state, and the `no-state' pool's support. Effectively, I am treating the `no-state' label as a separate, independent state, which is not pooled towards the state-level effect average. The remaining individual-level coefficients are still informed by these users' preferences. Making out-of-sample predictions I then use the linear predictor for the observations with known states, effectively discarding the no-state effect%\footnote{There are alternatives to this formulation -- for instance we could retain the two separate linear predictors, but not include a separate `no-state' effect. This would effectively let the `no-state' pool inform the intercept in full, without netting-out the difference with the rest of the sample. I consider this approach unwise: composition effects in this particular pool of users could lead to distortions of the baseline rate of support. Imagine for instance that all of these users would belong to one particularly polarised state -- say California or Wyoming. If their state of residency was known, its effect would largely be captured by the respective state-level random effect, and the intercept would be largely untouched. We recover similar behaviour by introducing a `no-state' fixed effect. However, were we to let this pool of `masked' Californians or Wyomingites inform the intercept, our out-of-sample predictions would become skewed according to the unknown states' respective bias. According to a principle of caution, we opt to net-out the unknown-state effects as described.}
.

\subsubsection{Alleviating Attenuation Bias} \label{sec:attenuation}
Post-stratified state-level estimates of vote share generated from the above model will tend to display attenuation bias, performing poorly in `tail areas'. The size of the attenuation bias is roughly proportional to the size of the difference between the true state-level performance of the candidate, and the candidates' average performance across states. 

Attenuation bias arises as a direct result of modeling choices. Parsimonious models typically perform well for the average state, at the cost of large attenuation bias in tail states. Consider as an example the impact of assuming the relationship between individual-level covariates and voting propensity is constant across states. Similarly, regularising coefficients via partial-pooling will foster attenuation bias by design, as tail effects are smoothed towards the size of average effects. As a result, ``tail areas'', which are typically the product of various ``tail effects'', are themselves smoothed towards the ``average state''.\\ 

\noindent One might therefore be tempted to fit a ``deep'' model \cite{goplerud2024re}, which makes few parametric assumptions and considers the full set of complex interactions to generate state-level estimates. One issue with this approach is computational tractability, which is part of a trilemma: under typical resources constraints, one cannot have all three of a) flexible modeling;  b) simulations from well-behaved posterior distributions; c) fast fitting times which make frequent (e.g. daily) model updates viable. Beyond that, there are concerns related to over-fitting to unrepresentative samples. I have discussed already issues with LLMs' tendency to exaggerate relationships due to reliance on stereotypes, which affects the representative quality of the \texttt{PoSSUM} data, and make structured regularisation desirable. Unobservable selection effects into online samples also play a role.\\%This is true of all applications where MrP is used to explicitly account for the selection process in order to make representative inference. There are however some selection effects that cannot be corrected for - see \cite{cerina2023artificially} for an example of exogenous selection; and some which we might fail to correct for due to model misspecification. As such we are typically well-served by making conservative modeling choices -- and hence favouring parsimonious, regularised models. \\

\noindent One way to address the attenuation bias in a parsimonious, conservative, and tractable manner is to selectively relax regularisation using interaction effects between area-level and individual-level covariates. These have to be sensibly informed by prior knowledge in the data generating process. In the context of vote share estimates, we wish to relax the regularisation pressure at the area-level proportionally to the level of attenuation bias, which we expect to be large wherever a given candidate is most / least dominant, relative to their average performance. A good predictor for this expectation is the candidates' past performance in the given area, which is one component of the state-level predictor. We further know that individual-level vote-choice is driven primarily by past-vote effects. By interacting the individual-level past vote and the area-level past vote-share estimate for the candidate at hand, we can achieve the desired effect:

\begin{equation*}
     \mu_{ij} = \cdots + \sum_{k_j} \beta_{\{k_j,j\}} \times  z_{ \{\mbox{\tiny state\_id[i]},k_j\}} + \zeta_{\{\mbox{\tiny vote20\_id[i]},j\}} \times \nu_{\{\mbox{\tiny state\_id[i]},j\}} 
     + \cdots ;\\
\end{equation*}

\noindent where $k_j$ represents the index of a column from the state-level predictor matrix $z$ which is used to predict choice $j$; $\beta_{\{k_j,j\}}$ is the fixed effect of that state-level predictor on choice $j$; $\nu$ is the matrix of past measurements of the dependent variable at the state-level (e.g. $2020$ vote share of choice $j$ in a given state), typically a subset of $z$;  and $\zeta_{\{v,j\}}$ is the gradient of the effect of choice $j$'s past state-level measure of choosing option $j$, for an individual who voted for option $j$ in $2020$.\\ 

\noindent The interaction allows us to account for an additional gradient in the effect of individual-level past-vote across states, allowing the state-level post-stratified estimates to escape, to some degree, regression to the mean. 

\subsubsection{Aggregating Polls} \label{sec:aggregate}
Aggregating \texttt{PoSSUM} samples is possible -- this has the potential to improve the accuracy of estimates for a given fieldwork period by leveraging information from previously fielded polls, and discounting these at an exponential rate \cite{linzer2013dynamic}. %It can be especially useful for generating reliable estimates for under-samples areas. 
This effect is achieved in the above model via the introduction of a fieldwork-date random effect with a random-walk prior: 

\begin{align*}
 \mu_{ij} = & \dots +  \eta^{P}_{\mbox{\tiny poll\_id}[i],j} + \dots \\
\eta^{P}_{pj}  \mid \eta^{P}_{p-1\mbox{ }j} \dots \eta^{P}_{1j} \sim & N( \eta^{P}_{p-1\mbox{ }j}, \sigma^P)\\
\sigma^P \sim & N^{+}(0,1).
\end{align*}

\noindent This effect is simplistic, in that it only captures national-level trends across fieldwork days, whilst in reality state trends over a campaign can vary. Larger state-level sample sizes could justify the inclusion of a more comprehensive state-level random walk with an informative covariance prior \cite{heidemanns2020updated}. %The data in principle allows for the estimation of daily trends, as opposed to fieldwork-period trends as above. This is however unwise given the preceding quota-sampling approach, in that unless one is able to account for all factors which explain sampling order due to the quota constraints, the resulting trends will be skewed -- the preferences of hard-to-sample users will dominate days towards the end of the fieldwork period. 

\noindent

\subsection{Estimation}
The Hierarchical Bayesian model is implemented in \texttt{Stan}\footnote{The code for every \texttt{Stan} model implemented in this paper is available on the \texttt{GitHub} repository.} \citep{carpenter2017stan, stan2018rstan}. To encourage well behaved MCMC sampling, covariates are standardised and random effects are estimated via non-centered parametrisation \cite{papaspiliopoulos2007general}. The BYM2 effects are specified according to non-generative improper priors \cite{morris2019bayesian,donegan2022flexible}. General recommendations for weakly-informative priors \cite{gelman2024default} are adhered to.

Models are fit separately for each of the $5$ polls fielded over the course of the election. A model including fieldwork-period random effects is also fit to an aggregate dataset obtained by stacking the silicon samples. %To evaluate the sensitivity of \texttt{PoSSUM} to levels of speculation, I fit the model to each silicon sample twice: once to the full data, including highly speculative records; once to the subset of data which excludes high-speculation inference (i.e. where every characteristic in the record complies with $S\leq80$).

Each model is ran for $8$ chains of $5,000$ iterations, with the first $4,750$ used as burn-in. The chains are thinned by a factor of $4$ to account for auto-correlation, and the \texttt{max\_treedepth} parameter is set to $15$ to allow convergence of the otherwise problematic spatial structure. This results in around $500$ posterior samples from each model.\\

\noindent \textbf{Posterior Predictive Distribution of Crosstabs} \hspace{10pt} Posterior simulations for each model parameter%, noted below by the index $s$ and the superscript $\sim$ ,
can be used to generate samples from the posterior predictive distribution of cell-level choice probabilities. Letting $c$ index the rows of the stratification frame, each representing a population cell, the posterior choice probability samples are derived as follows:

\begin{align*}
\tilde{\mu}_{scj} = & \tilde{\alpha}_{sj} + \\
 & \tilde{\lambda}_{s,\mbox{\tiny state\_id}[c],j} + \\
 & %\eta^{P}_{\mbox{\tiny poll\_id}[i],j} + 
 \tilde{\eta}^{A}_{s,\mbox{\tiny age\_id}[c],j} + \tilde{\eta}^{H}_{s,\mbox{\tiny income\_id}[c],j} + \\
 & \tilde{\gamma}^{G}_{s,\mbox{\tiny sex\_id}[c],j} + \tilde{\gamma}^{R}_{s,\mbox{\tiny race\_id}[c],j} + \tilde{\gamma}^{V}_{s,\mbox{\tiny vote20\_id}[c],j} + \\
 & \sum_{k_j} \tilde{\beta}_{\{s,k_j,j\}} \times  z_{\{\mbox{\tiny state\_id}[c],k_j\}} + \\
& \tilde{\zeta}_{\{s,\mbox{\tiny vote20\_id}[c],j\}} \times \nu_{\{\mbox{\tiny state\_id}[c],j\}} ; \\[0.5cm]
\tilde{\pi}_{scj} = & \frac{\exp (\tilde{\mu}_{scj})}{\sum_j \exp(\tilde{\mu}_{scj})}.
\end{align*}

\noindent Post-stratification is then necessary to obtain population-, crosstab- and state-level estimates of choice-shares. Let $\mathcal{F}(c)=f$ be a function assigning each cell \(c\) to a crosstab index \(f\) -- each \(c\) lies in exactly one crosstab, and the crosstabs are exhaustive and mutually exclusive. At the population-level, every cell belongs to the same crosstab $F(c)=f^\star \mbox{ } \forall \mbox{ } c$. It follows that posterior samples from a post-stratified choice-share estimate, for a given crosstab $f$, are obtained as follows:

\begin{equation*}
   \tilde{\pi}_{sfj}
   \;=\;
   \frac{\sum_{c:F(c)=f} \tilde{\omega}_{sc} \,\tilde{\pi}_{scj}}{\sum_{c: F(c)=f} \tilde{\omega}_{sc}};
\end{equation*}
\noindent where $\omega_{sc}$ represents a sample from the posterior predictive distribution of stratification-frame weights (obtained from the model described in Section \ref{sec::extend}), and $\pi_{scj}$ is a posterior sample from the cell-level choice probabilities described above.

\section{Evaluation} \label{sec:results}

\noindent Pre-election opinion polls are useful if they provide an accurate snapshot of the distribution of winning odds across candidates. I score \texttt{PoSSUM}'s \textbf{electoral predictive power} by comparing estimates against observed returns at the electoral college district\footnote{These are the constituencies which independently allocate electoral college votes to elect the US President. They typically correspond to US States, with the exceptions of Nebraska and Maine for which they match congressional districts.}  level. The evaluation framework involves statistical measures that reflect essential dimensions of \texttt{PoSSUM}'s estimate quality: i. average error at the point-estimate is captured by $\mbox{Bias} = \frac{1}{n}\sum_i(\hat{y}_{i} - y_{i} )$; ii. the average size of the error is given by $\mbox{RMSE} = \sqrt{\frac{1}{n} \sum_i (y_i - \hat{y}_i )^2}$; %the degree to which predictions and observations are linearly related is described by Pearson's correlation coefficient
%$\rho = \frac{\sum_i \left( \hat{y}_i - \bar{\hat{y}} \right) \left( y_i - \bar{y} \right)}{\sqrt{\sum_i \left( \hat{y}_i - \bar{\hat{y}} \right)^2} \sqrt{\sum_i \left( y_i - \bar{y} \right)^2}}$; 
iii. the silicon estimates' ability to correctly order the observations on a line is measured by the Spearman's rank correlation coefficient $\rho \;=\; \frac{\sum\limits_i \Bigl( R_{\hat{y}_i} - \overline{R_{\hat{y}}} \Bigr)\,\Bigl( R_{y_i} - \overline{R_y} \Bigr)}
{\sqrt{\sum\limits_i \Bigl(R_{\hat{y}_i} - \overline{R_{\hat{y}}}\Bigr)^2}\,\sqrt{\sum\limits_i \Bigl(R_{y_i} - \overline{R_y}\Bigr)^2}}$, where $R$ denotes the rank of a given observation / prediction; iv. $\mbox{Coverage}_{90\%} = \frac{1}{n} \sum_{i=1}^n [ \hat{y}^{5\%}_i \leq y_i \leq \hat{y}^{95\%}_i ]$ measures the calibration of the estimates' credibility intervals. To measure the agreement between \texttt{PoSSUM}'s estimated silicon density and that implied by traditional polls I use the overlap coefficient $\mbox{OVL}[p, \hat{p}] 
= \int_{\mathcal{Y}} \min\bigl(p(y),\, \hat{p}(y)\bigr)\, dy$.\\ %This is the preferable measure of distributional agreement in the context of polling estimates, as the interest lies not in penalising mismatches in distributional shapes, but in quantifying the degree to which two competing approaches would come to the same conclusion as to the winner of a particular contest. $\mbox{OVL}$ further possesses the virtue of being clearly interpretable as the proportion of shared density across two distributions.\\

\noindent Beyond the ability to predict election results, \texttt{PoSSUM} seeks to enable dynamic measurement of public opinion. To establish \texttt{PoSSUM} amongst state-of-the-arts polling methodologies I must prove it enables \emph{novel learning} and that its estimates are \emph{human-aligned} and \emph{time-sensitive}. I will present an assessment of novelty and alignment exclusively for the last pre-election \texttt{PoSSUM} poll, upon which I build my final pre-election predictions, and use the full set of polls fielded during the campaign to establish time sensitivity.\\

\noindent \textbf{Crosstab-level Comparisons} \hspace{10pt} I wish to establish \texttt{PoSSUM}'s properties at multiple levels of analysis (e.g. by age, gender, education, etc.). Exact election returns at the crosstab level are unavailable. To remedy this, during the campaign I collect a benchmark crosstab-level dataset of polls. The collection protocol is as follows: i) I monitor \url{https://projects.fivethirtyeight.com/polls/president-general/2024/} in search for newly published polls duting the campaign; ii) a poll is identified as a candidate for collection if it shares at least $1$ fieldwork date with \texttt{PoSSUM}; iii) the poll is discarded if crosstabs are behind a paywall; iv) the poll is discarded unless crosstabs are presented in an easily accessible and transcribable format (either pdf, excel, html, or similar). Amazon Textract is used to convert these pdfs into csv files, and these are manually inspected to correct discrepancies; v) weighted or unweighted counts from the crosstabs are stored -- polls that provide percentages without the ability to recover counts are discarded. After harmonisation, enough polling data is available to compare \texttt{PoSSUM} with reference pollsters at the levels of gender, ethnicity, age and education-level. A comparison is also made at the level of $2020$ past vote, though here the available polls are far fewer (see Figures \ref{fig:poll_coverage_nat} to \ref{fig:poll_coverage_vote}). The database of reference polls is available on the \texttt{GitHub} repository. \\

\noindent \textbf{Novel learning} is operationalised as the ability to correctly capture the \emph{direction} and \emph{magnitude} of changes over past election results. I limit the analysis to a single outcome -- namely the $Republican - Democrat$ margin, $\mathcal{M} = \pi_{j = R} - \pi_{j = D}$. Novelty can be assessed at multiple levels of analysis: at the national and area-level, I compare estimates against results known with certainty via official election returns; at the crosstab level (e.g. by age, gender, education, etc.) the expanded Stratification Frame (see Section \ref{sec::extend}) can be aggregated at the relevant levels of analysis to provide $2020$ vote benchmarks. 

Learning an optimal \emph{direction} of change in public opinion implies the ability to minimise the \emph{probability of misdirection}: take $d = \mbox{sign}(\mathcal{M}^{2024} - \mathcal{M}^{2020})$ to be the observed direction of change, and $\tilde{d}_s = \mbox{sign}(\tilde{\mathcal{M}}_s^{2024} - \mathcal{M}^{2020})$ to be a single draw from the posterior distribution of predicted changes. The misdirection probability can be estimated as $ \widehat{\Pr}(d \neq \tilde{d}) = \mid d - \frac{1}{S}\sum_s H(\tilde{d}_s) \mid $, where $H$ is the Heaviside step function.

An optimal learner of change must correctly capture direction as well as the \emph{magnitude} of change. The \emph{bias} of the learner's change prediction provides us with a comprehensive evaluation metric: take $\Delta =\mathcal{M}^{2024} - \mathcal{M}^{2020}$ to be the observed change in the margin; $\hat{\Delta} = \hat{\mathcal{M}}^{2024} - \mathcal{M}^{2020}$ is the point estimate of the predicted change; then $bias =\hat{\Delta} - \Delta$.\\

\noindent \textbf{Human alignment} can be demonstrated by verifying that the observed election outcomes remain consistent with the predictions derived from the \texttt{PoSSUM} posterior. In other words, the empirical data do not provide sufficient evidence to reject \texttt{PoSSUM} as a plausible data-generating process. We can cast this as a hypothesis test: treat the \texttt{PoSSUM} posterior as the null hypothesis (our best prior guess at the result), and posit as the alternative hypothesis that the true result deviates from that distribution. We then compute a \emph{p-value} -- the probability that, under the null distribution, we observe a more statistically extreme value than the election result -- to assess how plausible the observed result is given \texttt{PoSSUM}'s posterior distribution. To handicap the null, I calculate a one-sided \emph{p-value} for the side of the posterior distribution (relative to its median) that contains the observed value. The Monte-Carlo estimate of this p-value is calculated as follows:

\[
\hat{p} = 
\begin{cases}
\displaystyle \frac{1}{S} \sum_s \bigl[ \tilde{\mathcal{M}}_s \geq \mathcal{M}_{\text{obs}} \bigr], & \text{if } \mathcal{M}_{\text{obs}} \geq \mathrm{median}\{\mathcal{M}_s, \dots, \mathcal{M}_{S}\}, \\[1em]
\displaystyle \frac{1}{S} \sum_s \bigl[ \tilde{\mathcal{M}}_s \leq \mathcal{M}_{\text{obs}} \bigr], & \text{if } \mathcal{M}_{\text{obs}} < \mathrm{median}\{\mathcal{M}_1, \dots, \mathcal{M}_S\}.
\end{cases}
\]

\noindent \textbf{Time sensitivity} can similarly be established by comparing temporal dynamics in \texttt{PoSSUM} estimates with those measured by the polling average. Again I rely on Spearman's rank correlation coefficient over time -- large and statistically significant correlation with the polling average at multiple levels of analysis would imply the \texttt{PoSSUM} estimates are \emph{time-sensitive}. 

Unfortunately, given the setup of this set of \texttt{PoSSUM} polls, I expect the time-sensitivity test to be largely inconclusive at this stage. \texttt{PoSSUM} polls were in the field for just $5$ separate instances during the election campaign, a relatively small number to assess temporal congruence with polling-average dynamics. Moreover, the sample size of each \texttt{PoSSUM} poll was relatively small (approximately $n = 1,000$ in each instance) and smoothed over time (fieldwork dates spanning approximately $1$ week) making it challenging to establish patterns of over-time change with high confidence. I will therefore present a single measure of temporal congruence between \texttt{PoSSUM} and the polling average at this stage -- namely the estimated probability that the temporal correlation between the two is positive -- $\widehat{\Pr}(\rho^{\tau} > 0) = \frac{1}{S} \sum_s [\tilde{\rho}^\tau_s>0]$, where $\rho^{\tau}$ indicates the temporal Spearman rank correlation.\\

\subsection{Electoral Predictive Power}

\begin{figure}
    \centering
    \includegraphics[width=1\linewidth]{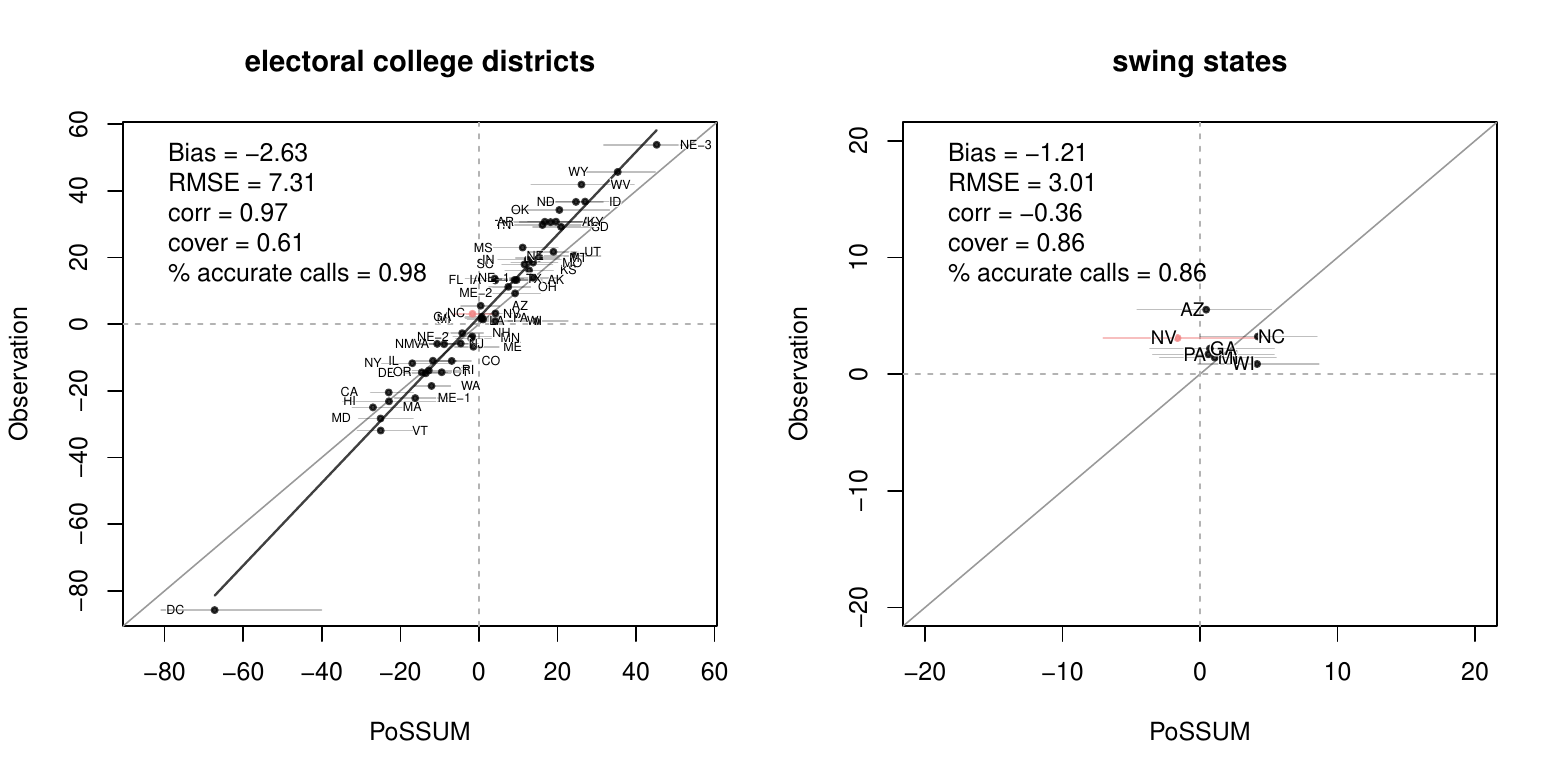}
    \caption{State-level predictive power on the Republican - Democrat margin. Training data includes highly speculative records. Model fit to the final \texttt{PoSSUM} poll, fielded from the $17^{th}$ to the $26^{th}$ of October.}
    \label{fig:RDmargin_Speculative_10.26}
\end{figure}

\noindent \textbf{Performance on the $R-D$ Margin} \hspace{10pt}
Figure \ref{fig:RDmargin_Speculative_10.26} presents the comparison of estimates of the $Republican - Democrat$ margin from \texttt{PoSSUM}'s last campaign poll, conducted between October $17^{th}$ and $26^{th}$, against the observed state-level election results. The poll is based on a sample of $1,056$ synthetic responses -- a relatively small sample by traditional standards, and tiny relative to previous MrP efforts on US data \cite{wang2015forecasting,lauderdale2020model}, it provides above-average informational value.\\% I focus on the margin as it allows to determine the winner of each electoral constituency in a two-party system -- predictive power on the margin translates to the overall election results. \\

\noindent \texttt{PoSSUM} estimates correctly predicted the Republican win, by calling $98\%$ of electoral college districts correctly -- the state of Nevada being the single discordant note. The Spearman rank correlation between point-estimates and observations for the Republican - Democrat margin is generally high, at $0.97$, indicating \texttt{PoSSUM}'s ability to accurately order the electoral college districts by the difference in voting intention between Republicans and Democrats.  The somewhat elevated RMSE at around $7.3$ percentage points emerges as a result of significant attenuation bis, does not affect the ability to call states correctly one way or another, and concentrates the error amongst non-competitive states. On average, margin estimates are characterised by an anti-Trump bias worth $2.6$ percentage points, a perhaps surprising finding given the perception of $\mathbb{X}$ users as being substantially more right-leaning. Coverage of the estimates is unsatisfactory low, with a mere $61\%$ of observed margins falling in the $90\%$ credibility intervals. 

\noindent RMSE on the seven competitive swing-states is well contained, at around $3$ points -- attenuation in less competitive states is responsible for a large overall RMSE. Anti-Trump bias is similarly reduced here to $1.2$ points, and is concentrated in the western states of Arizona and Nevada -- the crosstab comparisons against state of the arts pollsters in Figure \ref{fig:dynamic_ethnicity_comparison} reveal \texttt{PoSSUM}'s inability to capture the Hispanic shift towards Republicans on the last pre-election poll to be the likely cause. Coverage recovers to acceptable levels  for these states, with Arizona being the sole state falling outside of the $90\%$ credibility interval. \texttt{PoSSUM} is unable to rank swing states by their Republican - Democrat margin, as evidenced by the negative Spearman correlation coefficient.  \\

\begin{figure}
    \centering
    \includegraphics[width=1\linewidth]{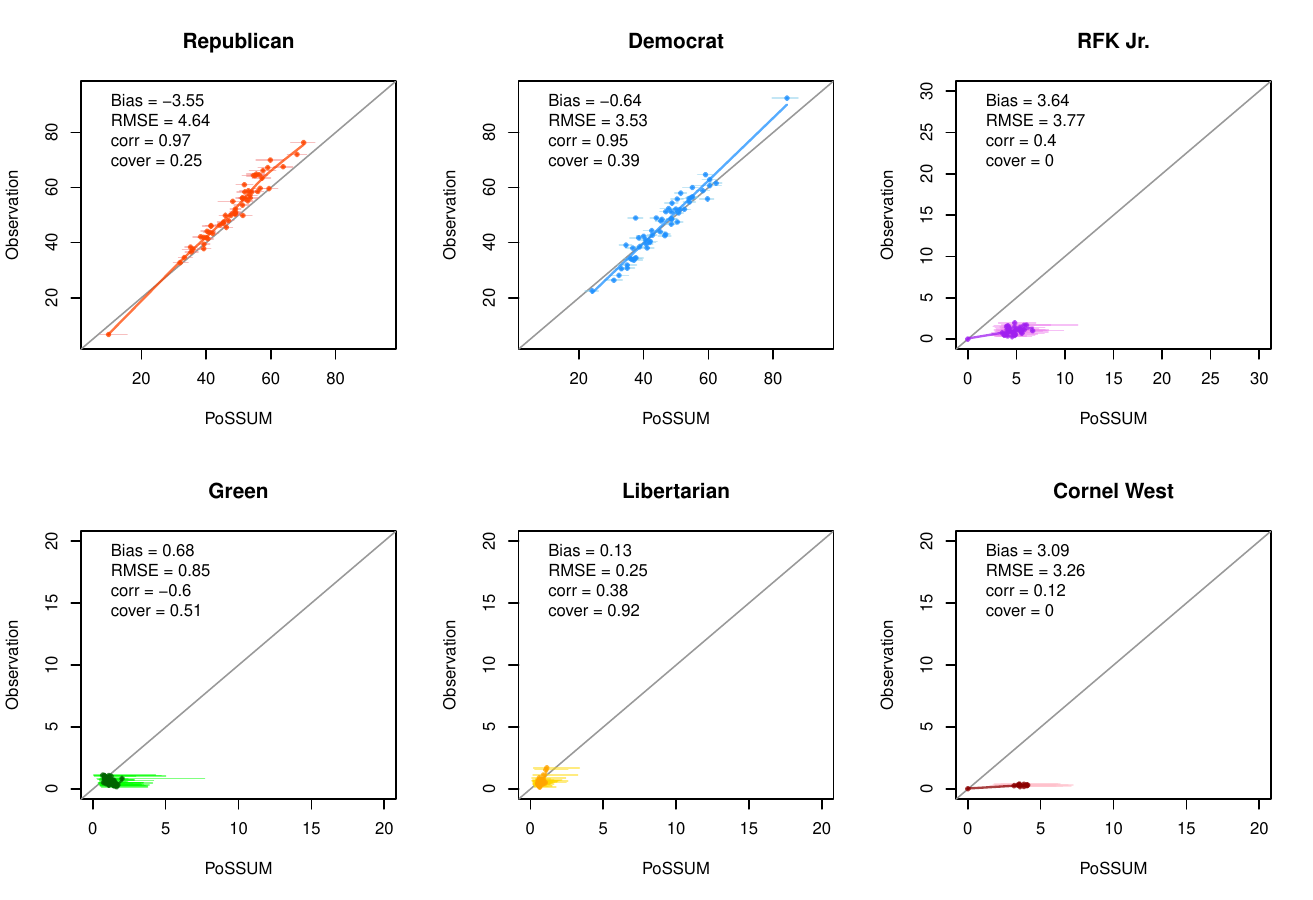}
    \caption{State-level predictive power on vote share by candidate. Training data includes highly speculative records. Model fit to pooled dataset of $5$ polls, fielded from the $15^{th}$ of August to the $26^{th}$ of October.}
    \label{fig:VoteShare_Speculative}
\end{figure}

 \noindent \textbf{Third-Party Under-Performance} \hspace{10pt}
 Figure \ref{fig:VoteShare_Speculative} presents the party-wise vote share estimates, generated from a model fit to an aggregate sample of every poll fielded during the campaign, from August $15^{th}$ to October $27^{th}$, for a total of $4,982$ synthetic responses. I use this aggregate sample to attempt a meaningful appraisal of performance on third-parties, whose voters are otherwise too scarce on any given poll to appropriately characterise performance. With the exception of Libertarians, whose state-level vote share is estimated with low RMSE and close-to-perfect coverage, third party prediction poses a challenge for \texttt{PoSSUM}. There is large positive bias in the state-level vote share of RFK Jr. ($+3.6\%$) and Cornel West ($+3\%$). A number of speculative reasons for this over-estimation -- such as lack of context available to the LLM with regards to Kennedy's sui-generis late-endorsement of Trump, or the LLM's prior beliefs about Dr. West's chances as a political candidate given his quasi-celebrity status -- can be articulated. The systematic partitioning of third-party error according to likely sources is beyond the scope of this paper, though I will discuss these further in Section \ref{sec:discussion}, outlining a clear set of hypotheses to be tested under an experimental setting in future research.

This bias is large enough to have played a consequential role in the predictions. For example in the pivotal state of Wisconsin, where the observed Republican - Democrat margin was a mere $0.8\%$, and third-parties ended up getting less than $1\%$ of the vote, PoSSUM estimates the margin to be anywhere between $1\%$ and $6.6\%$. This is a likely consequence of the over-estimation of West and Kennedy, tallying respectively between $1.5\%$ to $4.8\%$, and between $2\%$ and $5.4\%$. In this scenario third-parties effectively act to draw votes away from the two major parties. Whilst Kennedy's vote is generally considered to be contested by both Republicans and Democrats, West's appeal is limited to Democrats, and as such his over-estimation worked as the equivalent of depressing Democratic turnout.\\ %Why is this concerning ? Because \texttt{PoSSUM} may have reached the right conclusion -- Trump winning Wisconsin -- in part through the wrong inference -- Democrats fleeing to West. One could argue this is still picking up something useful -- namely plausible democratic voters fleeing away from the party. This sort of reasoning holds in the peculiar two-party system in the US, though it provides few assurances as to generalisable performance in traditional multi-party systems. \\

\noindent \textbf{Comparing with SoTA Pollsters} \hspace{10pt} I obtain a collection of state-level polls from the polling aggregator \texttt{FiveThirtyEight}(\url{https://projects.fivethirtyeight.com/polls/data/president_polls_historical.csv}). Out of the universe of state-level polls fielded since the $25^{th}$ of July $2024$, I retain the most recent poll related to a Trump v. Harris matchup for each state and pollster. A total of $464$ state-level polls remain, fielded by $137$ pollsters, for $48$ electoral college districts. To perform an apples-to-apples comparison with \texttt{PoSSUM}'s posterior samples for state-level vote-share estimates, and account for their probability distribution, I generate an equivalent amount of samples from a Dirichlet distribution (conjugate posterior for the party-choice probability), with $\bm{\alpha} = N_1 \dots N_J$, where $N_j \mbox{ } \forall j \in \{1,...,J\}$ is the number of declared supporters of candidate $j$.\\

\noindent The final comparison involves calculating and comparing the performance metrics described above for the latest pre-election polls fielded by both \texttt{PoSSUM} and the reference pollster. Not every pollster fields in every district, hence the performance comparison will encompass a different set of electoral college districts for each pollster. Presumably pollsters select into fielding in specific states, either because they can assure good performance to clients due to special knowledge, or because the states in question are of broad public interest. Hence the comparison should be broadly favourable to these rival methodologies.\\

\begin{figure}
    \centering
    \includegraphics[width=\linewidth]{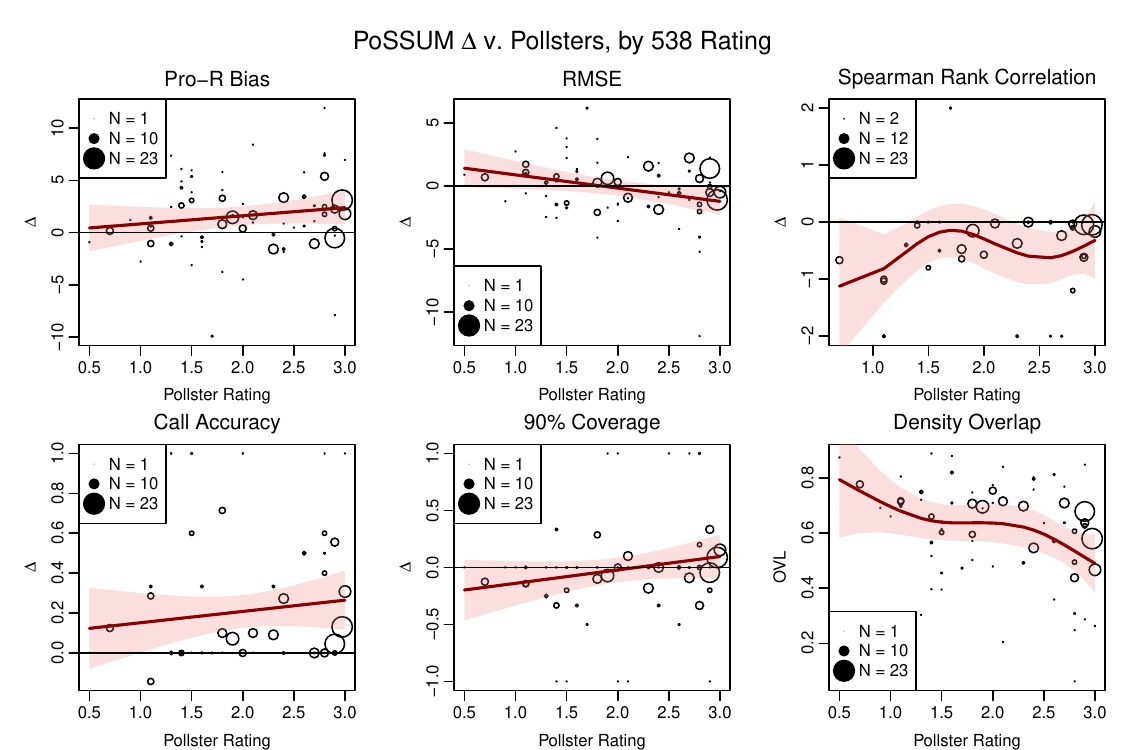}
    \caption{\texttt{PoSSUM} performance difference over reference pollsters ($\Delta = \texttt{PoSSUM} - \mbox{Pollster}$) across multiple evaluation metrics by \texttt{FiveThirtyEight} pollster ratings. Each comparison is limited to the latest pre-election polls fielded by both parties, in the electoral college districts where the reference pollster has fielded during the campaign.}
    \label{fig:Delta_comparison}
\end{figure}

\begin{figure}
    \centering
    \includegraphics[width=\linewidth]{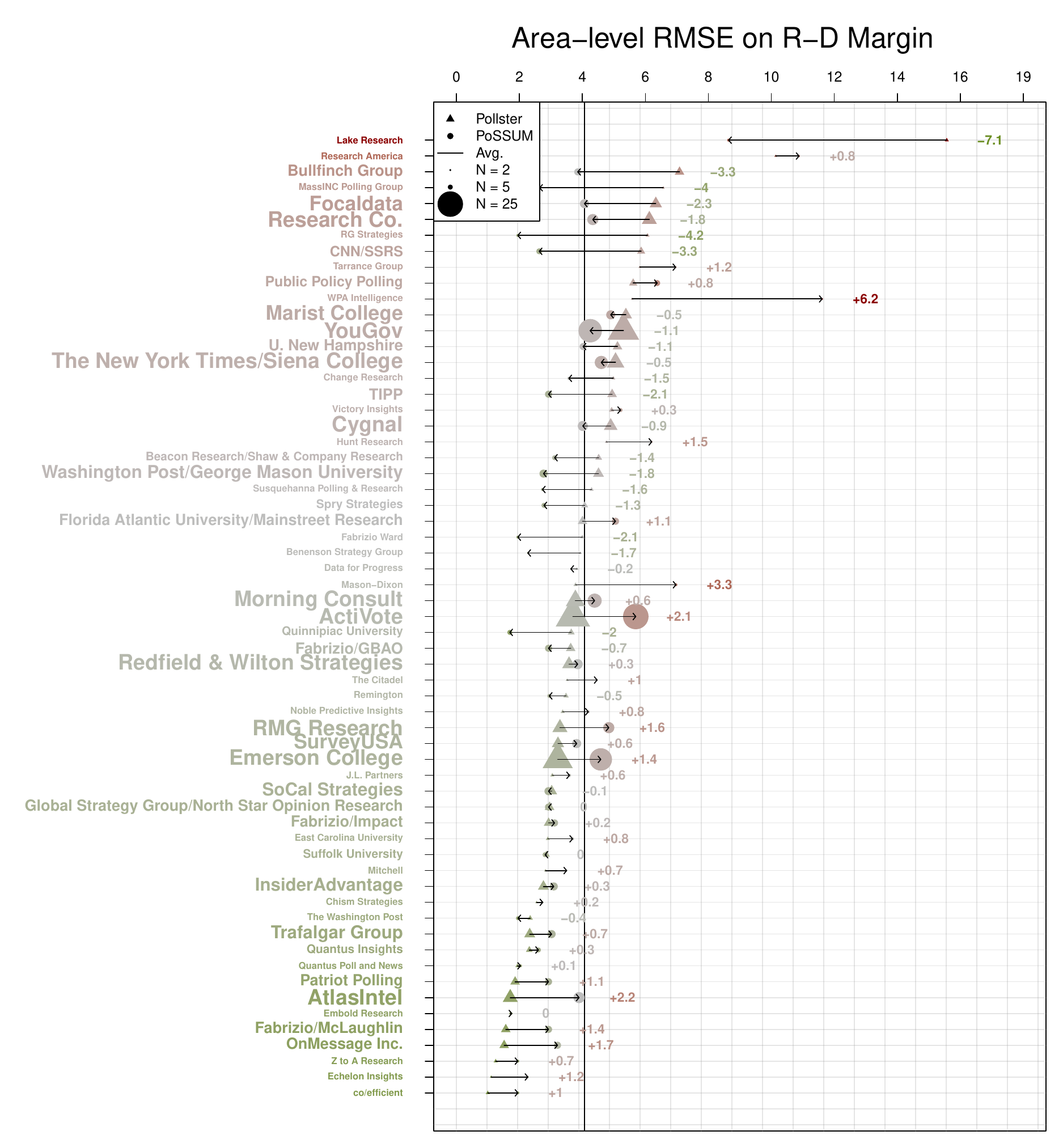}
    \caption{Comparison of RMSE on the area-level Republican margin for each reference pollster ($\blacktriangle$) v. \texttt{PoSSUM} ($\large\bullet$). Arrow length reflects the difference between estimates, and arrowheads point toward \texttt{PoSSUM}. Pollsters are listed from highest (top) to lowest (bottom) RMSE. The red–green color scale indicates worse–better performance relative to the average. \texttt{PoSSUM}'s RMSE difference ($\Delta$) is displayed to the right of each comparison: green if \texttt{PoSSUM}'s RMSE is lower than the reference pollster, and red if higher. Symbol and label sizes are proportional to the number of areas compared. Only pollsters with data from more than one area are included.}
    \label{fig:RMSE_comparison}
\end{figure}

\noindent Figure \ref{fig:Delta_comparison} presents the performance difference on the Republican - Democrat margin $\Delta$ between \texttt{PoSSUM} and the reference pollsters, by \texttt{FiveThirtyEight} pollster rating. Despite non-negligible anti-Republican Bias (Figure \ref{fig:VoteShare_Speculative}), \texttt{PoSSUM} is generally more Republican-leaning than other pollsters. This difference has a clear gradient: pollsters rated lowest by \texttt{FiveThirtyEight} had bias indistinguishable from \texttt{PoSSUM}, whilst state-of-the-arts pollsters significantly favoured Democrats on average. The distribution of this difference is detailed in Figure \ref{fig:Bias_comparison}. The lower anti-Republican Bias translated to \texttt{PoSSUM} obtaining lower average error (RMSE) than state-of-the-arts pollsters, though \texttt{PoSSUM}'s RMSE performance was generally indistinguishable from that of the average pollster. Figure \ref{fig:RMSE_comparison} presents the distribution of \texttt{PoSSUM}'s $\Delta$. A few notable comparisons, ordered in terms of the rival pollster's performance, worst to best: the AI pollster's RMSE on the state-level margin was on average $3.3$ points lower than CNN/SSRS; $0.5$ points lower than Marist and NYT/Siena; $1.1$ points lower than YouGov; $1.8$ points lower than Washington Post / George Mason. It was however $0.6$ points larger than Morning Consult's; $2.1$ points larger than ActiVote; $1.4$ points greater than Emerson; $0.7$ points greater than Trafalgar; $2.2$ points greater than AtlasIntel and $1.4$ points greater than Fabrizio/McLaughlin .

%It is challenging to fairly compare \texttt{PoSSUM} with rival pollsters. \texttt{PoSSUM} is alone in generating estimates of preferences for every electoral college district throughout the campaign. By necessity then a comparison will tend to favour reference pollsters, who presumably only field polls in states where they can guarantee a relatively high level of performance. On the other hand it can be argued that more frequently polled states are the most competitive / more directly of public interest -- hence good performance in frequently polled states is, in some sense, all that matters. Moreover there exists a dramatic resource-gap between traditional pollsters and \texttt{PoSSUM}. Table \ref{tab:cost} presents the 

\noindent 

\pagebreak

\subsection{Novel Learning, Human Alignment and Time-Sensitivity}

\begin{figure}
    \centering
    \includegraphics[width=\linewidth]{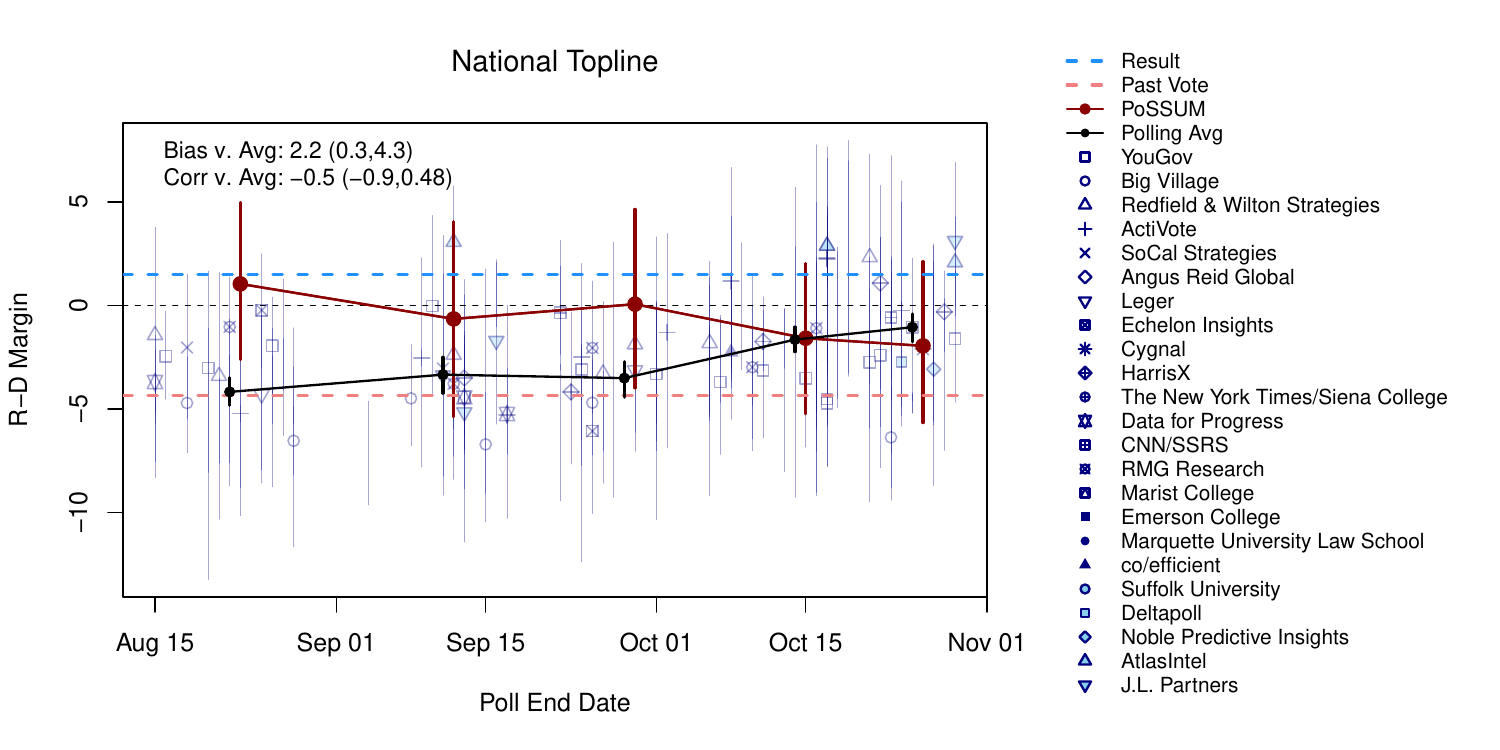}
    \caption{National-level \texttt{PoSSUM} estimates over the course of the campaign, shown alongside individual polls overlapping \texttt{PoSSUM}'s fieldwork periods, the aggregated polling average for each \texttt{PoSSUM} fieldwork window, and the observed $2020$ and $2024$ outcomes.}
    \label{fig:dynamic_nat_comparison}
\end{figure}

\begin{figure}
    \centering
    \includegraphics[width=\linewidth]{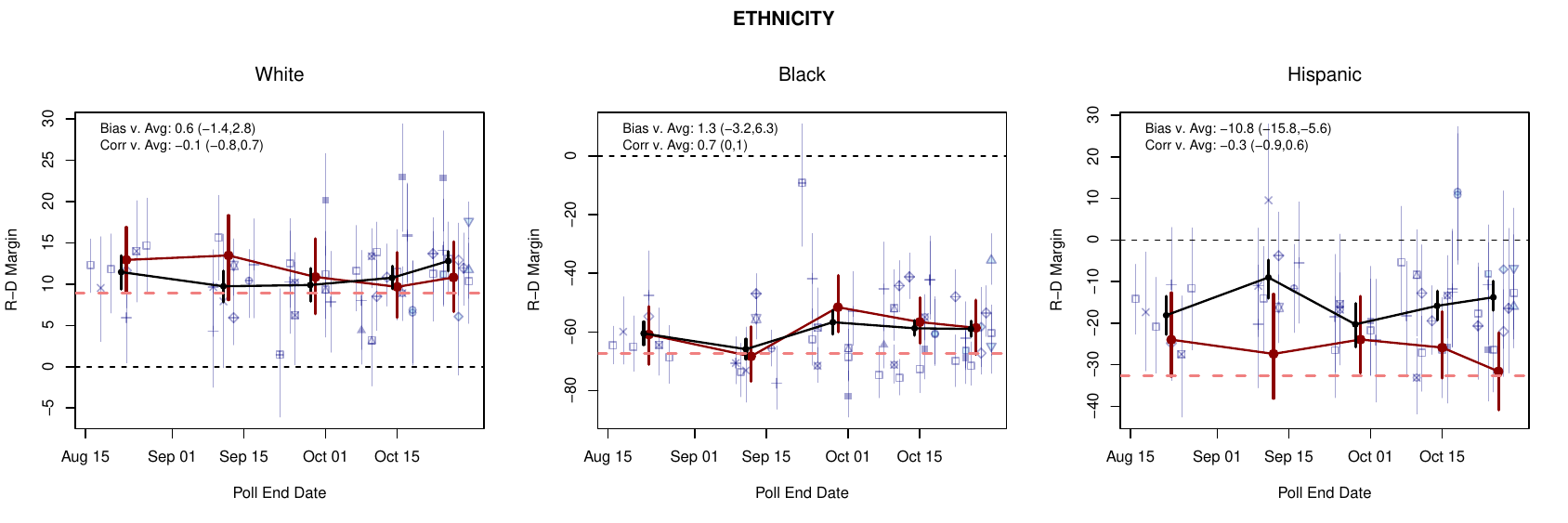}
    \caption{Ethnicity-level \texttt{PoSSUM} estimates over the course of the campaign, shown alongside individual polls overlapping \texttt{PoSSUM}'s fieldwork periods, the aggregated polling average for each \texttt{PoSSUM} fieldwork window, and the reference preferences from $2020$.}
    \label{fig:dynamic_ethnicity_comparison}
\end{figure}

Figures \ref{fig:dynamic_nat_comparison}, \ref{fig:dynamic_ethnicity_comparison} and  Figures \ref{fig:dynamic_gender_comparison} to \ref{fig:dynamic_age_comparison} in the Appendix present an assessment of the novel learning, human alignment and time-sensitivity of \texttt{PoSSUM} estimates, with respect to observed election results and polling averages. Table \ref{tab:possum_consistency} summarises these results, according to the metrics presented at the beginning of this section. To interpret the results in context, the table also presents a calculation of the same metrics for state-of-the-arts pollster \texttt{YouGov}.\\

\noindent I find \textbf{strong evidence in favour of \texttt{PoSSUM}'s ability to generate novel learning}. The evidence in favour of this is  as follows: a) \texttt{PoSSUM}'s point estimate for the change in support is in the correct direction for all crosstabs under considerations, with the exception of those aged $65+$ -- and even here \texttt{PoSSUM} correctly predicts this is the only crosstab not to experience positive change on the Republican margin; b) the probability of incorrectly predicting the direction of the change for any given crosstab is generally small, proportional to the size of the observed change, and well below random chance for most crosstabs -- again with the exception of $65+$ voters where we see a toss-up; d) the crosstab-level bias of the predictions is below $4\%$ for most crosstabs, a state-of-the-arts level of performance at the crosstab level. Exceptions to this level of error, varying in degrees of severity, are seen in Hispanics, those in the $18-25$ and $35-44$ age brackets, and those who have obtained college degrees. 

When comparing \texttt{PoSSUM} and \texttt{YouGov}'s performance on these crosstabs, we see point-estimate error of similar or greater magnitude for \texttt{YouGov}, coupled with a higher frequency of misdirection (largely driven by greater anti-Republican bias in the topline). The higher frequency and sample-size of \texttt{YouGov} polling over the campaign translate to narrower prediction intervals, which leading to more extreme misdirection probabilities -- tending to $0$ when the point-estimate is in the right direction, and tending to $1$ vice-versa. It is notable that \texttt{YouGov} fails to capture any significant dynamics in the $65+$ demographic.\\

\noindent There is \textbf{moderate evidence favouring human alignment} of \texttt{PoSSUM}'s estimates with observed outcomes. At the $10\%$ significance level, we fail to reject the \texttt{PoSSUM} posterior for all crosstabs except Hispanics, the $18-25$ and $25-44$ age brackets, and college-educated voters; the national topline estimate similarly appears statistically different at the $10\%$ level. By contrast, at the more relaxed $5\%$ threshold, we find that both the national topline and the $35-44$ bracket remain statistically compatible with the \texttt{PoSSUM} distribution.

In absolute terms, most of the \texttt{PoSSUM} crosstab estimates are closely aligned with the observed data. Notable exceptions include the topline result and the $35-44$ cohort, which lie near the extreme edge of plausibility, and the observed shifts in Hispanic and $18-25$ voters (albeit with caveats), which appear misaligned. Comparing \texttt{PoSSUM} to \texttt{YouGov}, the former generally yields larger \(p\)-values---only the Female and Hispanic crosstabs produce lower \(p\)-values under \texttt{YouGov}---suggesting \texttt{PoSSUM}'s estimates are more congruent with actual results. Indeed, the observed margins for topline, Black, $35-44$, and $65+$ groups fall completely outside \texttt{YouGov}'s plausible range. In a complex electoral context, \texttt{PoSSUM}'s learned dynamics thus appear more closely tethered to human beings on the ground than those derived from \texttt{YouGov}’s methods.\\

\noindent The test aimed at \textbf{establishing the time-sensitivity of \texttt{PoSSUM} estimates is inconclusive.} At face value, the probability of a positive temporal correlation is below random expectations for nearly every crosstab, except for the Female, Black, and $25-34$ groups---and of these, the result is decisively positive only for the Black crosstab. By contrast, a similar test applied to the state-of-the-art pollster \texttt{YouGov} shows similarly lackluster performance, with only the Black, Hispanic, $35-44$, and $65+$ estimates exhibiting a positive temporal correlation.  Several factors may explain this broadly negative performance. First, the method used to compute the correlation may have been primed to report negative results. Specifically, the reference polling average itself is suspect, inasmuch as it is formed by averaging polls conducted at different points in time, overlapping only partially with \texttt{PoSSUM}'s fieldwork. This procedure likely injects noise and smooths the reference benchmark. Similarly, the benchmark derived from \texttt{YouGov} data is also noisy, arising from aggregating multiple polls during each \texttt{PoSSUM} fieldwork window, with potential dependencies among polls that are assumed to be independent (e.g., \texttt{YouGov} polls for CBS or Yahoo vs.\ those for The Economist). Moreover, aggregating data based on partial overlap with \texttt{PoSSUM} fieldwork periods likely exacerbates this noise, further smoothing and confounding temporal patterns. Although introducing these noisy estimates does not greatly diminish our capacity to evaluate the final \texttt{PoSSUM} fieldwork snapshot against a polling average or the \texttt{YouGov} measurements, it complicates the assessment of changes over time. The resulting compounded noise in the benchmark introduces uncertainty that impedes a conclusive evaluation of \texttt{PoSSUM}'s time-sensitivity in relation to observed outcomes.\\

\noindent In general, \texttt{PoSSUM} exhibits some difficulties with Hispanic voters, for whom there is a $17.8$ percentage-point shortfall in the estimated margin shift, a discrepancy without a clear immediate explanation. The lack of novel learning on Hispanic voters in the final polling wave directly contributed to weaker performance in Nevada and Arizona, whereas more accurately captured swings among Black voters enabled correct predictions in Georgia and North Carolina. By contrast, although the error for the $18-25$ age bracket is even larger at $19.5$ points relative to the polling average, this issue is less concerning. Part of the discrepancy can be attributed to inconsistencies in how various pollsters define and aggregate age categories (see the footnote in Table \ref{tab:possum_consistency}). As illustrated in Figure~\ref{fig:dynamic_edu_comparison}, many polls combine the $25-34$ and $18-24$ brackets, and those that do not show alignment with \texttt{PoSSUM} throughout the campaign. Finally, there are smaller underestimations of the Republican margin for the $35-44$ bracket and for college-degree voters, at $5.9$ and $10$ percentage points, respectively. Although these discrepancies lack a clear proximate cause, their magnitudes are less pronounced than those for Hispanics and the $18-25$ bracket.

% Define commands for tick/cross
\newcommand{\cmark}{{\checkmark}}
\newcommand{\xmark}{{\textsf{x}}}
\definecolor{bloodred}{rgb}{0.6, 0, 0}

\begin{table}[htp!]
\centering
\scalebox{1}{
\begin{threeparttable}
\begin{tabular}{r|cc|ccc|c|c}
\hline
\hline
\emph{Crosstab} & $\Delta^{PoSSUM}_{R-D}  $ & $\Delta^{Avg.}_{R-D}   $   &$\hat{d} = d$  & $\Pr(\hat{d} \neq d)$ & \emph{bias}  & \emph{p-value} & $\Pr(\rho^{\tau} > 0)$\\
\midrule
  \hline
National         &  2.5 & 5.9 & \cmark & 0.143 & -3.3 & \textcolor{bloodred}{0.087}& \textcolor{bloodred}{0.179} \\ 
Male             &  3.0 & 5.2 & \cmark & 0.127 & -2.1 & 0.238 & \textcolor{bloodred}{0.183} \\ 
Female            & 2.0 & 1.3 & \cmark & 0.256 & 0.7 & 0.393 & 0.552 \\ 
White             & 1.9 & 3.9 & \cmark & 0.222 & -2.0 & 0.214 & \textcolor{bloodred}{0.381} \\
Black             & 8.7 & 8.4 & \cmark & 0.060 & 0.4 & 0.462 & 0.931 \\ 
Hispanic             & 1.1 & 18.9 & \cmark & 0.419 & \textcolor{bloodred}{-17.8} & \textcolor{bloodred}{0.004} & \textcolor{bloodred}{0.270} \\ 
Age 18--25\tnote{1}         & 1.9 & 21.4 & \cmark & 0.417 & \textcolor{bloodred}{-19.5} & \textcolor{bloodred}{0.004} & \textcolor{bloodred}{0.423} \\ 
Age 25--34         & 8.2 & 11.8 & \cmark & 0.026 & -3.6 & 0.183 & 0.597 \\ 
Age 35--44         & 4.4 & 10.3 & \cmark & 0.135 & \textcolor{bloodred}{-5.9} & \textcolor{bloodred}{0.091} & \textcolor{bloodred}{0.435} \\ 
Age 45--54         &2.1 & 2.3 & \cmark & 0.313 & -0.2 & 0.480 & \textcolor{bloodred}{0.375} \\ 
Age 55--64         &2.0 & 0.9 & \cmark & 0.292 & 1.1 & 0.389 & \textcolor{bloodred}{0.440} \\ 
Age 65+            &  0.0 & -3.0 & \textcolor{bloodred}{\xmark} & \textcolor{bloodred}{0.500} & 3.1 & 0.175 & \textcolor{bloodred}{0.349} \\ 
No College Degree\tnote{2}      &0.7 & 0.7 & \cmark & 0.403 & 0.0 & 0.500 & \textcolor{bloodred}{0.464} \\ 
College Degree  &3.5 & 13.5 & \cmark & 0.048 & \textcolor{bloodred}{-10.0} & \textcolor{bloodred}{0.000} & \textcolor{bloodred}{0.226} \\ 

\hline
\hline

& $\Delta^{YouGov}_{R-D}  $ & $\Delta^{Avg.}_{R-D} \tnote{3}   $   &$\hat{d} = d$  & $\Pr(\hat{d} \neq d)$ & \emph{bias}  & \emph{p-value} & $\Pr(\rho^{\tau} > 0)$\\
\midrule
National           & 1.9 & 5.9 & \cmark & 0.018 & -4.0 & \textcolor{bloodred}{0.000} & \textcolor{bloodred}{0.377} \\ 
Male             & 3.1 & 5.7 & \cmark & 0.067 & -2.6 & 0.119 & \textcolor{bloodred}{0.391} \\ 
Female            &  1.1 & 1.4 & \cmark & 0.286 & -0.3 & 0.558 & \textcolor{bloodred}{0.359} \\ 
White             &  2.6 & 4.3 & \cmark & 0.036 & -1.7 & 0.188 & \textcolor{bloodred}{0.256} \\ 
Black             &  0.5 & 10.6 & \cmark & 0.446 & \textcolor{bloodred}{-10.1} & \textcolor{bloodred}{0.000} & 0.746 \\ 
Hispanic             & 13.9 & 20.2 & \cmark & 0.000 & \textcolor{bloodred}{-6.3} & \textcolor{bloodred}{0.069} & 0.706 \\ 
Age 18--25\tnote{1}         &13.6 & 23.4 & \cmark & 0.000 & \textcolor{bloodred}{-9.8} & \textcolor{bloodred}{0.006} & \textcolor{bloodred}{0.313} \\ 
Age 35--44         &-1.4 & 13.3 & \textcolor{bloodred}{\xmark} & \textcolor{bloodred}{0.692} & \textcolor{bloodred}{-14.6} & \textcolor{bloodred}{0.000} & 0.567 \\ 
Age 45--54         & 3.8 & 1.5 & \cmark & 0.056 & 2.3 & 0.196 & \textcolor{bloodred}{0.276} \\ 
Age 55--64         & -0.8 & 1.4 & \textcolor{bloodred}{\xmark} & \textcolor{bloodred}{0.617} & -2.3 & 0.181 & \textcolor{bloodred}{0.292} \\ 
Age 65+            & 5.2 & -5.3 & \textcolor{bloodred}{\xmark} & \textcolor{bloodred}{0.966} & \textcolor{bloodred}{10.5} & \textcolor{bloodred}{0.000} & 0.569 \\ 
\hline
\hline
\end{tabular}
\begin{tablenotes}
\footnotesize
\item[1] Note that few pollsters collect data for the $18-25$ category, and in general age categories are grossly misaligned. The reference polls for \texttt{PoSSUM's} age categories are picked to include any age category which has midpoint distance which is $\leq 5$ years from \texttt{PoSSUM's} category. For the $18-25$ category, this produces a severe smoothing bias -- if one nets-out polls which also overlap with the $25-34$ category, the polling average looks much more similar -- though still greater on average -- than the \texttt{PoSSUM} estimate for this crosstab. \texttt{YouGov}'s natural age categories are $[18-29, 30-44, 45-64, 65+]$, hence here too there is a fair degree of misalignment -- no reference comparison for the $25-34$ category is possible. 
\item[2] \texttt{YouGov} does not provide a breakdown by college education. No other single alternative pollster is consistent enough in their crosstab reporting to provide a satisfactory amount of comparison points.  
\item[3] The polling average against which I compare the \texttt{YouGov} data is calculated excluding \texttt{YouGov} -- hence it is slightly different from the average I compare the $\texttt{PoSSUM}$ crosstabs against. 
\end{tablenotes}
\end{threeparttable}
}
\caption{Assessment of \emph{novel learning} (Direction, Probability of Mis-Direction $\Pr(\hat{d}\neq d)$, Prediction Bias \emph{err}), \emph{human exchangeability} (\emph{p-value} with $H_0: \texttt{PoSSUM}$ posterior distribution) and \emph{time-sensitivity} (chance of positive correlation across \texttt{PoSSUM} fieldwork dates $\Pr(\rho^\tau > 0)$) in \texttt{PoSSUM}'s final pre-election poll. In \textcolor{bloodred}{red} are highlighted instances of inability to correctly learn novel preferences ( wrong point estimate direction, higher-than-chance probability of mis-direction $\Pr(\hat{d} \neq d) > 0.5$ , or prediction bias $\pm 4$), incompatibility with \texttt{PoSSUM}'s posterior distribution ($p-value \leq 0.1$), and worse-than-chance ability capturing of temporal dynamics ($\Pr(\rho^\tau > 0)<0.5$).}
\label{tab:possum_consistency}
\end{table}

%\begin{figure}
%    \centering
%    \includegraphics[width=\linewidth]{results_plots/dynamic_exchangeability_vote2020.pdf}
%    \caption{Past-vote level \texttt{PoSSUM} estimates over the course of the campaign, shown alongside individual polls overlapping \texttt{PoSSUM}'s fieldwork periods, the aggregated polling average for each \texttt{PoSSUM} fieldwork window.}
%    \label{fig:dynamic_nat_comparison}
%\end{figure}

\pagebreak

\section{Discussion} \label{sec:discussion}
This paper has introduced \texttt{PoSSUM}, an end-to-end protocol for unobtrusive polling of social-media users leveraging multimodal LLMs. Tested during the 2024 Presidential Election campaign, it showed that relatively small, calibrated silicon samples — derived from unstructured user data on $\mathbb{X}$ — can produce valid insights into public opinion. \texttt{PoSSUM} accurately predicted state-level outcomes and identified subgroup preferences consistent with official election results and traditional public opinion polls. These findings demonstrate how combining silicon samples with targeted prompting and structured statistical modeling can capture nuanced shifts in public sentiment beyond an LLM's training cut-off. At the same time, the results highlight persistent challenges, such as underestimating third-party support, reliance on stereotypes and other machine-biases, and the limitations of relying on a single unrepresentative social-media platform. Despite these hurdles, the \texttt{PoSSUM} framework offers a promising path to fully automated, unobtrusive public-opinion polling, capturing granular opinion dynamics of real people without human intervention. The following paragraphs discuss remaining limitations and avenues for future research.\\

\noindent \textbf{Outstanding Issues} Accurately classifying the ethnicity of Hispanic individuals in social media data presents significant challenge. LLMs' reliance on stereotypical cues \cite{choenni2021stepmothers} may have hindered \texttt{PoSSUM}'s ability to understand the substantial right-shift in this group. It is plausible that right-leaning Hispanics may have been more likely coded as ``white'', both due to their lack of stereotypical presentation, and due to the LLMs' tendency to assume minorities do not support Republicans. Developing more sophisticated methods for identifying the ethnicity of social-media users beyond superficial markers remains an important area of future research.

Regarding educational attainment, a decision was made at the outset of the campaign to substitute college education with income as the optimal socio-economic indicator in both quotas and modeling. This decision was motivated by anecdotal evidence during the piloting phase, which seemed to suggest the LLM overstated the correlation between not holding a college degree and supporting Donald Trump. In hindsight, this choice may have negatively affected model performance, and it is regrettable that a more rigorous comparative assessment was not conducted. Future studies should directly evaluate the LLM biases associated with imputing income v. education levels from social-media data.

The suboptimal performance observed at the extremes of the age distribution appears attributable, at least in part, to limited sampling coverage (Figure \ref{fig:unfilled_quotas}). Specifically, \texttt{PoSSUM} exhibited difficulties in adequately populating the youngest and oldest age brackets. One strategy to mitigate this shortfall is to supplement samples from $\mathbb{X}$ with data from other social media platforms. For instance, incorporating \texttt{TikTok} could improve coverage of younger users, whereas \texttt{Facebook} might yield better representation among older populations.\\

%Would there have been consistency 

%If the LLM was offered a choice between Republicans, Democrats and stay-home only, would it have acted in the same way (i.e. putting these 
%# hypothetical west voters amongst the non-voters) or would they have tipped towards Harris ? If so, then PoSSUM's 
%# correct call may have ultimately been a fluke. 

\noindent \textbf{Sources of Third-Party Error} \hspace{10pt} Though rigorous experimental evidence beyond the scope of this paper is necessary to unambiguously tease-out the causes of the large third-party miss, a number of competing hypotheses can be set out: 

i. \emph{selection effects} -- \texttt{PoSSUM}'s reliance on users discussing politics or trends on $\mathbb{X}$ increases the chance of over-sampling highly engaged users. Queries explicitly seeking third-party related discussions are liable to inflate the share of third-party voters in the pool, relative to their objective share in the population. The very nature of $\mathbb{X}$ as a ``free speech platform'' is likely to encourage alternative political discussion; 

ii. \emph{parroting of outdated training data} -- the LLM's own training data likely includes polling data from before and around the training cut-off (October $2023$). At that time third-parties looked highly competitive. It follows that the baseline rate of labeling users as likely voters of one candidate or another may be biased towards the polling average at the training cutoff period; 

iii. \emph{lack of context}: RFK's endorsement of Trump was not included in the background prompting module, nor was it to be found in the LLM's training data. Moreover it was a sui-generis endorsement, in that RFK remained on the ballot in many states. It follows that wherever Kennedy was on the ballot, the LLM took him as a legitimate competitor, whilst in the minds of voters he was not viable;

iv. \emph{failure of contextualising a novel candidate}: for candidates without previous political history or much polling, but with an established track-record of policy positions, we can hypothesise the LLM works by placing the candidate on some latent space to put them in relation to other candidates, and to the kinds of voters who may cast a ballot in their favour. Failure to generalise candidate placement on this latent space can lead to poor estimation of vote-choice preferences. 

Systematic testing of the hypotheses above requires an experimental setup to uncover counterfactuals -- could the LLM have labeled the users differently under different stimuli ? Back-testing on \texttt{PoSSUM} data would likely be unrepresentative of the LLM's behaviour during the fieldwork period, owing to the temporal instability of the underlying LLM models \cite{bisbee2023synthetic} -- hence we cannot hope to learn much more about \texttt{PoSSUM}'s third-party under-performance using the available data. However the hypotheses outlined above can provide a solid foundation to a study of LLM performance on recognising voters of minor parties in multi-party systems. \\

\noindent \textbf{LLM Identity \& Bias} \hspace{10pt} The \emph{neutral annotator} approach taken in this paper is somewhat at odds with the agent-based silicon surveying which dominates the literature. The approach is motivated by seeking out the most objective read of a social media user's timeline, rather than simulate behaviour anew. Under uncertainty however, synthetic responses for specific users rely on the LLM's best guess, which is often affected by its underlying personality. It follows that a reasonable critique of the \emph{neutral annotator} approach is that the LLM's default personality is politically biased \cite{rozado2024political}, and that these biases could affect the annotations. An initial defence against this critique is that neutrality might be the preferable default option under uncertainty, given the absence of similar systematic analyses of bias for alternative LLM personalities (i.e. is the ``expert forecaster'' personality more or less biased than the default ?). Future work should attempt to optimise prompting architecture to minimise annotation biases correlated with relevant features. \\

\noindent \textbf{The Pollster's Critique} \hspace{10pt} Traditional public opinion researchers are skeptical of the claim that \texttt{PoSSUM} is meaningfully tethered to real, on-the-ground dynamics. This is despite \texttt{PoSSUM}'s effort to link silicon samples to real-life $\mathbb{X}$ users and the text they generate, and further despite LLMs' proven track-record in labelling political text \cite{tornberg2024large,gilardi2023chatgpt,cerina2023artificially}.  These critics may find it difficult to attribute the model's accurate prediction of the $2024$ election to mere chance, or to dismiss these findings as statistical / computational artifacts: the intricate patterns of change captured by \texttt{PoSSUM} -- in both direction and magnitude of preferences, as well as alignment with observed human attitudes across multiple crosstabs, challenge such skepticism. 

A more substantive critique concerns the allocation of ``credit'' for this success: to what extent did the methodological decisions (i.e., treating $\mathbb{X}$ users as a panel, stratified sampling according to demographic quotas, extracting survey content from each user’s posting history, and applying MrP to the silicon samples) materially influence the protocol's predictive accuracy? Conversely, it remains possible that the base LLM learnings alone might have produced comparable results, had a simpler demographic-prompting approach (e.g., Argyle et al. \cite{argyle2023out}) been employed on a somewhat representative sample. This would be troublesome, and certainly worthy of further investigation, as it would undermine the generalisability of the performance observed in this experiment.

A key limitation to the present protocol is its non-repeatability. As a proprietary model, \texttt{gpt-4o} has been updated in ways that are not publicly documented, making exact replication of \texttt{PoSSUM}'s $2024$ polling exercise effectively impossible. Bisbee \cite{bisbee2023synthetic} has highlighted similar concerns regarding evolving LLM capabilities. Moreover, in any future attempts at replication, the $2024$ results may already be incorporated (via hard-coding or other data leakage) into the LLM’s knowledge base, confounding efforts to isolate genuine predictive performance.

To disentangle the various sources of predictive success, future research should more rigorously examine the model's individual-level ability to reconstruct respondents' preferences and attributes from digital trace data. One viable approach is to match traditional survey data with digital traces and measure the resultant error rates. Even if some degree of individual-level mismatch persists, this need not be detrimental for every application. For agent-based modeling studies that rely on faithfully simulating individual-level interactions to guarantee realistic emergent phenomena \cite{rossetti2024social}, greater fidelity at the respondent level might be crucial. In contrast, when the analytical target is an aggregate or crosstab-level estimate, capturing the broader group dynamics may suffice. Nonetheless, rigorously testing and enhancing alignment at the individual level would allow more granular inference and richer subgroup analyses, thereby expanding the utility of the approach.\\

\pagebreak
\pagebreak
\Urlmuskip=0mu plus 1mu\relax

\bibliography{bibliography.bib}

\pagebreak

\renewcommand \thepart{}
\renewcommand \partname{}
\appendix
\pagenumbering{arabic}
\setcounter{page}{1}
\renewcommand*{\thepage}{A\arabic{page}}
\addcontentsline{toc}{section}{Appendix} % Add the appendix text to the document TOC
\part{Appendix} % Start the appendix part
\parttoc % Insert the appendix TOC
\thispagestyle{empty}

\pagenumbering{arabic}
\counterwithin{table}{section}
\counterwithin{figure}{section}
\renewcommand{\thealgocf}{\thesection.\arabic{algocf}} % Format algorithms as A.1, A.2, etc.

\renewcommand{\thelstlisting}{\thesection.\arabic{lstlisting}} % Format listings as A.1, A.2, etc.

\setcounter{algocf}{0} % Restart algorithm counter
\setcounter{lstlisting}{0} % Restart listing counter

\begin{spacing}{1}

\section{\texttt{PoSSUM} Routines}

\begin{algorithm}[htbp]
\caption{Pseudo-code for the \texttt{get\_pool} routine.}\label{alg:get_pool}
\SetKwComment{Comment}{\# }{}

\KwIn{
  \begin{itemize}
  \vspace{-5pt}
    \item $\bm{q}$: optimal API search queries
      \vspace{-5pt}
    \item $\bm{w}$: weight of each query
  \end{itemize}
}
\KwOut{
  \begin{itemize}
    \vspace{-5pt}
    \item $\bm{\Upsilon}$: user-data object composed of profile info $\bm{\upsilon}$ and tweets $\bm{\mathcal{T}}$
  \end{itemize}
}

\SetKwProg{Routine}{Routine}{:}{}

\Routine{\texttt{get\_pool}}{
  $ K \gets \texttt{length}(\bm{q})$ \Comment*[r]{\textbf{Get:} number of queries }

  $\bm{\Upsilon} \gets \emptyset$ \Comment*[r]{\textbf{Initialize:} empty users object}
  
  \SetAlgoLined
  \For{$k = 1$ \KwTo $K$ }{ 
      $(\bm{\mathcal{T}},\bm{\upsilon})_{kt} \gets \mathbb{X}(\bm{q}_k,\bm{w}_k)$ \Comment*[r]{\textbf{Call:} sample of tweet-user pairs}
      $\bm{\Upsilon} \gets \bm{\Upsilon} \cup (\bm{\mathcal{T}},\bm{\upsilon})_{kt}$ \Comment*[r]{\textbf{Store:} newly observed users}
  }
}
\end{algorithm}

\begin{lstlisting}[caption={Search terms for tweets related to candidates involved in the US 2024 presidential election.},captionpos=t,label={lst:search_query_politics},framexleftmargin=0cm,xleftmargin=0cm]
query <- 
  "(
    Kamala OR VP OR KamalaHarris OR                                                                          # Democratic candidate terms
    MAGA OR Trump OR realDonaldTrump OR                                                                      # Republican candidate terms
    Robert Kennedy OR RFK OR RobertKennedyJr OR RFKJr OR KennedyShanahan24 OR Kennedy24 OR                   # RFK terms
    Cornel West OR Dr. West OR CornelWest OR                                                                 # Cornel West terms
    Jill Stein OR DrJillStein OR                                                                             # Green candidate terms
    ChaseForLiberty                                                                                          # Libertarian candidate terms 
    )"
  -from:VP -from:KamalaHarris -from:realDonaldTrump  -from:RobertKennedyJr -from:CornelWest 
  -from:DrJillStein -from:ChaseForLiberty                                                                # Don't sample candidate profiles
  "
\end{lstlisting}

\begin{algorithm}
\caption{Pseudo-code for the \texttt{poll\_users} routine.}\label{alg:poll_users}
\SetKwComment{Comment}{\# }{}

\KwIn{
  \begin{itemize}
    \vspace{-5pt}
    \item $\tau$: temporal filter function
      \vspace{-5pt}
    \item $\bm{\Upsilon}$: users object database
      \vspace{-5pt}
    \item $\mathcal{P}^{\mathcal{E}}$: entity filter
      \vspace{-5pt}
    \item $E$: list of acceptable entities
      \vspace{-5pt}
    \item $\mathcal{P}^{\mathcal{G}}$: geographic filter
      \vspace{-5pt}
    \item $G$: list of acceptable geographies
      \vspace{-5pt}
    \item $\bm{\mathcal{F}} \gets (\bm{\mathcal{F}}^{x}, \bm{\mathcal{F}}^{y})$: list of independent and dependent features
      \vspace{-5pt}
    \item $\mathcal{P}^{\mathcal{\phi}}$: feature extraction prompt
      \vspace{-5pt}
    \item $(\bm{X}^\mathcal{Q},\bm{\omega}^\star,\bm{\omega}^\prime = \bm{0})$: acceptable features, expected frequency, and sample counter 
      \vspace{-5pt}
    \item $m$: number of tweets per user
  \end{itemize}
}
\KwOut{
  \begin{itemize}
        \vspace{-5pt}
    \item $\bm{Z}$: survey object with extracted features
  \end{itemize}
}

\SetKwProg{Routine}{Routine}{:}{}

\Routine{\texttt{poll\_users}}{
\vspace{5pt}
  $\bm{\Upsilon}^\star \gets \bm{\Upsilon}[\tau(\bm{t}) = \mbox{\texttt{TRUE}} \lor \mathcal{G} \neq \emptyset]$ \Comment*[r]{\textbf{Filter:} recent + valid location}
$ N \gets \texttt{length}(\bm{\Upsilon}^\star)$ \Comment*[r]{\textbf{Get:} number of valid users }
  $\bm{Z} \gets \emptyset$ \Comment*[r]{\textbf{Initialize:} empty survey object}

  \vspace{10pt}
  \SetAlgoLined
  \For{$i = 1$ \KwTo $N$}{
    \vspace{2.5pt}
    $e_i \gets \mbox{GPT}\left\{ \mathcal{P}^{E}(\Upsilon_i) \right\}$ \Comment*[r]{\textbf{Call:} GPT entity filter}
        \vspace{10pt}
    \If{$e_i \in E$}{        
        \vspace{2.5pt}
        $g_i \gets \mbox{GPT}\left\{ \mathcal{P}^{\mathcal{G}}(\Upsilon_i) \right\}$ \Comment*[r]{\textbf{Call:} GPT geographic filter}
           \vspace{10pt}
        \If{$g_i \in  G$}{         
            \vspace{2.5pt}
            $\bm{X}_i \gets \mbox{GPT}\left\{ \bm{\mathcal{P}}^{\phi}(\Upsilon_i,\bm{\mathcal{F}}^x) \right\}$ \Comment*[r]{\textbf{Call:} GPT quota filter}
            \vspace{10pt}

            \If{$\bm{X}_i \in \bm{X}^\mathcal{Q} \mbox{ } \cup  \mbox{ } \omega^\prime_i < \omega^\star_i$ }{
            \vspace{2.5pt}
                
                $\bm{\mathcal{T}}_i^{+} \gets \mathbb{X}(\Upsilon_i)$ \Comment*[r]{\textbf{Call:} sample last $m$ tweets}

                $\bm{z}_i \gets \mbox{GPT}\left\{ \bm{\mathcal{P}}^{\phi}(\Upsilon_i,\bm{\mathcal{T}}_i^{+},\bm{\mathcal{F}}) \right\}$ \Comment*[r]{\textbf{Call:} GPT extraction}
                \vspace{10pt}
                $\bm{Z} \gets \bm{Z} \cup \bm{z}_i$\Comment*[r]{\textbf{Store:} survey object}
                \vspace{10pt}
                $\omega^\prime_i \gets \omega^\prime_i  +1$ \Comment*[r]{\textbf{Update:} sample quota counter}
} } } } 
}
\end{algorithm}

\pagebreak

\begin{figure}
    \centering
    \includegraphics[width=1\linewidth]{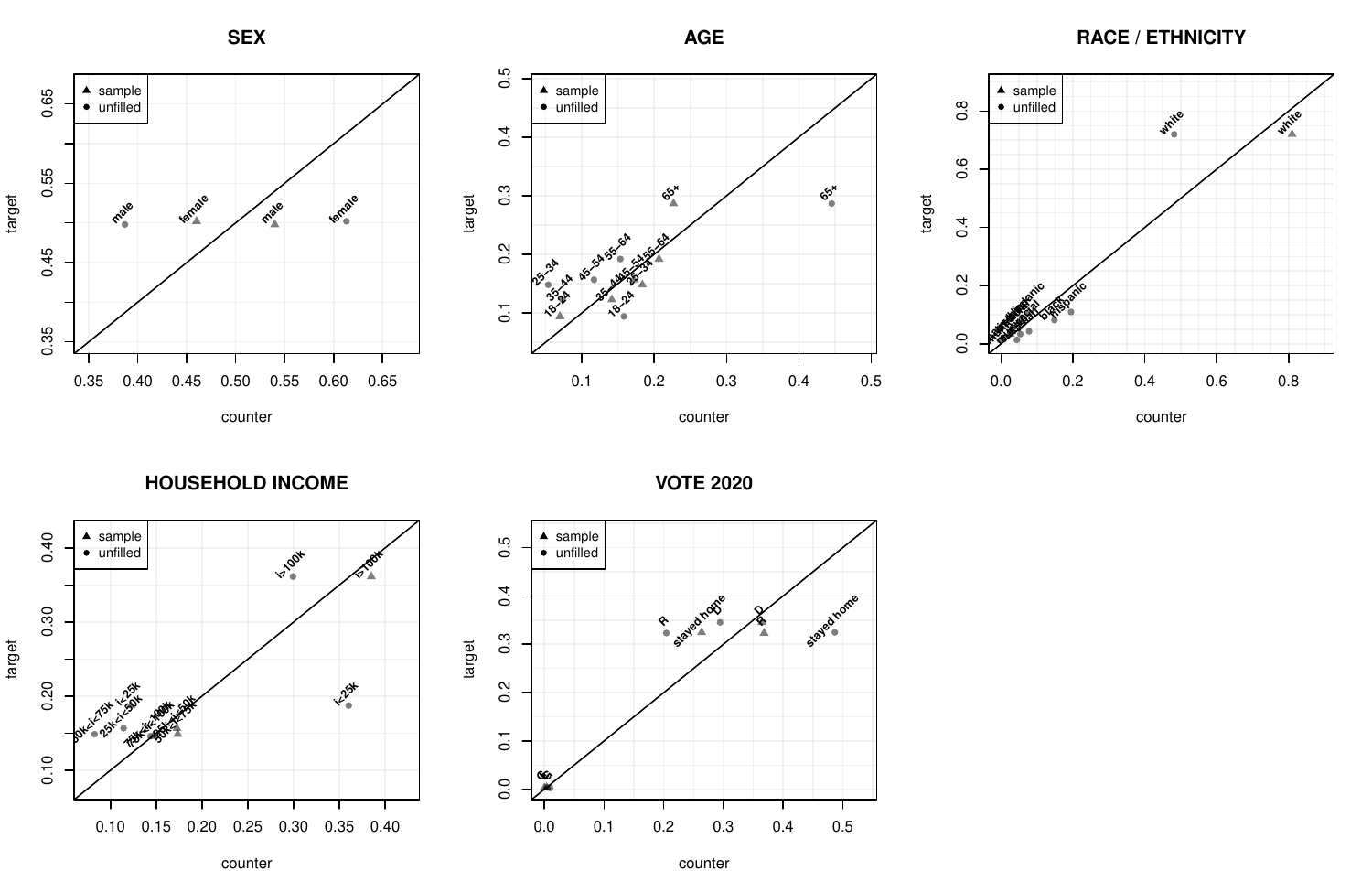}
    \caption{Sub-group prevalence by filled / unfilled quotas (x-axis) v. target (y-axis), for the final \texttt{PoSSUM} poll, fielded from the $17^{th}$ to the $26^{th}$ of October.}
    \label{fig:unfilled_quotas}
\end{figure}

\pagebreak

\begin{algorithm}[htpb]
\caption{Pseudo-code for the \texttt{build\_prompt} routine.}\label{alg:build_prompt}
\SetKwComment{Comment}{\# }{}

\KwIn{
  \begin{itemize}
        \vspace{-5pt}
    \item $\mathcal{B}$: background information
          \vspace{-5pt}
    \item $\bm{\mathcal{F}}$: features identified for extraction
          \vspace{-5pt}
    \item $\mathcal{M}$: mould wrapper
          \vspace{-5pt}
    \item $\mathcal{I}$: instructions
          \vspace{-5pt}
    \item $\bm{\Upsilon}$: user data
  \end{itemize}
}
\vspace{5pt}
\KwOut{
  \begin{itemize}
        \vspace{-5pt}
    \item $\mathcal{P}$: unified prompt
  \end{itemize}
}

\SetKwProg{Routine}{Routine}{:}{}
\Routine{\texttt{build\_prompt}}{
  $\mathcal{P} = \mathcal{B} \mathbin\Vert \mathcal{M}(\bm{\Upsilon}) \mathbin\Vert \mathcal{I}(\bm{\mathcal{F}})$ \Comment*[r]{\textbf{Concatenate:} to form unified prompt}
}
\end{algorithm}

\begin{algorithm}
\caption{Pseudo-code for the \texttt{temporal\_filter} routine.}\label{alg:temporal_filter}
\SetKwComment{Comment}{\# }{}
\KwIn{
  \begin{itemize}
  \vspace{-5pt}
    \item $\tau$: temporal filter function
    \vspace{-5pt}
    \item $\bm{\Upsilon}$: user data
        \vspace{-5pt}
    \item $\bm{t}$: user-wise collection time-stamp 
  \end{itemize}
}
\vspace{5pt}
\KwOut{
  \begin{itemize}
  \vspace{-5pt}
    \item $\bm{\Upsilon}^\star$: filtered user data including only recent users
  \end{itemize}
}

\SetKwProg{Routine}{Routine}{:}{}

\Routine{\texttt{temporal\_filter}}{
  $\bm{\Upsilon}^\star \gets \bm{\Upsilon}[\tau(\bm{t}) = \mbox{\texttt{TRUE}}]$ \Comment*[r]{\textbf{Filter:} exclude users based on criteria}

}
\end{algorithm}

\begin{algorithm}[htpb]
\caption{Pseudo-code for the \texttt{exclude\_null\_geography} routine.}\label{alg:nullgeo_filter}
\SetKwComment{Comment}{\# }{}

\KwIn{
  \begin{itemize}
    \vspace{-5pt}
    \item $\bm{\Upsilon}$: user data
    \vspace{-5pt}
    \item $\bm{\mathcal{G}}$: geographic information associated with users
  \end{itemize}
}
\vspace{5pt}
\KwOut{
  \begin{itemize}
    \vspace{-5pt}
    \item $\bm{\Upsilon}^\star$: filtered user data with non-null geographic information
  \end{itemize}
}

\SetKwProg{Routine}{Routine}{:}{}

\Routine{\texttt{exclude\_null\_geography}}{

  $\bm{\Upsilon}^\star \gets \bm{\Upsilon}[\bm{\mathcal{G}} \neq \emptyset]$ \Comment*[r]{\textbf{Filter:} exclude users with null geography}

}
\end{algorithm}

\begin{algorithm}[htpb]
\caption{Pseudo-code for the background-informed feature extraction routine.}\label{alg:bg_informed_prompt}
\SetKwComment{Comment}{\# }{}

\KwIn{
  \begin{itemize}
    \item $\mathcal{B}$: background information
        \vspace{-5pt}
    \item $\Upsilon$: user data
        \vspace{-5pt}
    \item $\mathcal{I}_1$: instructions to generate background-informed features
        \vspace{-5pt}
            \item $\bm{\mathcal{F}}_1$: empty features object
        \vspace{-5pt}
    \item $\mathcal{I}_2$: operation to extract features
        \vspace{-5pt}
    \item $\mathcal{M}$: mould wrapper
  \end{itemize}
}
\vspace{5pt}
\KwOut{
  \begin{itemize}
          \vspace{-5pt}
    \item $\bm{\mathcal{F}}_2$: background-informed features generated by GPT
            \vspace{-5pt}
    \item $\bm{X}_i$: extracted features for the user data
  \end{itemize}
}

\SetKwProg{Routine}{Routine}{:}{}

\Routine{\texttt{background\_informed\_feature\_extraction}}{
  $\mathcal{P}_1 = \mathcal{B} \mathbin\Vert \mathcal{I}_1(\bm{\mathcal{F}}_1)$ \Comment*[r]{\textbf{Concatenate:} to form $1^{st}$ prompt}
  $\bm{\mathcal{F}}_2 \gets \mbox{GPT}\left( \mathcal{P}_1 \right)$ \Comment*[r]{\textbf{Call:} GPT to generate bg-informed features}
  
  \vspace{10pt}
  $\mathcal{P}_2 = \mathcal{B} \mathbin\Vert \mathcal{M}(\Upsilon)  \mathbin\Vert \mathcal{I}_2(\bm{\mathcal{F}}_2)$ \Comment*[r]{\textbf{Concatenate:} to form $2^{nd}$ prompt}
  $\bm{X}_i \gets \mbox{GPT}\left( \mathcal{P}_2 \right)$ \Comment*[r]{\textbf{Call:} GPT to extract features}

}
\end{algorithm}
\FloatBarrier  % Ensures all floats above remain in this section

\pagebreak

\section{Prompting Architecture}

\begin{lstlisting}[caption={Excerpt from the Speculation Module of the prompt, defining how to assign and interpret speculation scores.}  ,captionpos=t,label={lst:speculation},xleftmargin=0cm]

For each selected symbol / category, please note the level of Speculation involved in this selection.
Present the Speculation level for each selection on a scale from 0 (not speculative at all, every single element of the user data was useful in the selection) to 100 (fully speculative, there is no information related to this title in the user data).
Speculation levels should be a direct measure of the amount of useful information available in the user data.
Speculation levels pertain only to the information available in the user data -- namely the username, name, description, location, profile picture and tweets from this user -- and should not be affected by additional information available to you from any other source. 
To ensure consistency, use the following guidelines to determine speculation levels:

0-20 (Low speculation): The user data provides clear and direct information relevant to the title. (e.g., explicit mention in the profile or tweets)
21-40 (Moderate-low speculation): The user data provides indirect but strong indicators relevant to the title. (e.g., context from multiple sources within the profile or tweets)
41-60 (Moderate speculation): The user data provides some hints or partial information relevant to the title. (e.g., inferred from user interests or indirect references)
61-80 (Moderate-high speculation): The user data provides limited and weak indicators relevant to the title. (e.g., very subtle hints or minimal context)
81-100 (High speculation): The user data provides no or almost no information relevant to the title. (e.g., assumptions based on very general information)

For each selected category, please explain at length what features of the data contributed to your choice and your speculation level.

\end{lstlisting}

\pagebreak

\begin{lstlisting}[caption={Example of Multi-feature Object.},captionpos=t,label={lst::features_example}, framexleftmargin=0cm,xleftmargin=0cm]
ind.features <- c(
  'ETHNICITY:
E1) white - individuals with origins in any of the original peoples of europe, including, for example, english, german, irish, italian, polish, and scottish -- as well as arab or middle-eastern with origins in any of the original peoples of the middle east or north africa, including, for example, lebanese, iranian, egyptian, syrian, iraqi, and israeli.
E2) black or african american - individuals with origins in any of the black racial groups of africa, including, for example, african american, jamaican, haitian, nigerian, ethiopian, and somali.
E3) hispanic or latino - includes individuals of mexican, puerto rican, salvadoran, cuban, dominican, guatemalan, and other central or south american or spanish culture or origin.
E4) asian - individuals with origins in any of the original peoples of central or east asia, southeast asia, or south asia, including, for example, chinese, asian indian, filipino, vietnamese, korean, and japanese.
E5) american indian or alaskan native or native hawaiian or pacific islander - individuals with origins in any of the original peoples of north, central, and south america, including, for example, navajo nation, blackfeet tribe of the blackfeet the indian reservation of montana, native village of barrow inupiat traditional government, nome eskimo community, aztec, and maya -- as well as individuals with origins in any of the original peoples of hawaii, guam, samoa, or other pacific islands, including, for example, native hawaiian, samoan, chamorro, tongan, fijian, and marshallese.
E6) multiracial - individuals who identify explicitly as belonging to more than one of the racial and ethnic groups above, such as biracial individuals with one white and one black parent, or those with a combination of asian and hispanic heritage, etc. mixed-race individuals often face unique social experiences, such as celebrating diverse cultural holidays, speaking multiple languages, and bridging different cultural perspectives within their families and communities.
\n',
  'AGE:
A1) under 18 years old
A2) 18 to 24 years old
A3) 25 to 34 years old
A4) 35 to 44 years old
A5) 45 to 54 years old
A6) 55 to 64 years old
A7) 65 or older
\n',
  'SEX:
S1) masculine sex - male
S2) feminine sex - female
\n',
  'INTEREST IN POLITICS:
I1) not interested at all in politics
I2) slightly interested in politics
I3) moderately interested in politics
I4) highly interested in politics
\n',
  'MARITAL STATUS:
M1) married - currently legally married and living with a spouse
M2) single - never married, including those who are legally separated
M3) divorced - legally divorced and not currently remarried
M4) widowed - spouse has passed away and not currently remarried
\n',
  "HIGHEST EDUCATIONAL QUALIFICATION:
Q1) completed education up to and including high school - high school diploma, vocational training, associate degree
Q2) completed education at the college or university level - bachelor's degree, master's degree, doctorate
\n",
  'HOUSEHOLD INCOME BRACKET:
H1) up to 25000 USD per year
H2) between 25000 and 50000 USD per year
H3) between 50000 and 75000 USD per year
H4) between 75000 and 100000 USD per year
H5) more than 100000 USD per year
\n',
  'GENERAL TRUST IN OTHER PEOPLE:
Tru1) always trust other people
Tru2) most of the time trust other people
Tru3) about half of the time trust other people
Tru4) some of the time trust other people
Tru5) never trust other people
\n',
  'PAYING ATTENTION TO THE 2024 PRESIDENTIAL ELECTION:
Att1) not paying attention at all to the 2024 Presidential election in the US
Att2) paying only a little attention to the 2024 Presidential election in the US
Att3) paying some attention to the 2024 Presidential election in the US
Att4) paying a lot of attention to the 2024 Presidential election in the US
\n',
                                                                      .
                                                                      .
                                                                      .
)
\end{lstlisting}

\FloatBarrier

\pagebreak

\section{Supplementary Results}

\begin{figure}[htpb]
    \centering
    \includegraphics[width=1\linewidth]{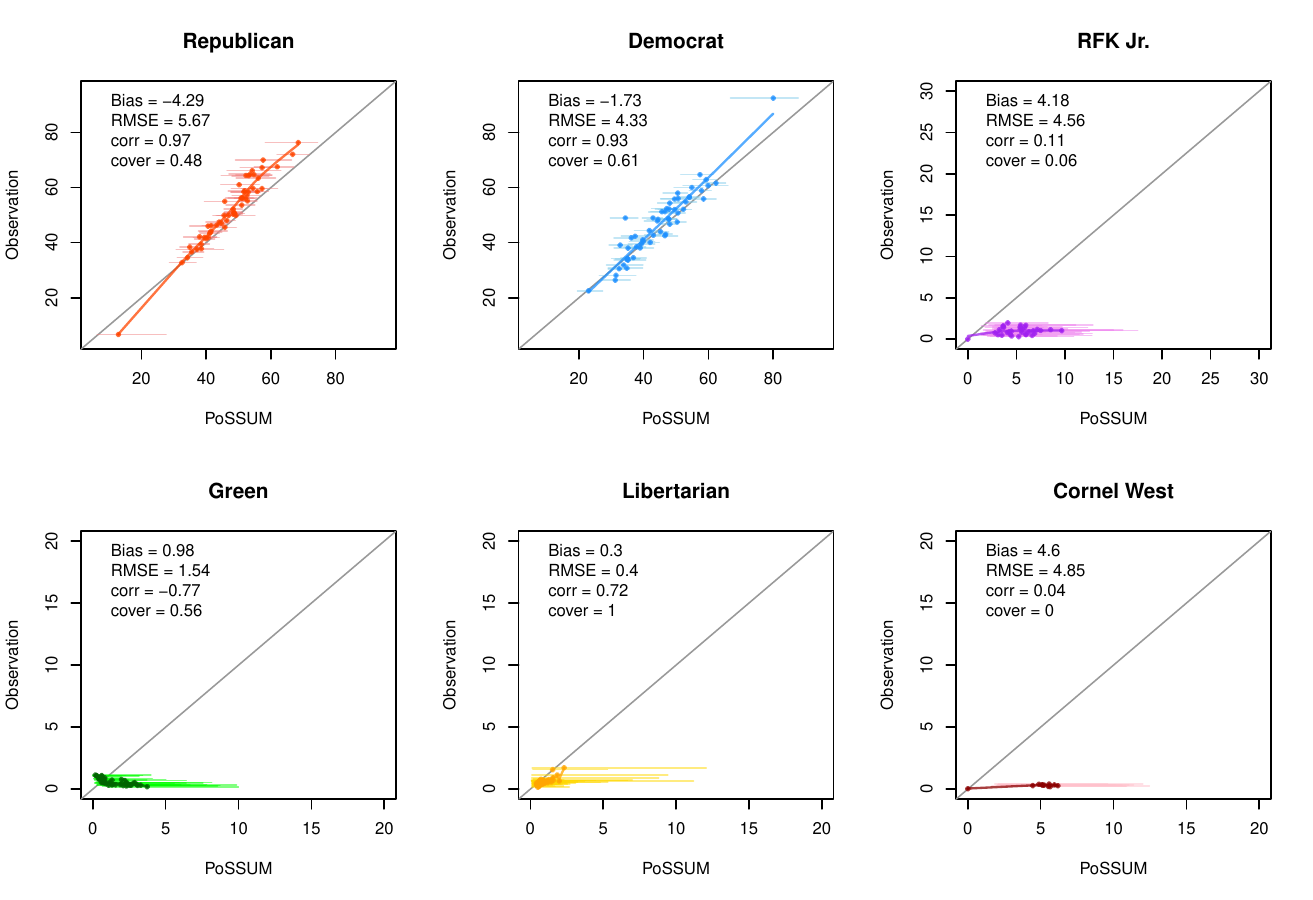}
    \caption{State-level predictive power on vote share by candidate. Training data includes highly speculative records. Model fit to the final \texttt{PoSSUM} poll, fielded from the $17^{th}$ to the $26^{th}$ of October.}
    \label{fig:VoteShare_Speculative_10.26}
\end{figure}

\begin{figure}[htpb]
    \centering
    \includegraphics[width=1\linewidth]{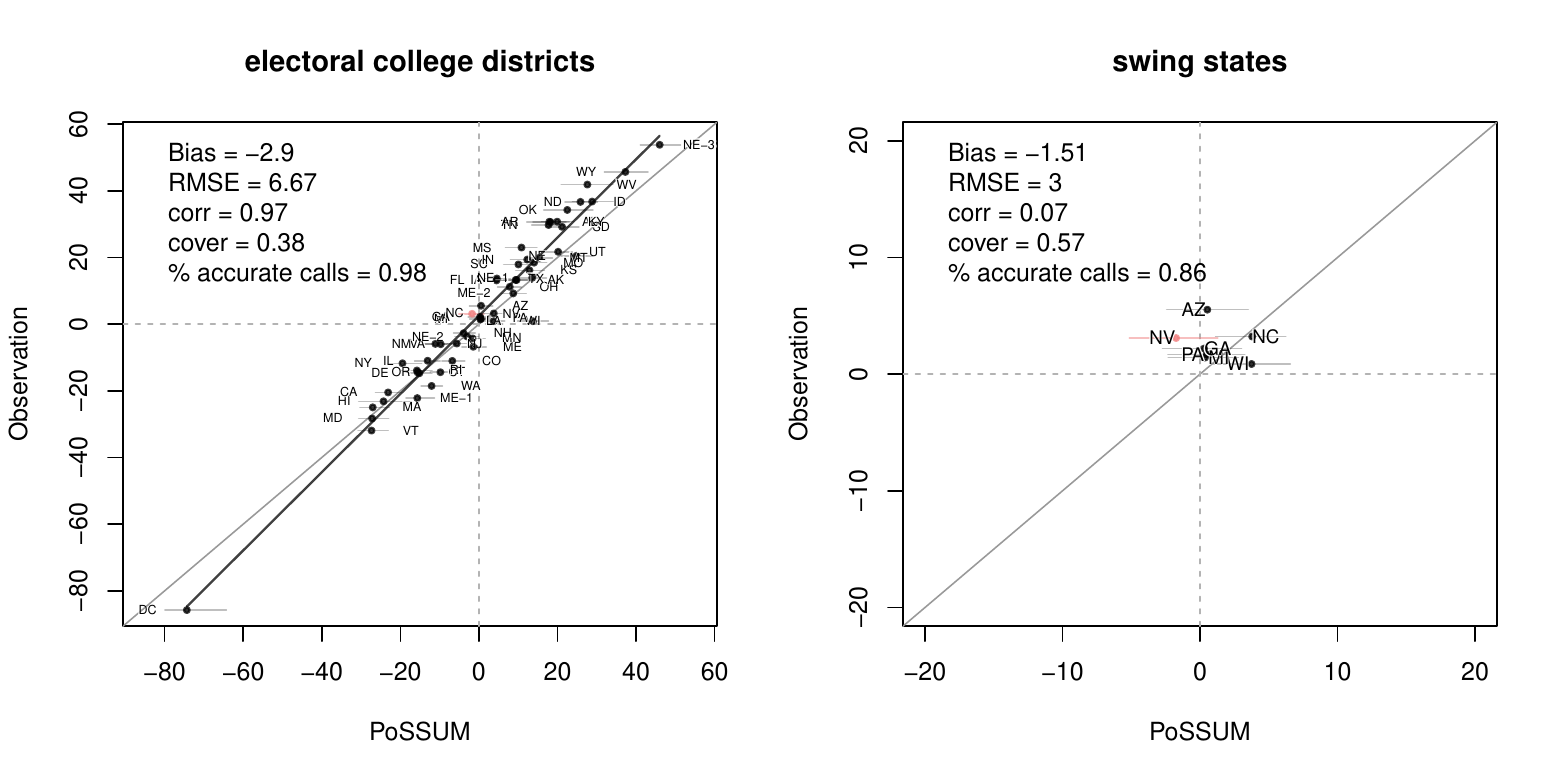}
    \caption{State-level predictive power on Republican - Democrat margin. Training data includes highly speculative records. Model fit to pooled dataset of $5$ polls, fielded from the $15^{th}$ of August to the $26^{th}$ of October.}
    \label{fig:RDmargin_Speculative}
\end{figure}

\begin{figure}[htbp]
    \centering
    \includegraphics[width=1\linewidth]{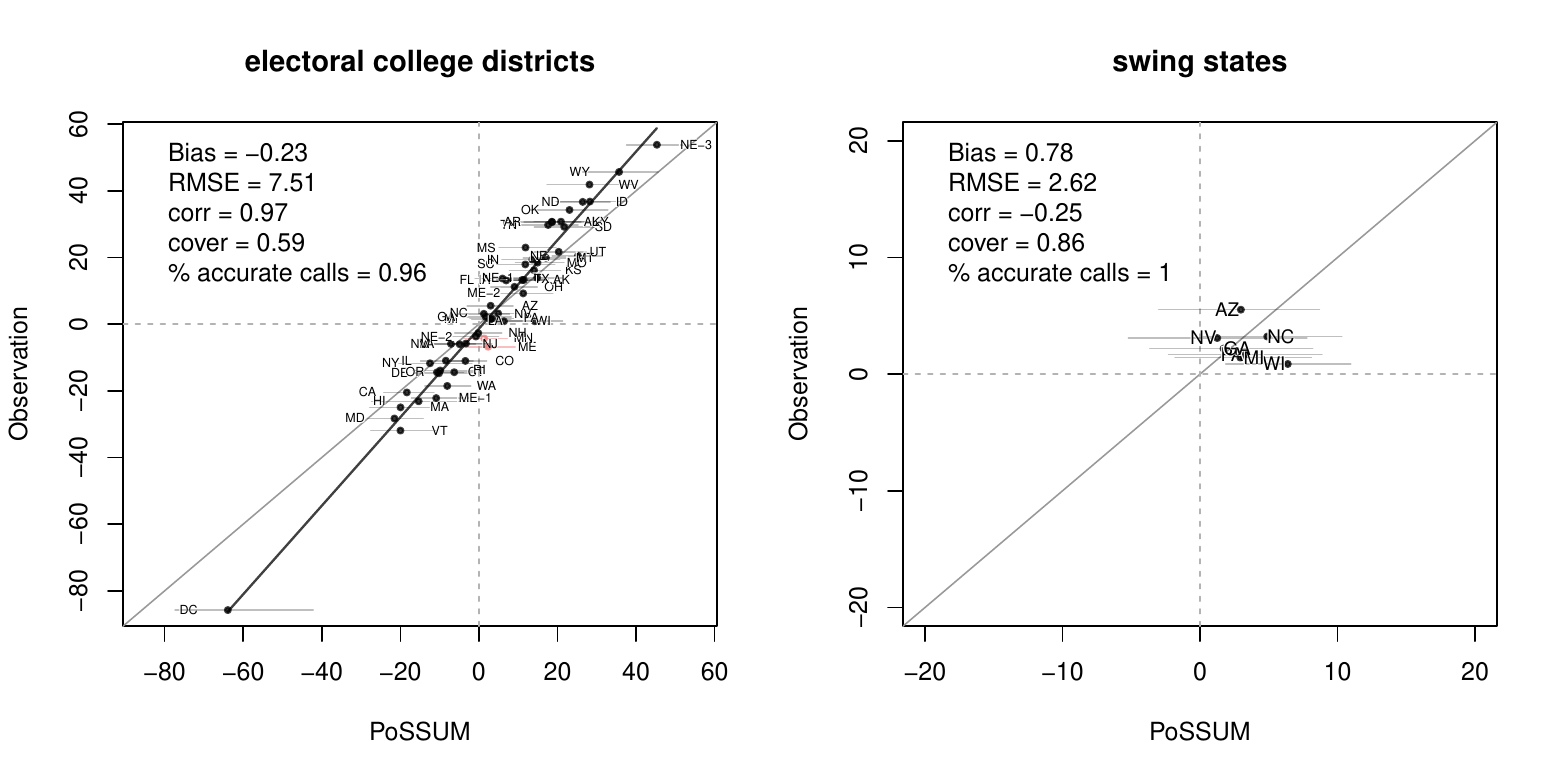}
    \caption{State-level predictive power on the Republican - Democrat margin. Training data does not include highly speculative records. Model fit to pooled dataset of $5$ polls, fielded from the $15^{th}$ of August to the $26^{th}$ of October. }
    \label{fig:RDmargin_Moderate}
\end{figure}

\begin{figure}[htbp]
    \centering
    \includegraphics[width=1\linewidth]{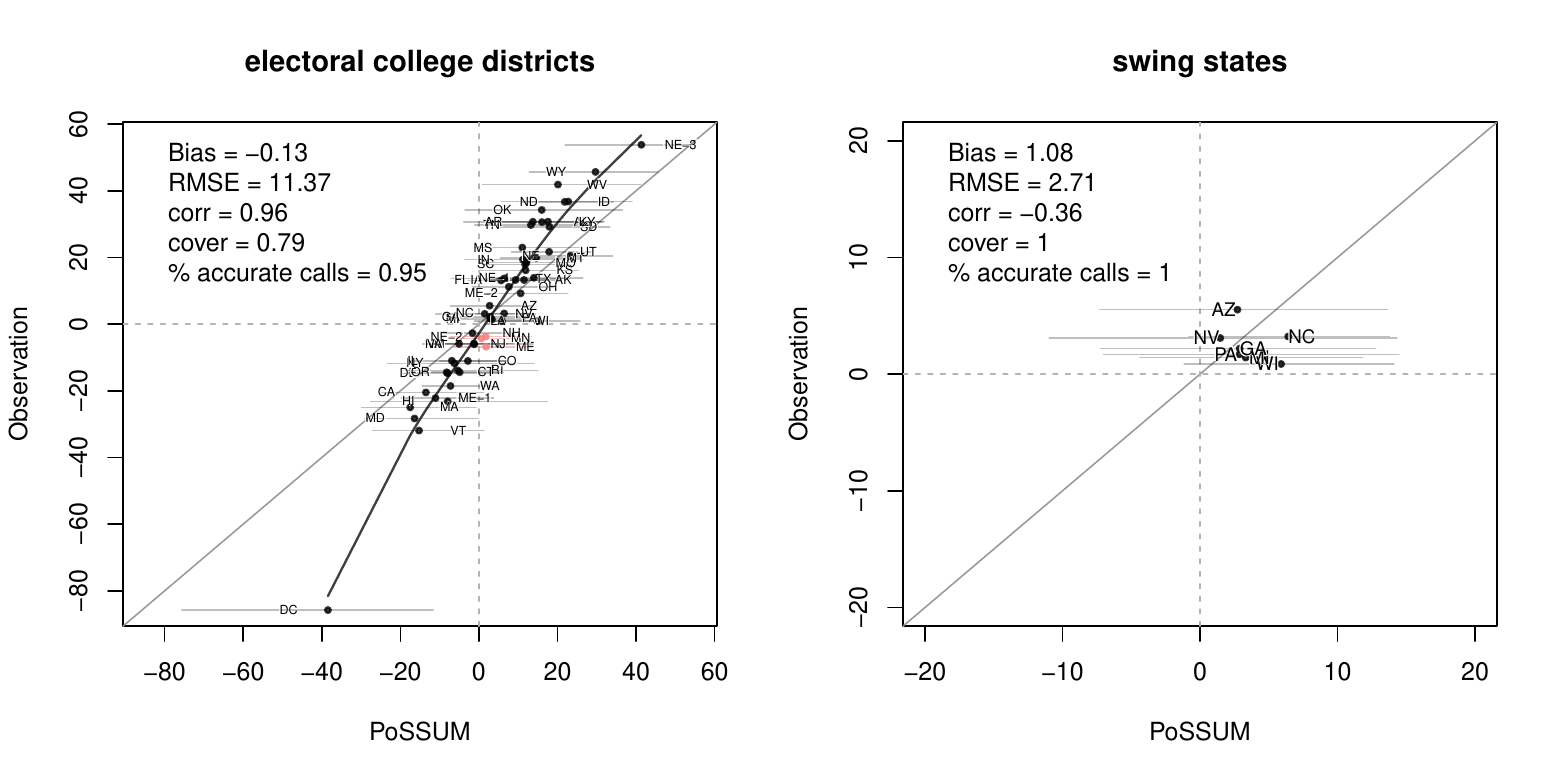}
    \caption{State-level predictive power on the Republican - Democrat margin. Training data does not include highly speculative records. Model fit to the final \texttt{PoSSUM} poll, fielded from the $17^{th}$ to the $26^{th}$ of October. }
    \label{fig:RDmargin_Moderate_10.26}
\end{figure}

\begin{figure}[htbp]
    \centering
    \includegraphics[width=1\linewidth]{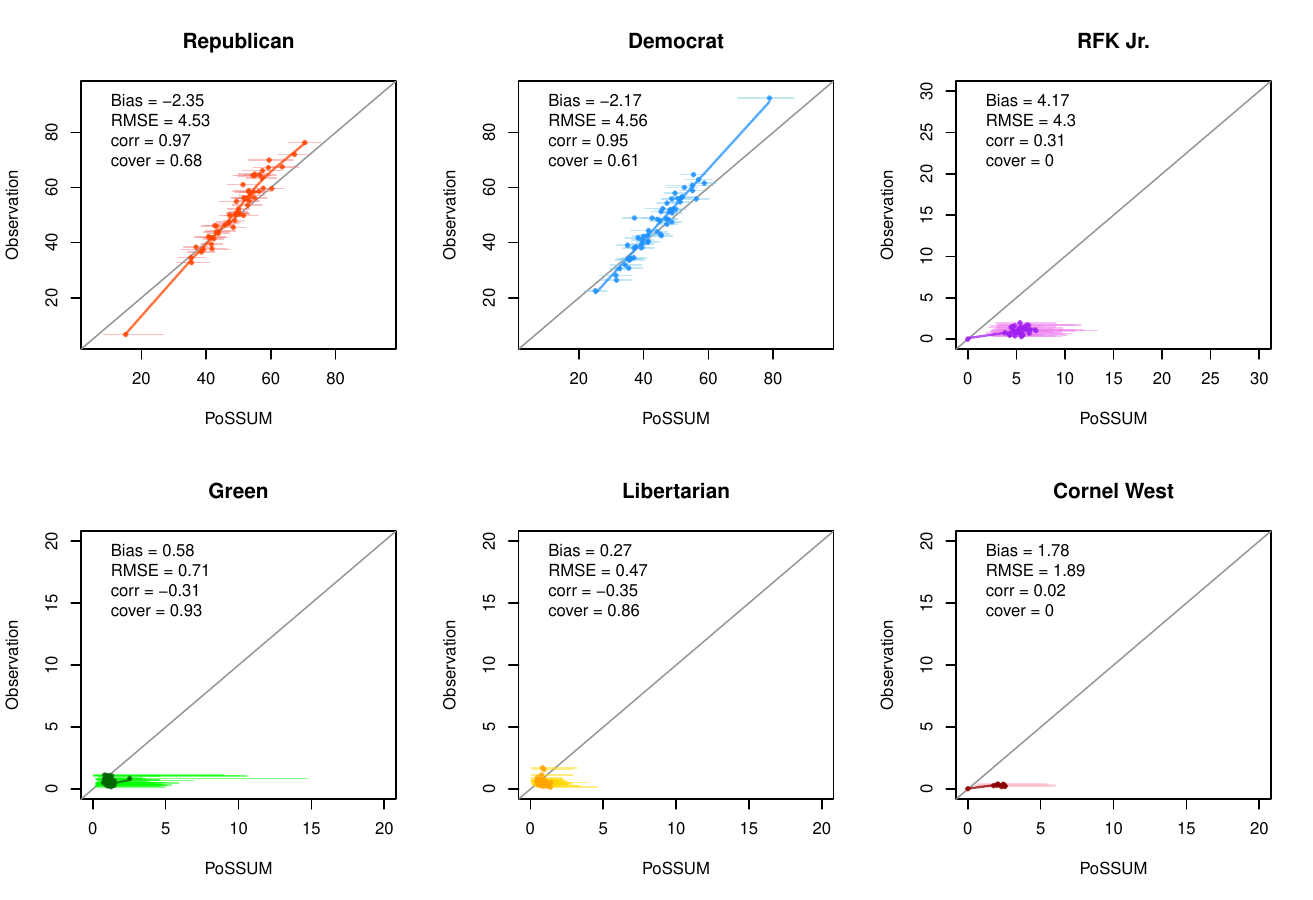}
    \caption{State-level predictive power on vote share by candidate. Training data does not include highly speculative records. Model fit to pooled dataset of $5$ polls, fielded from the $15^{th}$ of August to the $26^{th}$ of October.}
    \label{fig:VoteShare_Moderate}
\end{figure}

\begin{figure}[htbp]
    \centering
    \includegraphics[width=1\linewidth]{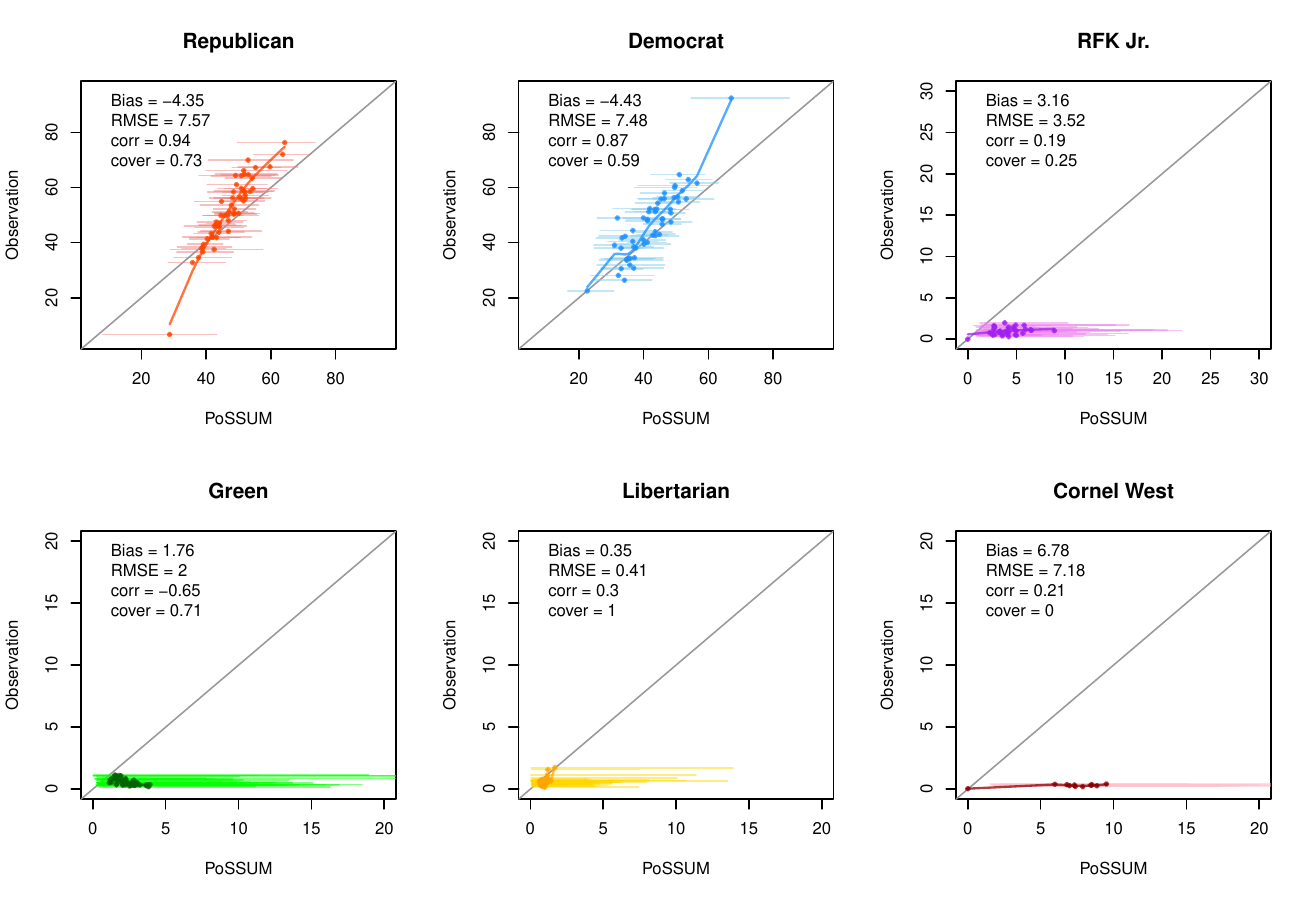}
    \caption{State-level predictive power on vote share by candidate. Training data does not include highly speculative records. Model fit to the final \texttt{PoSSUM} poll, fielded from the $17^{th}$ to the $26^{th}$ of October.}
    \label{fig:VoteShare_Moderate_10.26}
\end{figure}

\begin{figure}[htbp]
    \hspace*{-5em} % Adjust this value as needed
    \includegraphics[width=1.25\linewidth]{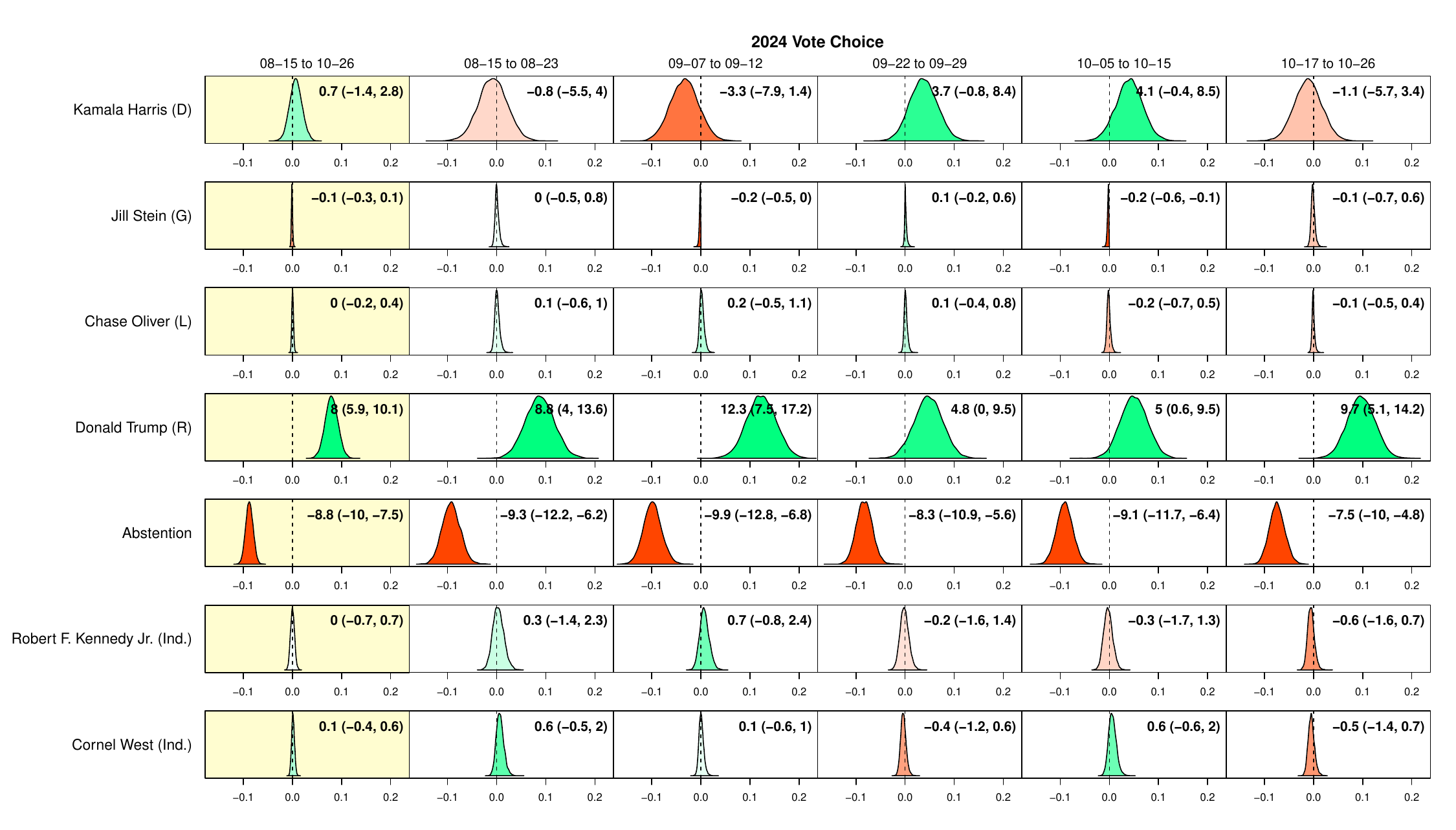}
    \caption{Posterior distribution of the effect of dropping highly speculative records on the raw training data, for $2024$ voting preferences amongst all synthetic samples. The highlighted leftmost column presents the effect on the aggregated complete sample of $5$ polls.}
    \label{fig:dropeffect_vote2024}
\end{figure}

%\begin{figure}[htbp]
%    \hspace*{-5em} % Adjust this value as needed
%    \includegraphics[width=1.25\linewidth]{results_plots/2nd_choice.pdf}
%    \caption{Distribution of $2^{nd}$ choice preferences, estimated fitting an MrP model to the pooled synthetic survey sample, using data from the $5$ polls fielded from the $15^{th}$ of August to the $26^{th}$ of October.}
%    \label{fig:2nd_choice}
%\end{figure}

\FloatBarrier

\pagebreak

\noindent \textbf{Learning from Highly Speculative Records} \hspace{10pt} When the MrP model is allowed to learn from highly speculative synthetic records, performance on objective predictive power indicators improves substantially, compared to when this is forbidden. This is demonstrated by a comparison of Figures \ref{fig:RDmargin_Speculative_10.26} and \ref{fig:RDmargin_Moderate_10.26}. 

It is challenging to pin down the exact mechanism through which this works. A substantial decrease in sample size as a result of moderating speculation is surely to be assigned some portion of the blame. Out of the $4,982$ users for whom \texttt{PoSSUM} inferred features during the campaign, $2,814$ are assigned at least one highly speculative feature necessary to fit the model, and are therefore dropped. The results of this are dramatic levels of attenuation bias ( Figure \ref{fig:RDmargin_Moderate_10.26} for a model fit excluding highly speculative records on a single fieldwork period, which enjoys only a tiny state-level sample size of $377$ users ).\\

%Attenuation bias seems more severe when highly speculative records are dropped (compare Figures \ref{fig:RDmargin_Speculative} and \ref{fig:RDmargin_Moderate}, as well as Figures \ref{fig:RDmargin_Speculative_10.26} and \ref{fig:RDmargin_Moderate_10.26}). This can be partially attributable to decreasing state-level sample size. The overall sample size cost of moderating speculation levels is high -- out of the $4,982$ users surveyed during the campaign, $2,814$ contain at least one highly speculative feature necessary to fit the model, and are therefore dropped. %The evidence for the beneficial effects of additional state-level samples is the following: the model fit to the full dataset (all $5$ polls, \ref{fig:RDmargin_Speculative}), generates a prediction for the last poll that has broadly greater predictive power than that of a model fit exclusively to the last poll  (Figure \ref{fig:RDmargin_Speculative_10.26}) -- and the gains in predictive power do not seem to come from the most populated states, but rather from the least sampled ones  (Figure \ref{fig:}). 
%This relationship holds at different dose-levels or sample size -- the models fit to the speculation-moderated poll are more accurate when the full sample is used, though still less accurate than their highly speculative counterparts.

\begin{figure}[htpb]
    \hspace*{-5em} % Adjust this value as needed
    \includegraphics[width=1.25\linewidth]{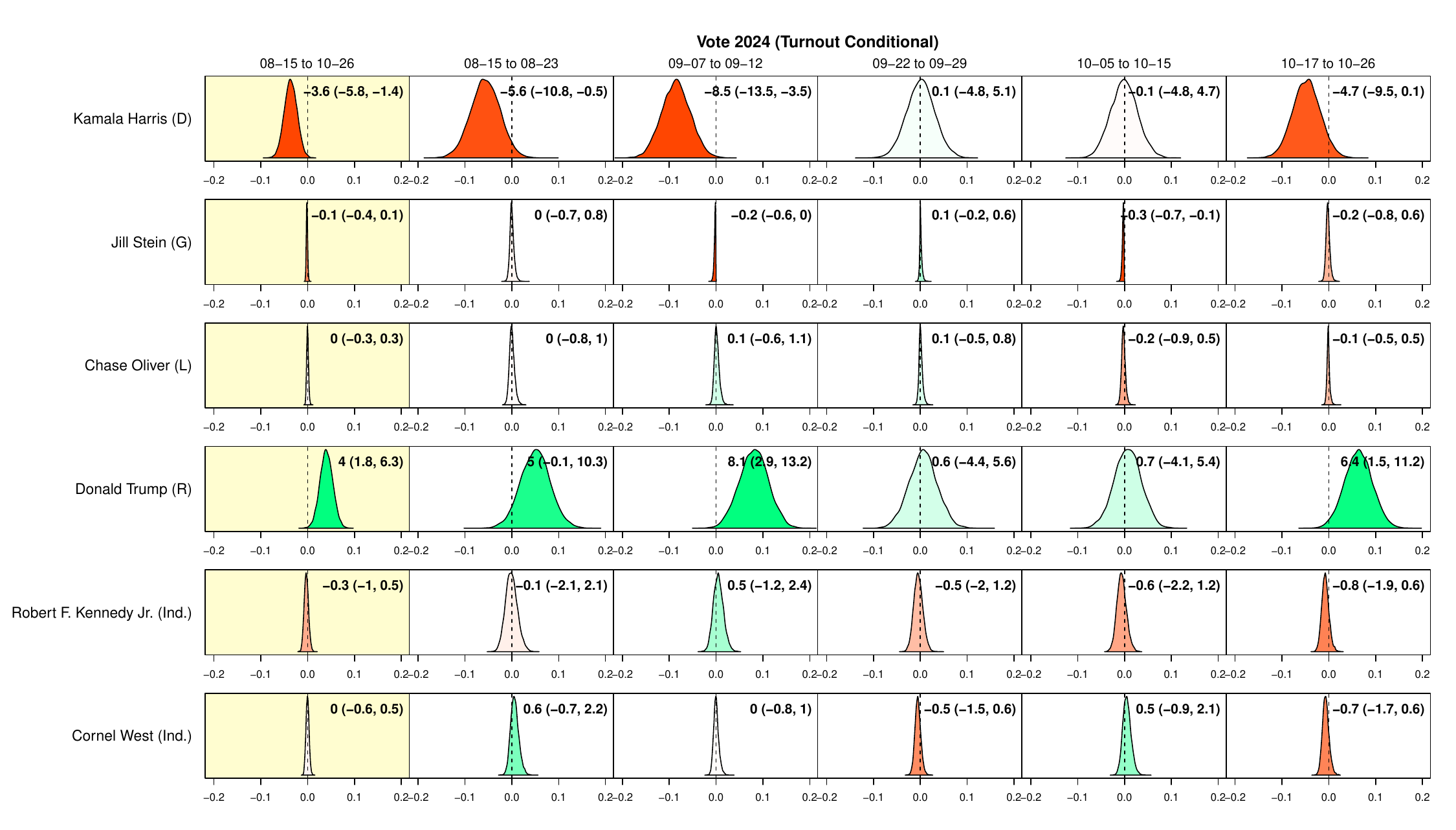}
    \caption{Posterior distribution of the effect of dropping highly speculative records on the raw training data, for $2024$ voting preferences amongst likely voters. The highlighted leftmost column presents the effect on the aggregated complete sample of $5$ polls.}
    \label{fig:dropeffect_vote2024turnout}
\end{figure}

\noindent Beyond the sheer effects of sample size, there are also changes in predictive accuracy due to the composition of users which are flagged as highly speculative. Figure \ref{fig:dropeffect_vote2024turnout} shows the effect of dropping highly speculative users on the distribution of $2024$ voting preferences amongst likely voters in the raw data sample. The effect of dropping highly speculative users is to turn the sample of likely voters substantially more Republican. Users with low turnout propensity (Figure \ref{fig:dropeffect_vote2024}), as well as some democrats, are more likely to be the subject of high-levels of speculation. The effect can be seen even after weighting,  as shown by lower anti-Republican bias in the MrP estimates trained on the moderately speculative sample. High-levels of speculation are associated with labeling users as medium-to-low income, white ethnicity and abstention in the $2020$ election. \\

\noindent The provocative intuition developed from this section is that the LLM's engagement in speculation is beneficial to preference estimation under the \texttt{PoSSUM} protocol. The fact that speculative records increase sample size does not in itself lead to better estimates -- if the estimates were mere noise, we would expect a drop in performance. It follows that the speculative records must contain some relevant auxiliary information, which is not directly acquirable from the underlying mould, but which is useful to address some of the underlying bias in the data. This is evidence in favour of the proposition that LLMs hold information to address bias in unrepresentative samples.\\

\noindent \textbf{Benefits of Speculation} \hspace{10pt} This paper presents a novel prompting strategy to obtain a self-reported \emph{speculation score} from the LLM. Speculation here is defined as the amount of information in the mould which is directly indicative of a category to which the user belongs to. I show that including highly speculative records, defined loosely as records for which at least one of the relevant variables attains a speculation score greater than $80\%$, is beneficial to the estimation of voting preferences. Future work should focus on uncovering the gradient of this benefit -- how sensitive are preferences to varying degrees of speculation ? When the LLM is speculating, it is in effect engaging in a similar process as multiple imputation. Unlike existing imputation algorithms, it is able to bring in knowledge external to the dataset at hand, which it has acquired during the training phase. An interesting question for future research is to identify when this external knowledge outperforms internal knowledge -- under what conditions does LLM imputation outperform classic multiple imputation models ?

\begin{figure}[htbp]
    \centering
    \includegraphics[width=\linewidth]{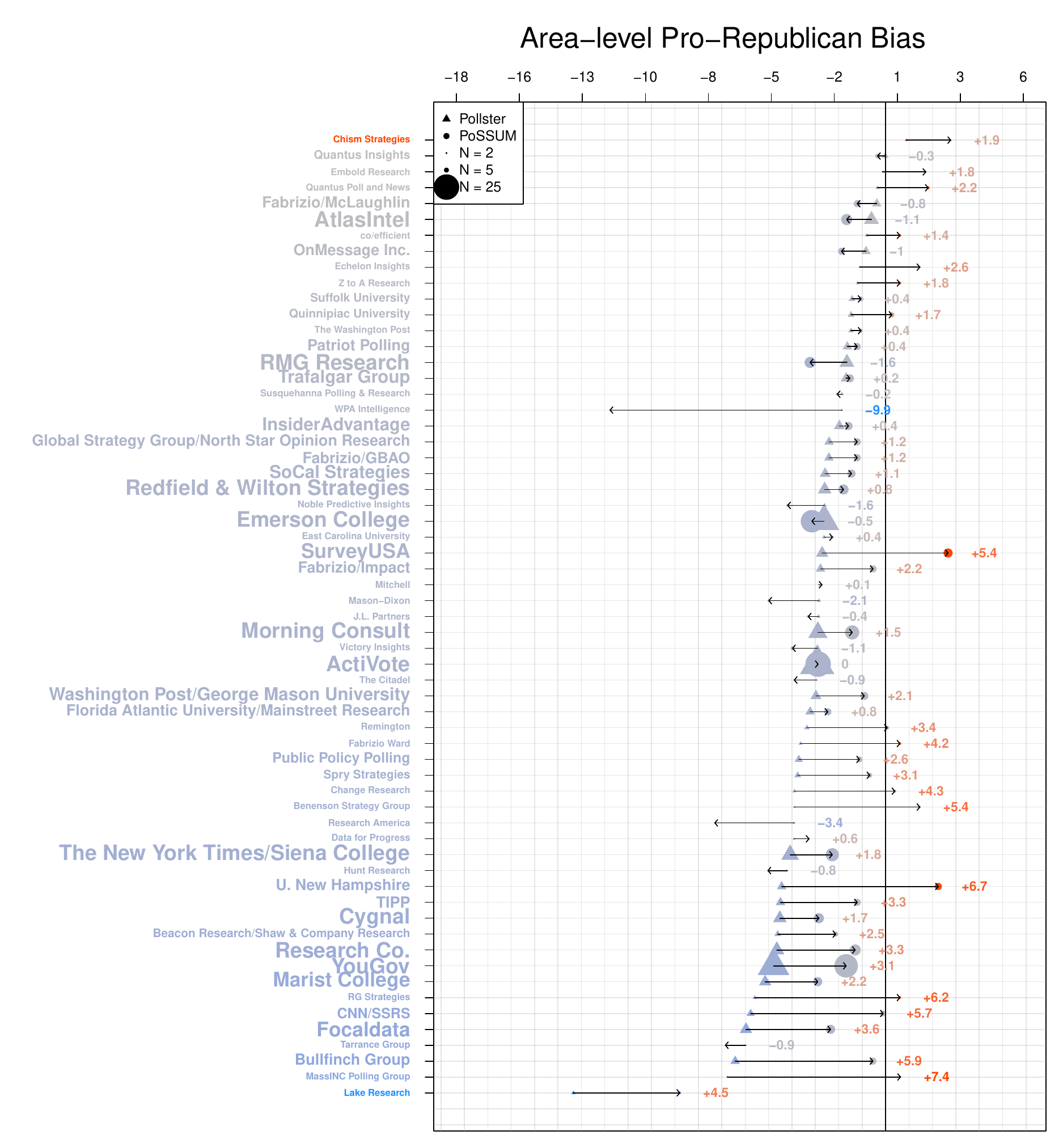}
    \caption{Comparison of Pro-Trump Bias on the area-level Republican margin for each reference pollster ($\blacktriangle$) v. \texttt{PoSSUM} ($\large\bullet$). Arrow length reflects the difference between estimates, and arrowheads point toward \texttt{PoSSUM}. Pollsters are listed from highest (top) to lowest (bottom) pro-Trump Bias. The blue–red color scale indicates lower-higher pro-Trump Bias relative to the average. \texttt{PoSSUM}'s Bias difference ($\Delta$) is displayed to the right of each comparison: red if \texttt{PoSSUM} favours the Republican more than the reference pollster, blue vice versa. Symbol and label sizes are proportional to the number of areas compared. Only pollsters with data from more than one area are included.}
    \label{fig:Bias_comparison}
\end{figure}

\begin{figure}[htbp]
    \centering
    \includegraphics[width=\linewidth]{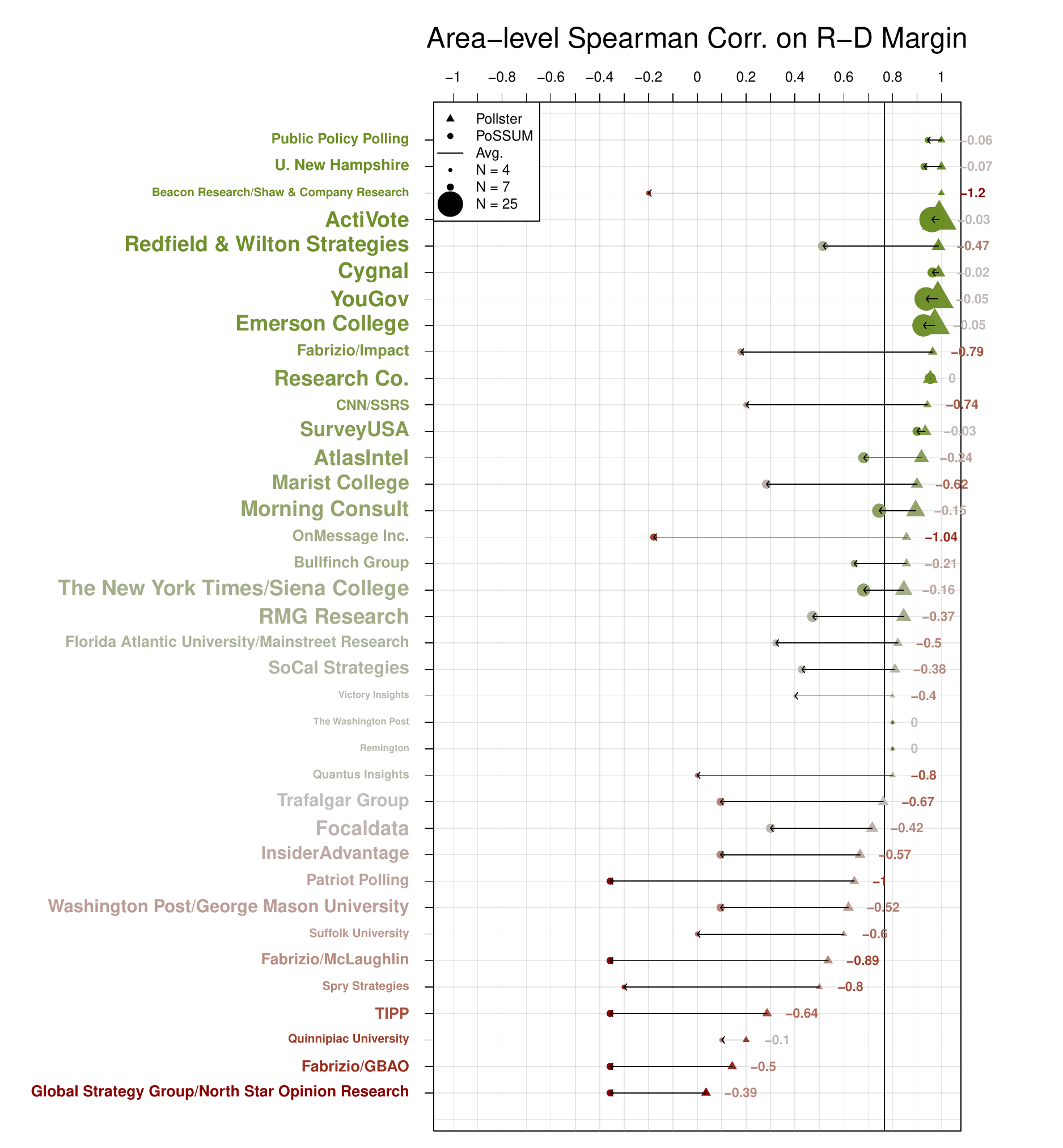}
    \caption{Comparison of Spearman Rank Correlation on the area-level Republican margin for each reference pollster ($\blacktriangle$) v. \texttt{PoSSUM} ($\large\bullet$). Arrow length reflects the difference between estimates, and arrowheads point toward \texttt{PoSSUM}. Pollsters are listed from highest (top) to lowest (bottom) Rank Correlation Coefficient. The red–green color scale indicates worse–better performance relative to the average. \texttt{PoSSUM}'s Rank Correlation difference ($\Delta$) is displayed to the right of each comparison: green if \texttt{PoSSUM}'s Rank Correlation is greater than the reference pollster, and red if smaller. Symbol and label sizes are proportional to the number of areas compared. Only pollsters with data from more than $3$ areas are included, as Rank Correlation comparisons tend to be unstable below that number.}
    \label{fig:Spearman_comparison}
\end{figure}

\begin{figure}[htbp]
    \centering
    \includegraphics[width=\linewidth]{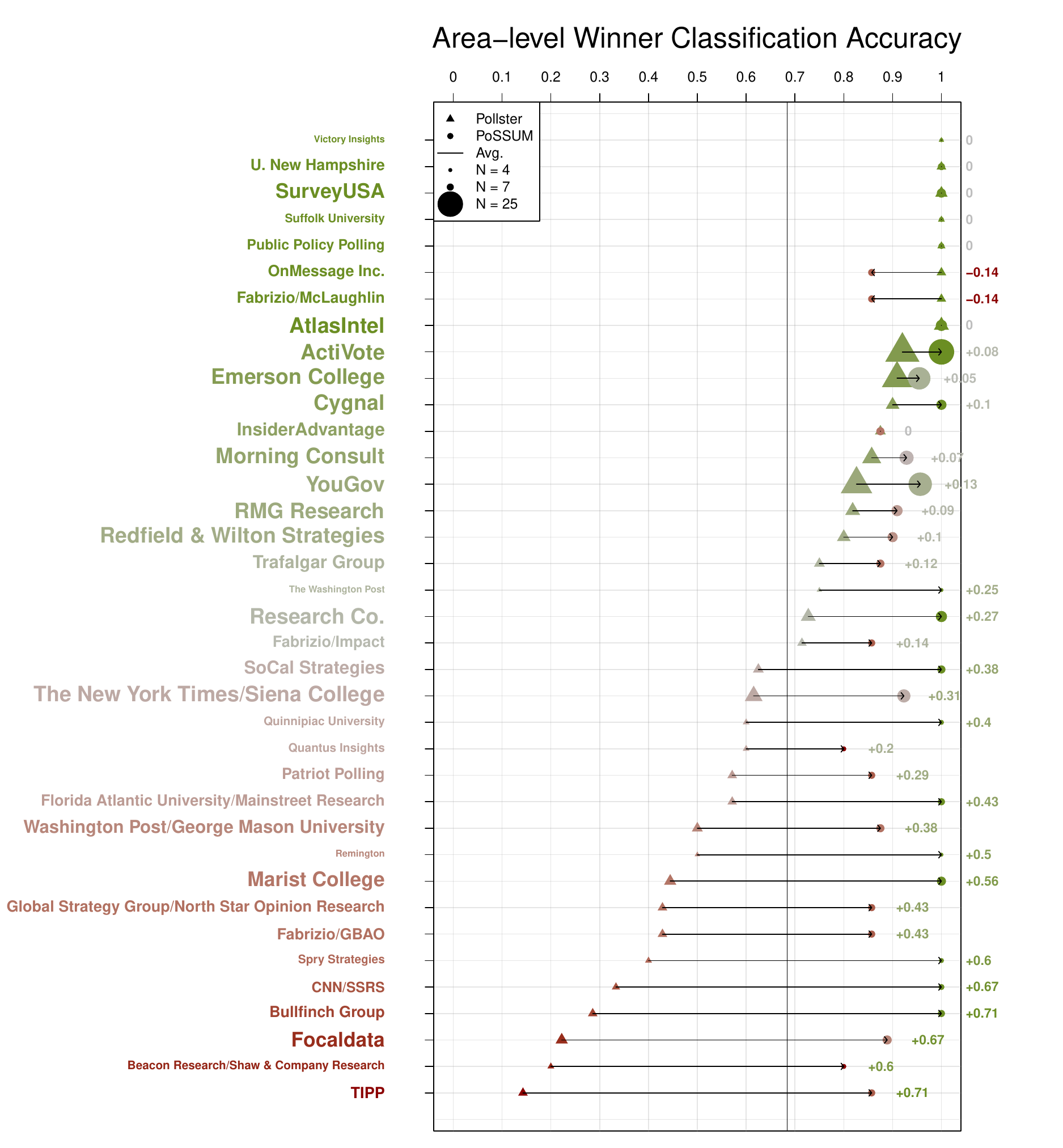}
    \caption{Comparison of Accuracy on winner prediction at the area-level for each reference pollster ($\blacktriangle$) v. \texttt{PoSSUM} ($\large\bullet$). Arrow length reflects the difference between estimates, and arrowheads point toward \texttt{PoSSUM}. Pollsters are listed from highest (top) to lowest (bottom) Accuracy. The red–green color scale indicates worse–better performance relative to the average. \texttt{PoSSUM}'s Accuracy difference ($\Delta$) is displayed to the right of each comparison: green if \texttt{PoSSUM}'s Accuracy is greater than the reference pollster’s, and red if lower. Symbol and label sizes are proportional to the number of areas compared. Only pollsters with data from more than $3$ areas are included, as Accuracy comparisons to be unstable below that number.}
    \label{fig:accuracy_comparison}
\end{figure}

\begin{figure}[htbp]
    \centering
    \includegraphics[width=\linewidth]{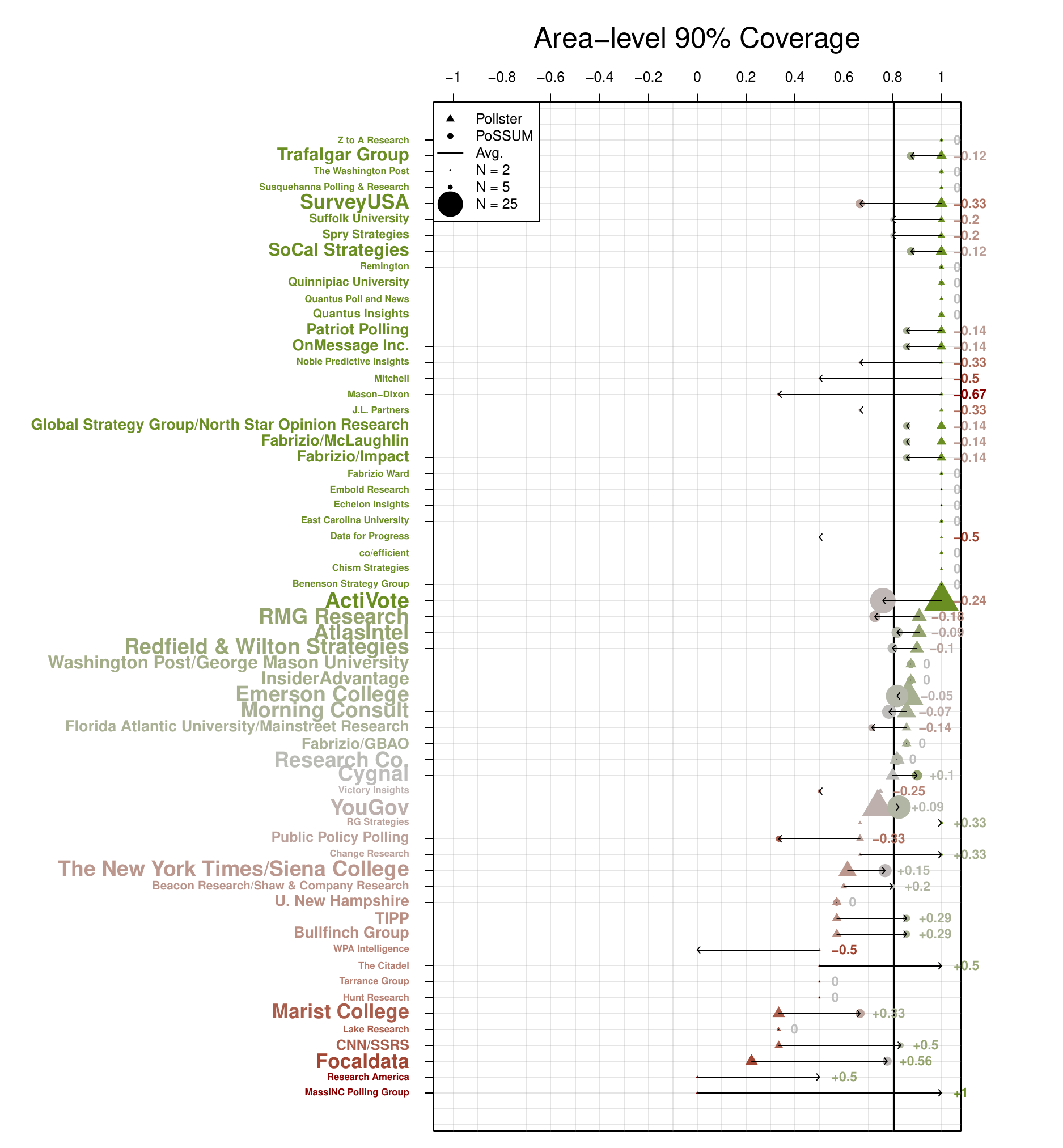}
    \caption{Comparison of Coverage ($90\%$) on the area-level Republican margin for each reference pollster ($\blacktriangle$) v. \texttt{PoSSUM} ($\large\bullet$). Arrow length reflects the difference between estimates, and arrowheads point toward \texttt{PoSSUM}. Pollsters are listed from highest (top) to lowest (bottom) Coverage. Stated Coverage is $90\%$, though over-coverage is not penalised in this comparison. The red–green color scale indicates worse–better performance relative to the average. \texttt{PoSSUM}'s Coverage difference ($\Delta$) is displayed to the right of each comparison: green if \texttt{PoSSUM}'s Coverage is greater than the reference pollster, and red if lower. Symbol and label sizes are proportional to the number of areas compared. Only pollsters with data from more than one area are included.}
    \label{fig:coverage_comparison}
\end{figure}

\begin{figure}[htbp]
    \centering
    \includegraphics[width=\linewidth]{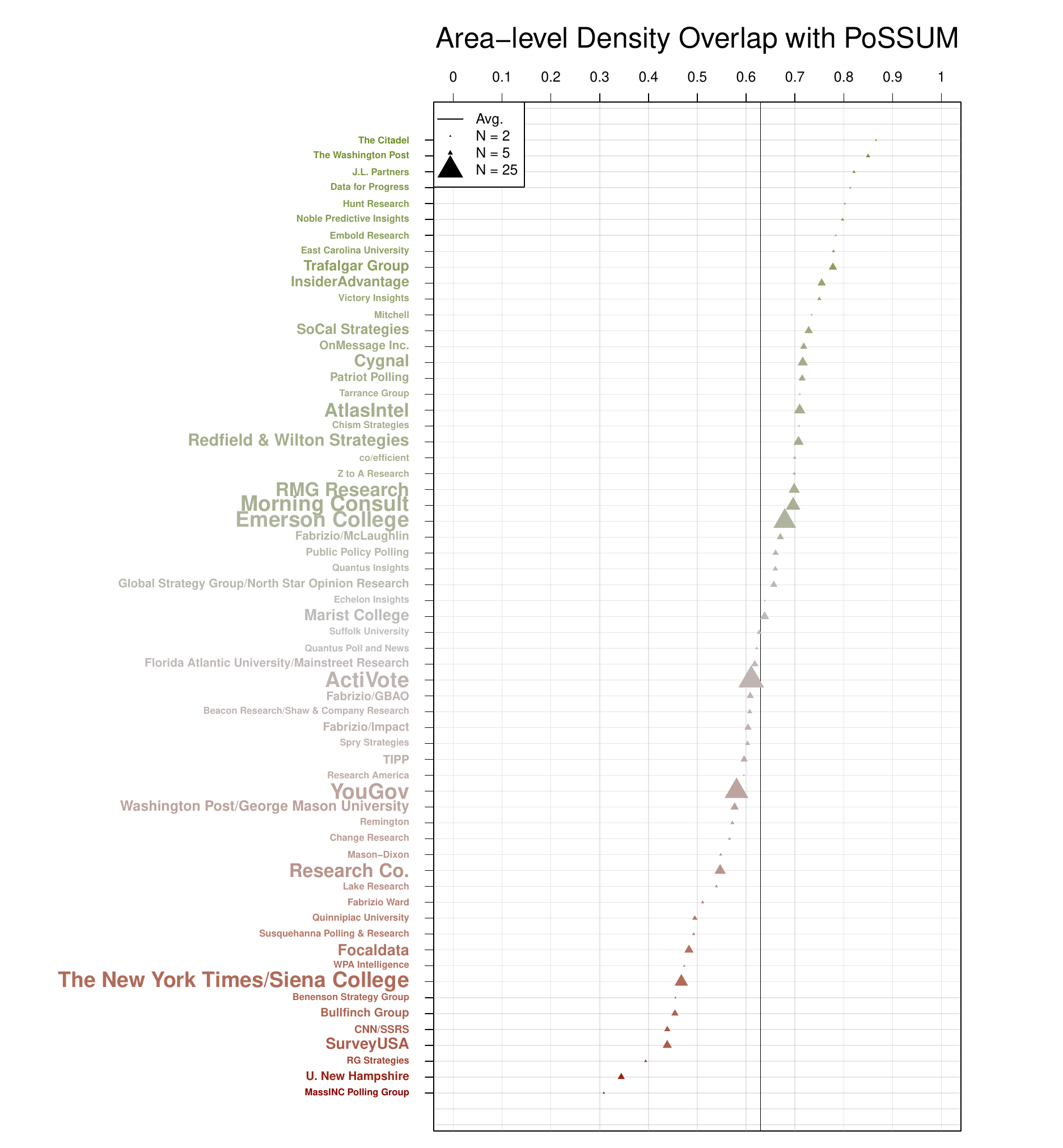}
    \caption{Density Overlap Coefficient (OVL) on the area-level Republican margin between each reference pollster and \texttt{PoSSUM}. The red–green color scale indicates lower-higher coverage relative to the average. Symbol and label sizes are proportional to the number of areas compared. Only pollsters with data from more than one area are included.}
    \label{fig:OVL_comparison}
\end{figure}
\pagebreak

\FloatBarrier

\newgeometry{left=0.5cm, right=0.5cm, top=2cm, bottom=2cm} 

\subsection{Pollster's Temporal Coverage by Crosstab}\label{sec:temporal_coverage_pollsters}

\begin{figure}[htbp]
    \centering
    \includegraphics[width=\linewidth]{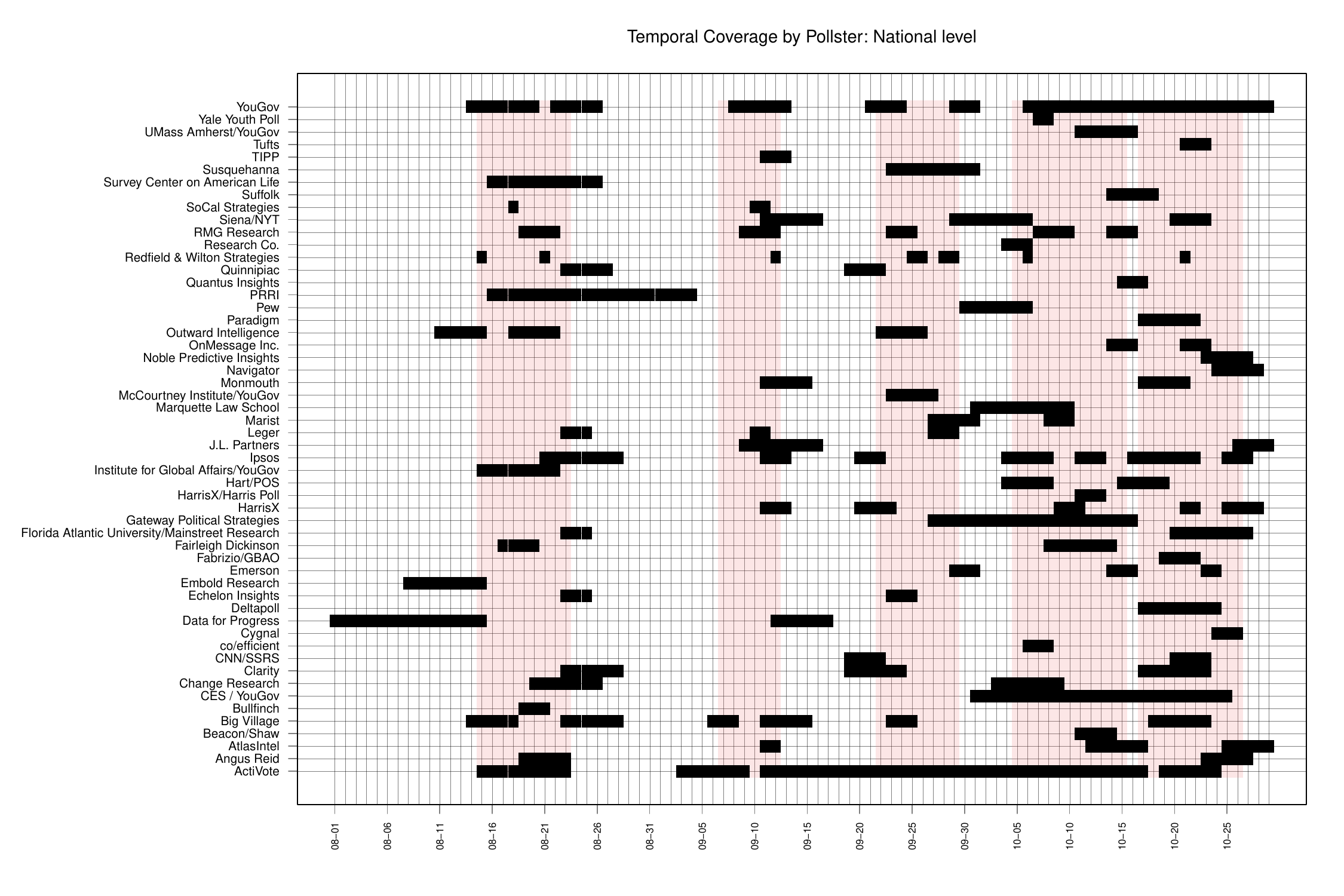}
    \caption{Temporal coverage of national-level reference pollsters. The light red shading marks the digital fieldwork period for \texttt{PoSSUM}, during which these polls were combined into the average used to evaluate \texttt{PoSSUM}'s results.}
    \label{fig:poll_coverage_nat}
\end{figure}

\begin{figure}[htbp]
    \centering
    \includegraphics[width=\linewidth]{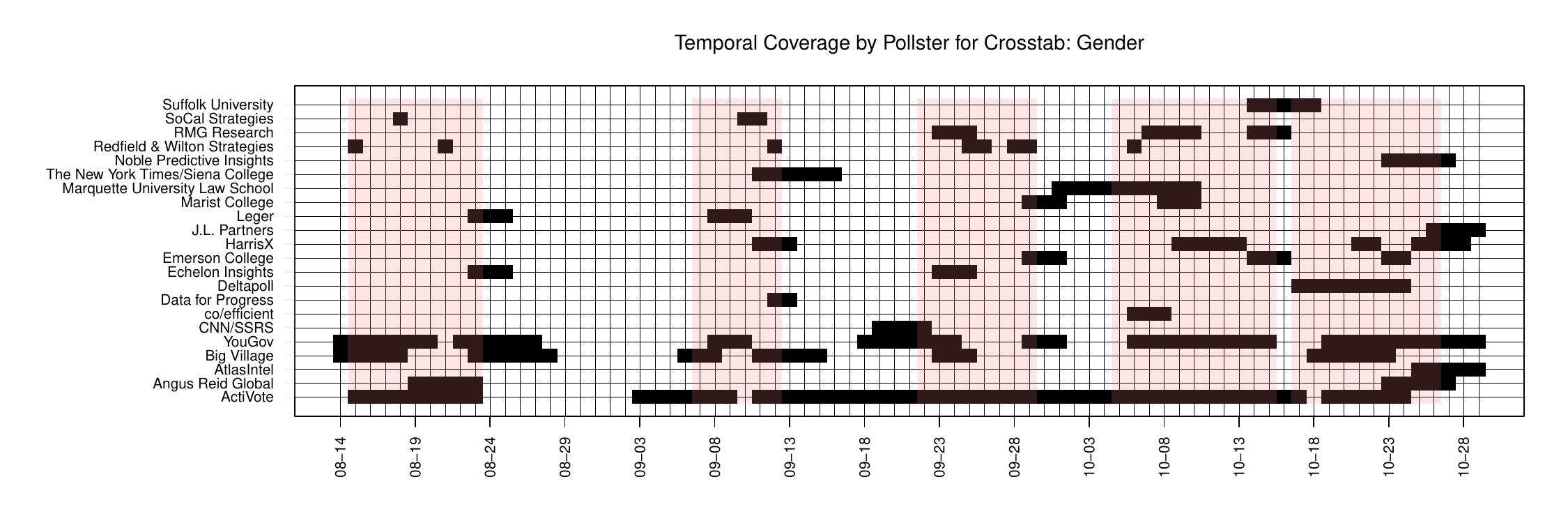}
    \caption{Temporal coverage of reference pollsters for gender-level crosstabs. The light red shading marks the digital fieldwork period for \texttt{PoSSUM}, during which these polls were combined into the average used to evaluate \texttt{PoSSUM}'s results.}
    \label{fig:poll_coverage_gender}
\end{figure}

\begin{figure}[htbp]
    \centering
    \includegraphics[width=\linewidth]{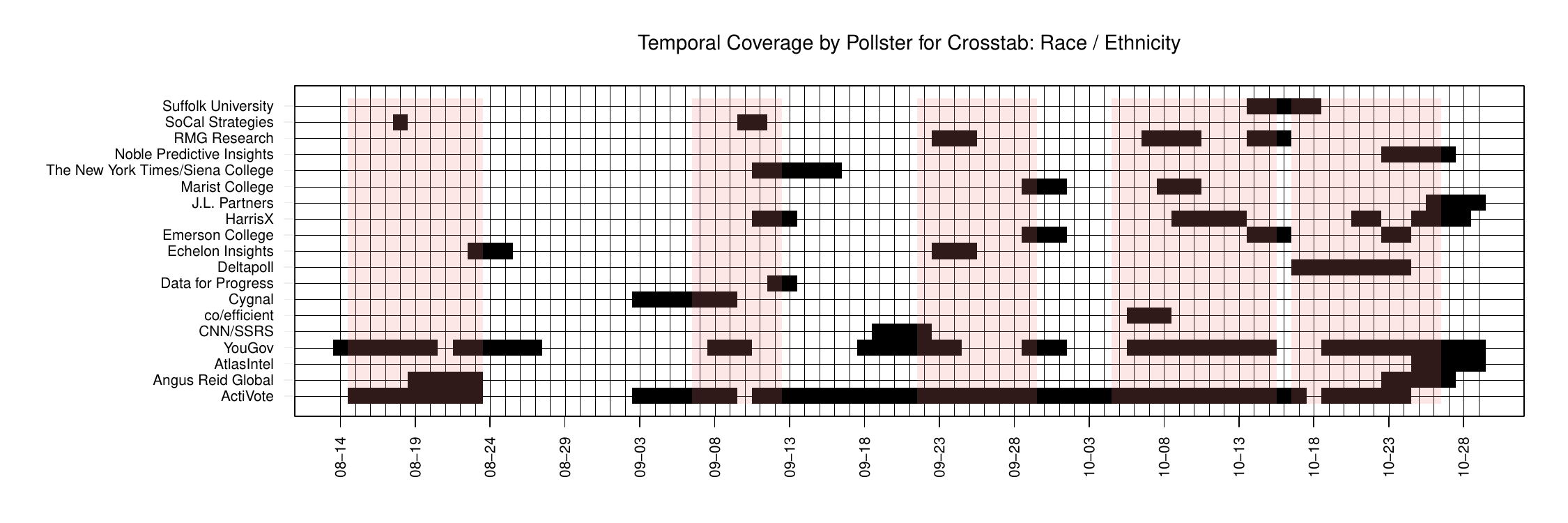}
    \caption{Temporal coverage of reference pollsters for race / ethnicity -level crosstabs. The light red shading marks the digital fieldwork period for \texttt{PoSSUM}, during which these polls were combined into the average used to evaluate \texttt{PoSSUM}'s results.}
    \label{fig:poll_coverage_race}
\end{figure}

\begin{figure}[htbp]
    \centering
    \includegraphics[width=\linewidth]{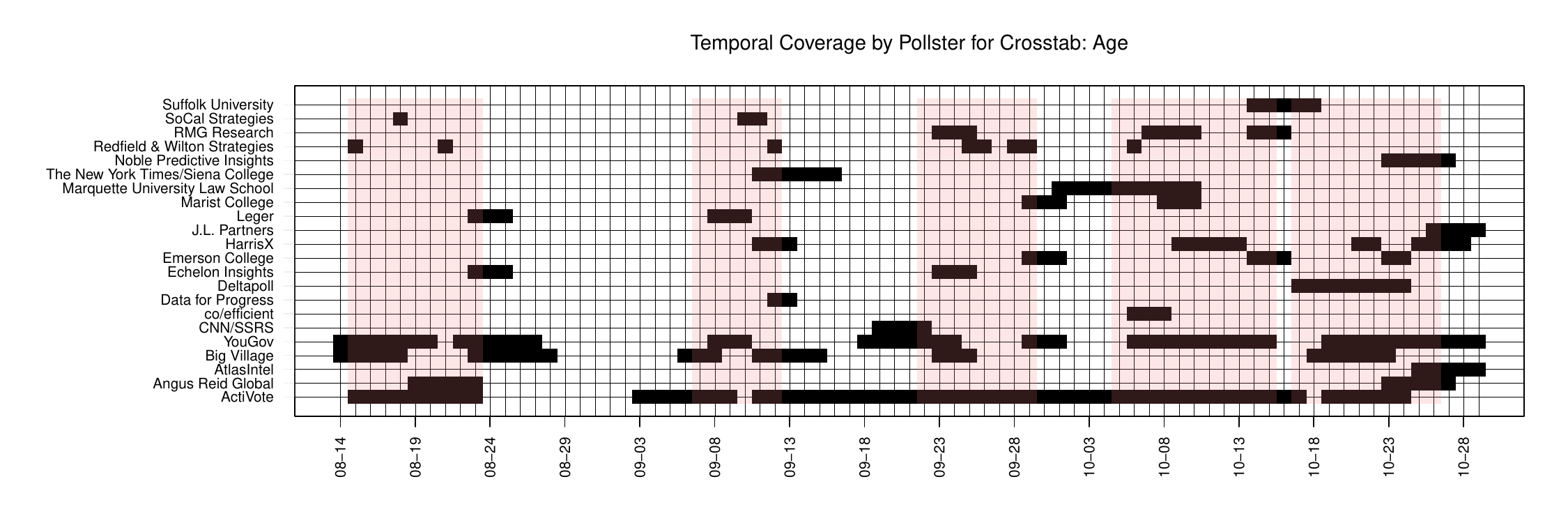}
    \caption{Temporal coverage of reference pollsters for age-group-level crosstabs. The light red shading marks the digital fieldwork period for \texttt{PoSSUM}, during which these polls were combined into the average used to evaluate \texttt{PoSSUM}'s results.}
    \label{fig:poll_coverage_age}
\end{figure}

\begin{figure}[htbp]
    \centering
    \includegraphics[width=\linewidth]{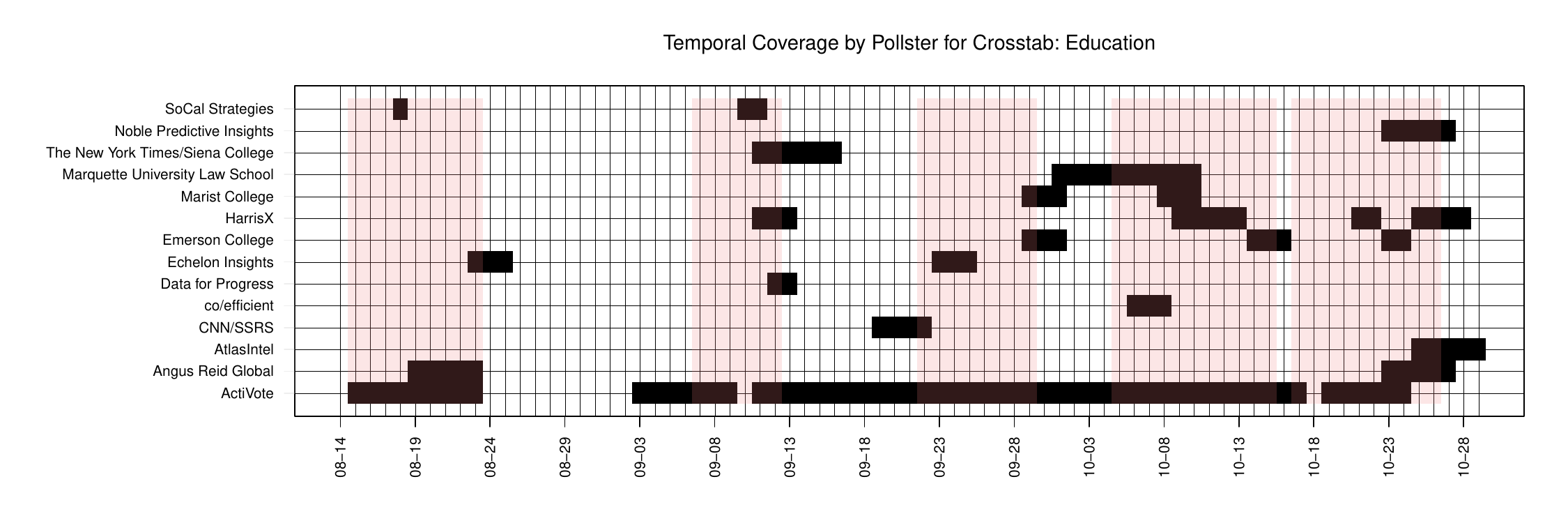}
    \caption{Temporal coverage of reference pollsters for education-level crosstabs. The light red shading marks the digital fieldwork period for \texttt{PoSSUM}, during which these polls were combined into the average used to evaluate \texttt{PoSSUM}'s results.}
    \label{fig:poll_coverage_education}
\end{figure}

\begin{figure}[htbp]
    \centering
    \includegraphics[width=\linewidth]{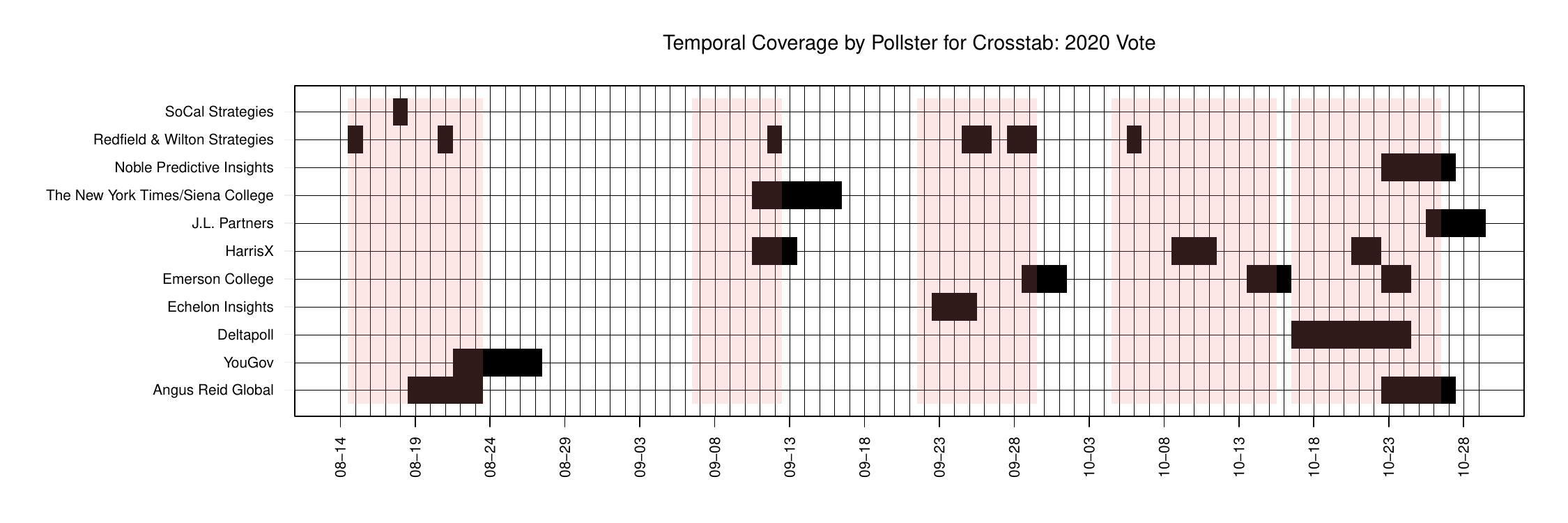}
    \caption{Temporal coverage of reference pollsters for $2020$ vote-level crosstabs. The light red shading marks the digital fieldwork period for \texttt{PoSSUM}, during which these polls were combined into the average used to evaluate \texttt{PoSSUM}'s results.}
    \label{fig:poll_coverage_vote}
\end{figure}

\restoregeometry

\pagebreak

\FloatBarrier

\subsection{Novel Learning, Human Alignment and Time-Sensitivity}

\begin{figure}[htpb]
    \centering
    \includegraphics[width=\linewidth]{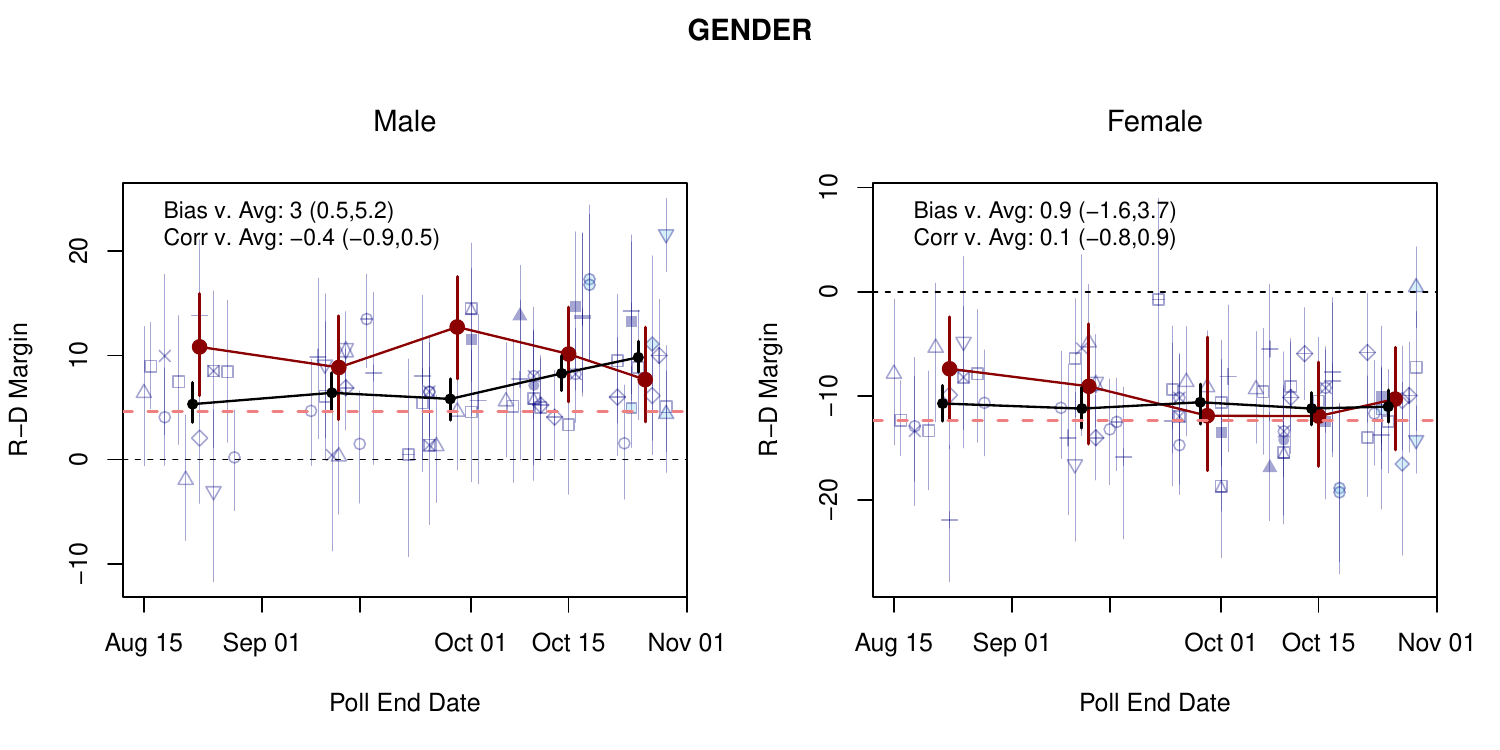}
    \caption{Gender-level \texttt{PoSSUM} estimates over the course of the campaign, shown alongside individual polls overlapping \texttt{PoSSUM}'s fieldwork periods, the aggregated polling average for each \texttt{PoSSUM} fieldwork window, and the reference preferences from $2020$.}
    \label{fig:dynamic_gender_comparison}
\end{figure}

\begin{figure}[htbp]
    \centering
    \includegraphics[width=\linewidth]{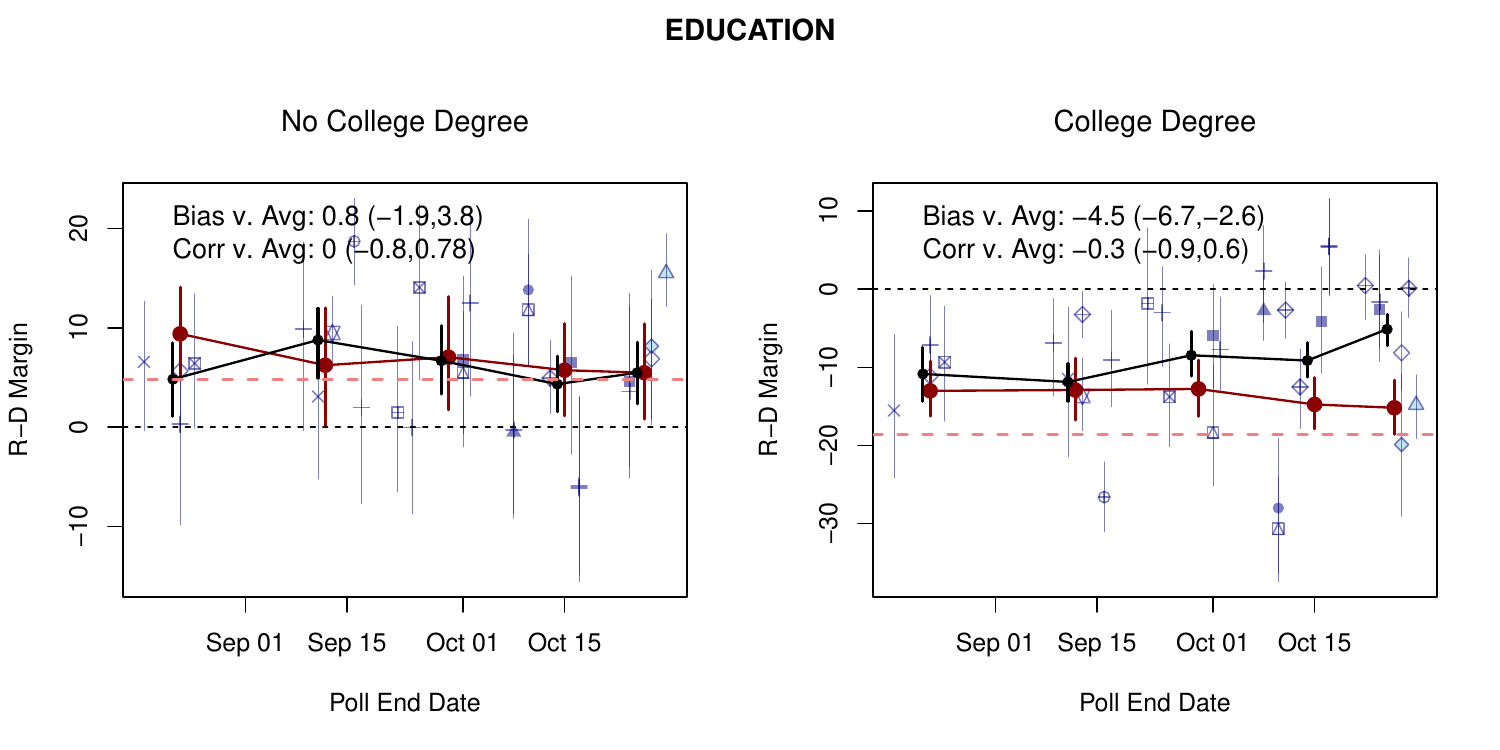}
    \caption{Education-level \texttt{PoSSUM} estimates over the course of the campaign, shown alongside individual polls overlapping \texttt{PoSSUM}'s fieldwork periods, the aggregated polling average for each \texttt{PoSSUM} fieldwork window, and the reference preferences from $2020$.}
    \label{fig:dynamic_edu_comparison}
\end{figure}

\begin{figure}[htbp]
    \centering
    \includegraphics[width=\linewidth]{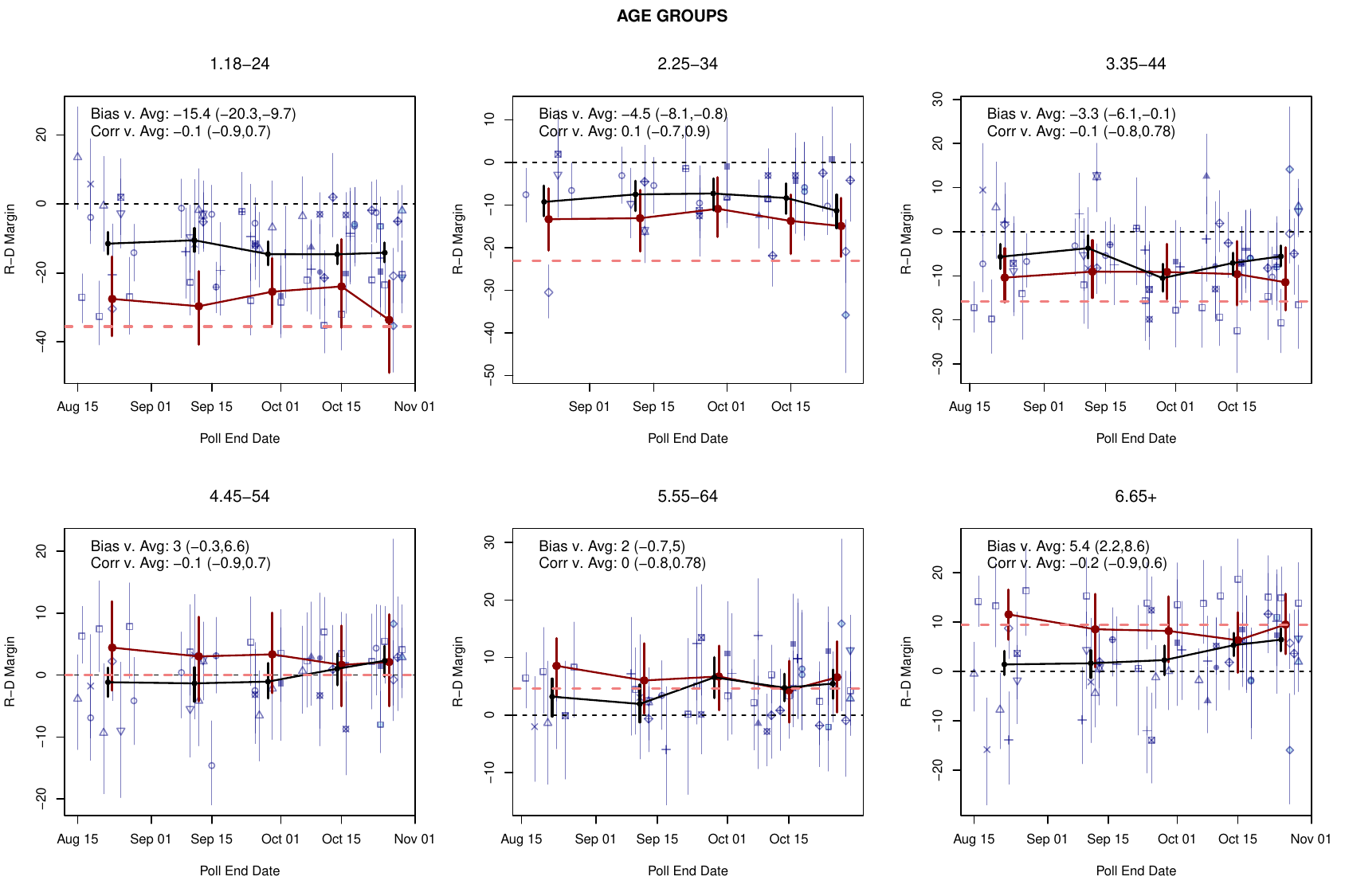}
    \caption{Age-group-level \texttt{PoSSUM} estimates over the course of the campaign, shown alongside individual polls overlapping \texttt{PoSSUM}'s fieldwork periods, the aggregated polling average for each \texttt{PoSSUM} fieldwork window, and the reference preferences from $2020$.}
    \label{fig:dynamic_age_comparison}
\end{figure}

\end{spacing}{}

\end{document}